\documentclass[PAPER, USenglish, cernpreprint, texmf, orcidlogo]{atlasdoc}
\usepackage{atlaspackage}
\usepackage{atlasparticle}
\usepackage{atlasjetetmiss}
\usepackage{atlasbiblatex}

\usepackage{atlasphysics}
\graphicspath{{logos/}{figures/}}

\usepackage{ANA-HMBS-2024-26-PAPER-defs}
\AtlasTitle{Search for decays of the Higgs boson into scalar particles decaying into four or six $b$-quarks using $pp$ collisions at $\sqrt{s}= 13\,\mathrm{TeV}$ with the ATLAS detector}

\AtlasAbstract{%
A search for exotic decays of the Higgs boson $H$ into new scalar or pseudoscalar particles that subsequently decay into $b$-quarks is presented. The search considers $ZH$ production with several decay scenarios for the Higgs boson: first to a pair of identical scalars, $H\rightarrow 2a \rightarrow 4b$, second to a pair of scalars with different masses ($m_{a1}<m_{a2}$), either directly, $H\rightarrow a_1a_2 \rightarrow 4b$, or via a longer decay chain, $H \rightarrow a_1a_2 \rightarrow 3a_1 \rightarrow 6b$. The analysis uses proton--proton collision data at $\sqrt{s} = 13$ TeV collected with the ATLAS detector at the Large Hadron Collider, corresponding to an integrated luminosity of $140~\mathrm{fb}^{-1}$. No significant excess above the Standard Model prediction is observed. The search sets upper limits at 95\% confidence level on the ratio of the Higgs boson production cross-section to the SM prediction times the branching ratio of Higgs bosons decaying into $4b$ or $6b$, between 4\% and 25\% for $\sigma(ZH)/\sigma_{\text{SM}}(ZH) \times \mathcal{B}$($H \rightarrow 2a \rightarrow 4b$), between 24\% and 38\% for $\sigma(ZH)/\sigma_{\text{SM}}(ZH) \times \mathcal{B}$($H \rightarrow a_1a_2 \rightarrow 4b$), and between 10\% and 20\% for $\sigma(ZH)/\sigma_{\text{SM}}(ZH) \times \mathcal{B}$($H \rightarrow a_1a_2 \rightarrow 3a_1 \rightarrow 6b$), depending on the masses of the scalar particles.
}

\AtlasRefCode{HMBS-2024-26}

\PreprintIdNumber{CERN-EP-2025-121}

\AtlasJournalRef{\PRD 112 (2025) 072005}
\AtlasDOI{10.1103/mzld-ldlt}

\AtlasCoverEgroupAnalysisTeam{atlas-hmbs-2024-26-analysis-team@cern.ch}

\hypersetup{pdftitle={ATLAS document},pdfauthor={The ATLAS Collaboration}}

\begin{document}

\renewcommand{\arraystretch}{1.1}

\maketitle

\tableofcontents

\section{Introduction}

Since the observation of the Higgs boson with mass $m_H$ near $125~\GeV$ by the ATLAS and CMS Collaborations~\cite{HIGG-2012-27,CMS-HIG-12-028}, a comprehensive program of measurements is being pursued to characterize the properties of this particle. The branching fraction of the Higgs boson into undetected beyond the Standard Model (BSM) particles are constrained by analyses of Higgs boson properties to approximately $\mathcal{B}_\text{u} \lesssim 12\%$~\cite{HIGG-2021-23,CMS-HIG-22-001}. Higgs boson decays are particularly sensitive to new physics due to the small total width ($\Gamma_H \approx 4~\MeV$). Even very small couplings to new particles can give sizable branching fractions~\cite{Curtin:2013fra} and can be compatible with available measurements of the Higgs boson couplings~\cite{HIGG-2021-23}.

Extensions of the Standard Model (SM) that include new light scalars, called $a$-bosons, can result in exotic Higgs boson decays $H \rightarrow 2a$. Such new light particles appear in theories with an extended Higgs sector~\cite{Dobrescu:2000yn,Ellwanger:2003jt,Dermisek:2005ar,Chang:2008cw,Morrissey:2008gm},
dark matter models~\cite{Silveira:1985rk,Pospelov:2007mp,Draper:2010ew,Ipek:2014gua,Martin:2014sxa},
models with a first-order electroweak phase transition~\cite{Kozaczuk:2019pet,Carena:2019une},
and theories of neutral naturalness~\cite{Craig:2015pha,Curtin:2015fna}.
Signatures of $H \rightarrow 2a$ can also arise in models with hidden-sector particles that are singlets under the SM gauge transformations~\cite{Curtin:2013fra,Robens:2015gla,Robens:2016xkb,Robens:2019kga,Bauer:2017ris}. In scenarios where the $a$-boson mixes with the Higgs boson and inherits its Yukawa couplings to fermions, decays of the $a$-boson into heavy fermions such as $b$-quarks are favored, and the process $H \rightarrow 2a \rightarrow (\bbbar)(\bbbar)$, shown in \Fig{\ref{fig:2a-4b}}, is expected to have a sizable branching fraction in the mass range $2 m_{b} < m_{a} < m_H/2$~\cite{Curtin:2013fra,Haisch:2018kqx}. Because of the scalar nature of the $a$-boson and the hadronization of the $b$-quarks, the results presented here do not depend on the parity assumed for the $a$-boson and can be interpreted in terms of Higgs boson decays into scalars or pseudoscalars.

Scenarios with multiple new hidden-sector particles, such as two real scalar singlets, referred to as $a_1$ and $a_2$ ($m_{a_1}<m_{a_2}$) have also been considered in the literature~\cite{Robens:2019kga}. In these scenarios, the Higgs boson can then decay into two different scalars, $H \rightarrow a_1 a_2 \rightarrow 4b$, shown in \Fig{\ref{fig:a1a2-4b}}. Another possibility is the successive decay into the lightest scalar state, $H \rightarrow a_1a_2 \rightarrow 3a_1 \rightarrow 6b$, which gives rise to a high multiplicity final state, shown in \Fig{\ref{fig:a1a2-6b}}. In models with two additional light scalars, the partial width of $a_2 \rightarrow a_1 a_1$ has to be much larger than any decay into SM fermions $a_2 \rightarrow f_{\mathrm{SM}} \bar{f}_{\mathrm{SM}}$ for the model to be consistent with measurements of Higgs boson couplings~\cite{Robens:2019kga}. As such, when $m_{a_1}<m_{a_2}/2$, only the decay $H\rightarrow a_1 a_2 \rightarrow 3 a_1$ is relevant. The search for models with multiple $a$-bosons can thus be effectively split into two independent channels: one with four $b$-quarks in the final state and one with six $b$-quarks in the final state, which are analyzed separately.

This paper presents a search for the exotic Higgs boson decay $H \rightarrow 2a/a_1a_2 \rightarrow 4b/6b$ over the mass range $12 < m_{a} < 60~\GeV$ using the full \RunTwo proton--proton ($pp$) collision data at $\sqrt{s}=13\,\TeV$ collected with the ATLAS detector at the Large Hadron Collider (LHC)~\cite{Evans:2008zzb}, corresponding to an integrated luminosity of 140~\ifb. The analysis targets the Higgs boson produced in association with a $Z$ boson ($ZH$), where the $Z$ boson decay provides a trigger and suppresses the $\ttbar$ backgrounds. The $WH$ associated production is not used since early results~\cite{HIGG-2017-05} show that this channel is less sensitive due to larger  backgrounds. Different analysis categories are defined depending on the decay mode of the $Z$ boson, which can be either into charged leptons (electrons $e$ or muons $\mu$), referred to as $2\ell$, or into neutrinos that leave a signature of transverse momentum imbalance (\met), referred to as $0\ell$. Due to the relatively low mass of the Higgs boson and the four- or six-body final state, the $b$-quarks tend to have low transverse momentum, typically below $\pt \lesssim 50\,\GeV$ and are challenging to trigger on. Therefore, the analysis uses the electron or muon triggers for the $2\ell$ category and uses \MET triggers for the $0\ell$ category. The analysis defines different event selections depending on the $Z$ boson decay mode: $2\ell$ or $0\ell$, and on the $b$-quark multiplicity: $4b$ or $6b$. The main backgrounds to the signal process are Drell--Yan production of charged leptons or neutrinos in association with heavy-flavor jets and $t\bar{t}$, which may also be produced in association with heavy-flavor jets.

Several novel reconstruction techniques are used to identify the $b$-hadron decay products. For $m_a \lesssim 20~\GeV$, the $a$-boson has a relatively large Lorentz boost and its decay products are collimated. Consequently, the $a\rightarrow \bbbar$ final state is reconstructed as a single jet that contains the hadronization products of the two $b$-quarks. While several techniques exist to resolve merged jets~\cite{Larkoski:2017jix}, most of them are only efficient in the case of high-mass and high-$\pt$ jets. The merged $a\to b\bar{b}$ jet from the decay products of a Higgs boson have low $\pt$. A dedicated algorithm is used to identify low-mass, merged, ``double $b$-quark'' jets ($B$-jets)~\cite{ATL-PHYS-PUB-2022-042}. For $m_a \gtrsim 25~\GeV$, the $b$-quark decays are typically well separated, so each reconstructed jet captures the hadronization of a single $b$-quark and standard algorithms are used to identify such jets ($b$-jets)~\cite{FTAG-2019-07,PERF-2015-09}. The sensitivity to high-multiplicity final states is also enhanced by the use of soft secondary vertices reconstructed outside of jets~\cite{FTAG-2023-02}. The search considers events with one or two $B$-jets, three or four $b$-jets, and up to one soft secondary vertex reconstructed outside of a jet, which are used to define several analysis categories.

\begin{figure}[!tbp]
\centering
\subfloat[]{\label{fig:2a-4b}
\includegraphics[width=0.32\linewidth]{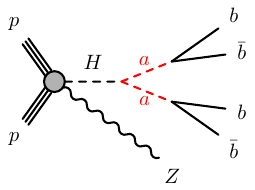}}%
\subfloat[]{\label{fig:a1a2-4b}
\includegraphics[width=0.32\linewidth]{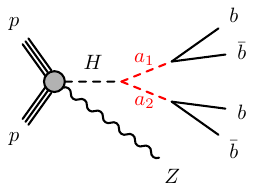}}%
\subfloat[]{\label{fig:a1a2-6b}
\includegraphics[width=0.32\linewidth]{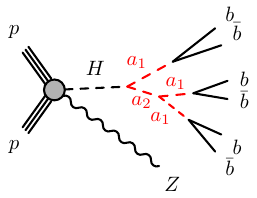}}\\
\caption{Representative Feynman diagrams for $ZH$ production with the \protect\subref{fig:2a-4b} $H \rightarrow 2a \rightarrow 4b$, \protect\subref{fig:a1a2-4b} $H \rightarrow a_1a_2 \rightarrow 4b$, and \protect\subref{fig:a1a2-6b} $H \rightarrow a_1a_2 \rightarrow 3 a_1 \rightarrow 6b$ processes considered in this search.}
\end{figure}

Similar searches in the $4b$ decay channel were performed by the ATLAS Collaboration~\cite{HIGG-2016-01,HIGG-2017-05,HDBS-2018-47} and the CMS Collaboration~\cite{CMS-HIG-18-026}. The latest search has placed 95\% confidence level (CL) upper limits on the branching fractions in the range of $1.1$ for $m_{a} = 20~\GeV$ to $0.36$ for $m_{a} = 60~\GeV$ using 138 \ifb\ of \RunTwo data at $\sqrt{s} = 13$~\TeV~\cite{CMS-HIG-18-026}. This result improves the sensitivity of previous analyses by using dedicated heavy-flavor reconstruction algorithms, particularly in the low mass regime ($m_a<20$~\GeV). It is also complementary to other searches for $H \rightarrow 2a$ decays performed by the ATLAS and CMS Collaborations using both $\sqrt{s} = 8~\TeV$ and $\sqrt{s} = 13~\TeV$ data in several final states including $\mu^+\mu^-\mu^+\mu^-$~\cite{EXOT-2013-15,EXOT-2016-22,HDBS-2018-55},
$\mu^+\mu^-\tau^+\tau^-$~\cite{HIGG-2014-02,CMS-HIG-17-029,CMS-HIG-18-006,CMS-HIG-18-024},
$\tau^+\tau^-\tau^+\tau^-$~\cite{CMS-HIG-14-019,CMS-HIG-18-006, HMBS-2024-25},
$b\bar{b}\mu^+\mu^-$~\cite{HIGG-2016-03,HDBS-2021-03,CMS-HIG-16-015,CMS-HIG-18-011}, $b\bar{b}\tau^+\tau^-$~\cite{CMS-HIG-17-024, CMS-HIG-22-007, HDBS-2021-07},
$\gamma\gamma\gamma\gamma$~\cite{EXOT-2013-24,CMS-HIG-21-003,CMS-HIG-21-016},
and $\gamma\gamma gg$~\cite{HIGG-2017-09}.

\section{ATLAS detector}

The ATLAS detector~\cite{PERF-2007-01} at the LHC covers nearly the entire solid angle around the collision point.\footnote{ATLAS uses a right-handed coordinate system with its origin at the nominal interaction point (IP)
in the center of the detector and the \(z\)-axis along the beam pipe.
The \(x\)-axis points from the IP to the center of the LHC ring,
and the \(y\)-axis points upwards.
Polar coordinates \((r,\phi)\) are used in the transverse plane,
\(\phi\) being the azimuthal angle around the \(z\)-axis.
The pseudorapidity is defined in terms of the polar angle \(\theta\) as \(\eta = -\ln \tan(\theta/2)\) and is equal to the rapidity
$ y = \frac{1}{2} \ln \left( \frac{E + p_z}{E - p_z} \right) $ in the relativistic limit.
Angular distance is measured in units of \(\Delta R \equiv \sqrt{(\Delta y)^{2} + (\Delta\phi)^{2}}\).}
It consists of an inner tracking detector surrounded by a thin superconducting solenoid, electromagnetic and hadronic calorimeters,
and a muon spectrometer incorporating three large superconducting air-core toroidal magnets.

The inner-detector system (ID) is immersed in a \qty{2}{\tesla} axial magnetic field
and provides charged-particle tracking in the range of \(|\eta| < 2.5\).
The high-granularity silicon pixel detector covers the vertex region and typically provides four measurements per track,
the first hit generally being in the insertable B-layer (IBL) installed before Run~2~\cite{ATLAS-TDR-19,PIX-2018-001}.
It is followed by the SemiConductor Tracker (SCT), which provides on average eight measurements per track.
These silicon detectors are complemented by the transition radiation tracker (TRT),
which enables radially extended track reconstruction up to \(|\eta| = 2.0\).
The TRT also provides electron identification information
based on the fraction of hits (typically 30 in total) above a higher energy-deposit threshold corresponding to transition radiation.

The calorimeter system covers the pseudorapidity range \(|\eta| < 4.9\).
Within the region \(|\eta|< 3.2\), electromagnetic calorimetry is provided by barrel and
endcap high-granularity lead/liquid-argon (LAr) calorimeters,
with an additional thin LAr presampler covering \(|\eta| < 1.8\)
to correct for energy loss in material upstream of the calorimeters.
Hadronic calorimetry is provided by the steel/scintillator-tile calorimeter,
segmented into three barrel structures within \(|\eta| < 1.7\), and two copper/LAr hadronic endcap calorimeters.
The solid angle coverage is completed with forward copper/LAr and tungsten/LAr calorimeter modules
optimized for electromagnetic and hadronic energy measurements, respectively.

The muon spectrometer (MS) comprises separate trigger and
high-precision tracking chambers measuring the deflection of muons in a magnetic field generated by the superconducting air-core toroidal magnets.
The field integral of the toroids ranges between \num{2.0} and \qty{6.0}{\tesla\metre}
across most of the detector.
Three layers of precision chambers, each consisting of layers of monitored drift tubes, cover the region \(|\eta| < 2.7\),
complemented by cathode-strip chambers in the forward region, where the background is highest.
The muon trigger system covers the range \(|\eta| < 2.4\) with resistive-plate chambers in the barrel, and thin-gap chambers in the endcap regions.

The luminosity is measured mainly by the LUCID--2~\cite{LUCID2} detector that records Cherenkov light produced in the quartz windows of photomultipliers located close to the beampipe.
Events are selected by the first-level trigger system implemented in custom hardware,
followed by selections made by algorithms implemented in software in the high-level trigger~\cite{TRIG-2016-01}.
The first-level trigger accepts events from the \qty{40}{\MHz} bunch crossings at a rate below \qty{100}{\kHz},
which the high-level trigger further reduces to record complete events to disk at about \qty{1}{\kHz}.
A software suite~\cite{SOFT-2022-02} is used in data simulation, in the reconstruction
and analysis of real and simulated data, in detector operations, and in the trigger and data acquisition
systems of the experiment.

\section{Data and simulation samples}

The analysis uses \pp\ collision data at a center-of-mass energy of $\sqrt{s} = 13$~\TeV collected with the ATLAS detector in \RunTwo of the LHC from 2015 to 2018,  corresponding to a total integrated luminosity of $140~\mathrm{fb}^{-1}$~\cite{DAPR-2021-01}. Only events for which the LHC beams were in stable-collision mode and all relevant subsystems were operational are used~\cite{DAPR-2018-01}.
Events with charged leptons are selected using single-lepton triggers with either a low \pt\ threshold and an isolation requirement, or a higher threshold but a looser identification criterion without any isolation requirement. For data collected in 2015 (2016--2018), the lowest \pt\ threshold in the single-muon trigger was 20 (26)~\GeV~\cite{TRIG-2018-01} and in the single-electron trigger was 24 (26)~\GeV~\cite{TRIG-2018-05}. Events without charged leptons are selected using \MET triggers whose threshold varied over time from 70 to 110~\GeV due to the increasing average effect of multiple inelastic interactions in the same and neighboring bunch crossings (\pileup).

Simulated signal and background samples are used to describe background sources, estimate signal efficiency and acceptance, design and optimize the analysis event selection, and model systematic uncertainties. Monte Carlo (MC) samples were produced using either the full ATLAS detector simulation~\cite{SOFT-2010-01} based on \GEANT~\cite{Agostinelli:2002hh}  or a faster simulation where the full \GEANT simulation of the calorimeter response is replaced by a parameterization~\cite{SOFT-2010-01}. The \pileup effects were modeled by overlaying each simulated hard-scattering event with inelastic $pp$ events generated with \PYTHIA[8.186]~\cite{Sjostrand:2007gs} using the \NNPDF[2.3lo] set of parton distribution functions (PDFs)~\cite{Ball:2012cx} and the A3 set of tuned parameters (tune)~\cite{ATL-PHYS-PUB-2016-017}. Simulated events are reweighted to match the \pileup conditions observed in the \RunTwo data sample. All simulated events are processed through the same reconstruction algorithms and analysis chain as the data. \Tab{\ref{tab:modeling_samples}} summarizes all the generated samples.

The signal event samples focus on Higgs bosons produced in association with a $Z$ boson ($ZH$). Two processes are included, $qq\rightarrow ZH$ and $gg\rightarrow ZH$, where the latter contributes about $10\%$ of the total production cross-section. The sample of Higgs bosons is generated at next-to-leading-order (NLO) accuracy in quantum chromodynamics (QCD) using \POWHEGBOX[v2]~\cite{Hamilton:2013fea,Hamilton:2015nsa,Alioli:2010xd,Nason:2004rx,Frixione:2007vw} with the \NNPDF[3.0nlo] PDF set ~\cite{Ball:2012cx, Ball:2014uwa} with mass $m_H = 125~\GeV$. The simulated Higgs boson production samples are normalized using dedicated higher-order cross-sections calculated at next-to-next-to-leading-order (NNLO) in QCD with NLO electroweak (EW) corrections~\cite{deFlorian:2016spz}. The Higgs boson is decayed into a pair of scalars that then each decay either into a pair of scalars or into a pair of $b$-quarks, where the decays are performed in the narrow-width approximation. Events are showered and hadronized with \PYTHIA[8.244] or \PYTHIA[8.245] for the $2\ell$ or $0\ell$ channels, respectively,~\cite{Sjostrand:2007gs} using the AZNLO tune to model the underlying event~\cite{STDM-2012-23}.

For the $0\ell$ channel, five signal samples are simulated for $H \rightarrow 2a \rightarrow 4b$ ($m_a = 12, 16, 20, 25,$~and~$30$~\GeV). For the $2\ell$ channel, signal simulation samples are produced for the same mass hypotheses and three more at higher masses ($m_a = 40, 50$,~and~$60$~\GeV). In addition, for the $2\ell$ case, three signal samples are produced for $H \rightarrow a_1a_2 \rightarrow 4b$, with mass hypotheses $(m_{a_1}, m_{a_2}) = (20,30), (40,60)$,~and~$(50,70)$~\GeV, and four for $H \rightarrow a_1a_2 \rightarrow 3 a_1 \rightarrow 6b$, with mass hypotheses $(m_{a_1}, m_{a_2}) = (15,50), (15,90), (20,70)$, and $(30,80)$~\GeV. The values of the mass hypotheses for the signal samples are chosen to cover the available parameter space given the mass resolution, which varies in the range 15--20\,\GeV\ depending on the properties of the final-state particles.

The main SM backgrounds are, in order of importance, Drell--Yan production with decays into charged leptons or neutrinos $Z/\gamma^{*} \rightarrow \ell\ell/\nu\nu$ produced in association with jets (\Zjets), and top-quark production ($\ttbar$ or single top quarks) where at least one of the $W$ bosons decays leptonically. Other backgrounds from diboson production and vector bosons produced in association with $\ttbar$ are also included. Processes such as triboson and four-top production are considered but found to be very small, and contributions from SM Higgs boson decays such as $H \rightarrow ZZ^* \rightarrow 4b$ are found to be negligible. For the $0\ell$ case, there is an additional background from events with multiple jets, referred to as QCD multijet background, which is estimated using a data-driven method.

The precision of the matrix element (ME) generators is NLO in QCD for most samples. All samples using \PYTHIA[8] for the parton shower (PS), hadronization, and multi-parton interactions (MPI) simulate the decays of $b$- and $c$-hadrons with \EVTGEN[1.6.0] and use the A14 tune~\cite{ATL-PHYS-PUB-2014-021} and the \NNPDF[2.3lo] PDF set.
Several simulation samples were produced with the \SHERPA[2.2] generator~\cite{Bothmann:2019yzt}. In this setup, NLO-accurate MEs for up to two partons and MEs with leading-order (LO) accuracy for up to four partons were calculated with the \textsc{Comix}~\cite{Gleisberg:2008fv} and \OPENLOOPS libraries~\cite{Cascioli:2011va}. They were matched with the \SHERPA parton shower~\cite{Schumann:2007mg} by using the \MEPSatNLO prescription~\cite{Hoeche:2011fd,Hoeche:2012yf,Catani:2001cc,Hoeche:2009rj} with the set of tuned parameters developed by the \SHERPA authors and based on the \NNPDF[3.0nnlo] set of PDFs~\cite{Ball:2014uwa}. The top-quark mass was set to $m_t = 172.5$~\GeV. Additional samples with alternative generators, parton shower models, and other model parameters, such as matching and resummation scales, were used to derive modeling uncertainties, as described in \Sect{\ref{sec:systematics}}. %

\begin{table}[tp]
\renewcommand{\arraystretch}{1.2}
\setlength{\tabcolsep}{3pt}
\centering
\caption{Sample generator setups.
The precision of the ME generator is NLO in QCD if no additional information is provided in parentheses, such as the number of partons at ME accuracy and [EW] for electroweak production. To treat the overlap between \ttbar and \tW diagrams, the sample using the diagram removal scheme is listed as [DR]. The higher-order cross-section used to normalize these samples is listed in the last column if different from the one of the generator. If no information is in this column, no higher-order k-factor is applied to this process. If multiple samples are listed for the same process, the first one corresponds to the one used for the predictions, while the subsequent ones are used to derive systematic uncertainties.
}
\footnotesize
\begin{tabular}{ccccc}
\toprule \midrule
\textbf{Process} & \textbf{ME generator} 		& \textbf{ME PDF} 					&	 \textbf{PS} 			& \textbf{Normalization}   		\\ \midrule
\multicolumn{5}{l}{Signal} \\
\midrule
$H \rightarrow 4b/6b$  & \POWHEGBOX[v2] 	& \NNPDF[3.0nlo] 				& \PYTHIA[8.244/5]		& NNLO+NLO (EW)~\cite{deFlorian:2016spz} \\
& \POWHEGBOX[v2] 	& \NNPDF[3.0nlo] 				& \HERWIG[7.04]	& NNLO+NLO (EW)~\cite{deFlorian:2016spz}			\\
\midrule
\multicolumn{5}{l}{Top} \\
\midrule
\ttbar	& \POWHEGBOX[v2] 	& \NNPDF[3.0nlo] 				& \PYTHIA[8.230]		& NNLO+NNLL~\cite{Beneke:2011mq,Cacciari:2011hy,Baernreuther:2012ws,Czakon:2012zr,Czakon:2012pz,Czakon:2013goa,Czakon:2011xx}			\\
& \POWHEGBOX[v2] 	& \NNPDF[3.0nlo] 				& \HERWIG[7.13]	& NNLO+NNLL~\cite{Beneke:2011mq,Cacciari:2011hy,Baernreuther:2012ws,Czakon:2012zr,Czakon:2012pz,Czakon:2013goa,Czakon:2011xx}		   	\\
& \MGNLO[2.6.0] 	& \NNPDF[3.0nlo] 				& \PYTHIA[8.230]		& NNLO+NNLL~\cite{Beneke:2011mq,Cacciari:2011hy,Baernreuther:2012ws,Czakon:2012zr,Czakon:2012pz,Czakon:2013goa,Czakon:2011xx}			\\
\ttbb 	& \POWHEGBOXRES 	& \NNPDF[3.1nnlo]nf4 	& \PYTHIA[8.230]		& --						\\
\tW 	& \POWHEGBOX[v2][DR]	& \NNPDF[3.0nlo] 			& \PYTHIA[8.230]		& NLO+NNLL
~\cite{Kidonakis:2010ux} 				   	\\
t-channel 	& \POWHEGBOX[v2] 	& \NNPDF[3.0nlo]nf4		& \PYTHIA[8.230]		& NLO~\cite{Aliev:2010zk,Kant:2014oha} 						\\
s-channel 	& \POWHEGBOX[v2] 	& \NNPDF[3.0nlo] 				& \PYTHIA[8.230]		& NLO~\cite{Aliev:2010zk,Kant:2014oha} 						\\
\midrule
\multicolumn{5}{l}{$Z/W$+jets} \\ \midrule
\Zjets 	& \SHERPA[2.2.11] (NLO [2j], LO [5j])	& \NNPDF[3.1nnlo]	& \SHERPA		& NNLO~\cite{Anastasiou:2003ds}						\\
& \MGNLO[2.6.5] (NLO [3j])	& \NNPDF[3.0nlo]	& \PYTHIA[8.240]		& NNLO~\cite{Anastasiou:2003ds}						\\
\Wjets 	& \SHERPA[2.2.11] (NLO [2j], LO [4j]) 	& \NNPDF[3.1nnlo] 	& \SHERPA		& NNLO~\cite{Anastasiou:2003ds}						\\
& \MGNLO[2.6.5] (NLO [3j])	& \NNPDF[3.0nlo]	& \PYTHIA[8.240]		& NNLO~\cite{Anastasiou:2003ds}						\\
\midrule
\multicolumn{5}{l}{Other} \\ \midrule
$VV$ (had.)  		& \SHERPA[2.2.1] 				& \NNPDF[3.0nnlo] 	& \SHERPA		& --						\\
$VV$ (lep.)  	 	& \SHERPA[2.2.2] 				& \NNPDF[3.0nnlo] 	& \SHERPA		& --						\\
$VV$ (lep.) + jj	& \SHERPA[2.2.2] (LO [EW])		& \NNPDF[3.0nnlo] 	& \SHERPA		& --						\\
\ttW 	& \MGNLO[2.3.3] 	& \NNPDF[3.0nlo] 				& \PYTHIA[8.210]		& NLO+NLO (EW)~\cite{deFlorian:2016spz}						\\
& \SHERPA[2.0.0] (LO [2j]) 	& \NNPDF[3.0nnlo] 	& \SHERPA					& NLO+NLO (EW)~\cite{deFlorian:2016spz}						\\
\ttbar{}$\ell\ell$ 	& \MGNLO[2.3.3] 	& \NNPDF[3.0nlo] 	& \PYTHIA[8.210]		& NLO+NLO (EW)~\cite{deFlorian:2016spz}						\\
& \SHERPA[2.0.0] (LO [1j]) 	& \NNPDF[3.0nnlo] 	& \SHERPA					& NLO+NLO (EW)~\cite{deFlorian:2016spz}						\\
\ttZ($qq,\nu\nu$) 	& \MGNLO[2.3.3] 	& \NNPDF[3.0nlo] 	& \PYTHIA[8.210]		& NLO+NLO (EW)~\cite{deFlorian:2016spz}						\\
& \SHERPA[2.0.0] (LO [2j]) 	& \NNPDF[3.0nnlo] 	& \SHERPA					& NLO+NLO (EW)~\cite{deFlorian:2016spz}						\\
\midrule \bottomrule
\end{tabular}

\label{tab:modeling_samples}
\end{table}

\paragraph{Generator-level categorization of simulated events}

Using a procedure similar to that described in Ref~\cite{HIGG-2020-23}, \ttbar+jets and \Zjets events are categorized depending on the flavor of associated jets. Generator-level particle jets are reconstructed from stable particles (mean lifetime $\tau > 3\times 10^{-11}$~s) with the \antikt algorithm~\cite{Cacciari:2008gp} with a radius parameter $R=0.4$ and are required to satisfy $\pt>15~\GeV$ and $|\eta|<2.5$. The flavor of a jet is determined by counting weakly-decaying $b$- or $c$-hadrons with $\pt>5$~\GeV within $\Delta R < 0.3$ of the jet axis. Events are labeled as $\ttbar+{\geq}1B$ or $Z+{\geq}1B$ if at least one additional jet, not originating from the decay of the \ttbar\ or \Zboson\ system, is matched to at least two $b$-hadrons and $\ttbar+{\geq}1b$ or $Z+{\geq}1b$ if at least one additional jet is matched to one $b$-hadron and no additional jets are matched to at least two $b$-hadrons. Similarly, events are labeled as $\ttbar+{\geq}1c$ or $Z+{\geq}1c$ if at least one additional jet is matched to at least one $c$-hadron and no jet is matched to a $b$-hadron. The remaining events, including those with no additional jets matched to a $b$-hadron or $c$-hadron, are labeled $\ttbar+$light or $Z+$light. Unless explicitly indicated, the category ${\geq}1c$ is combined with the light category.

\section{Object selection}
\label{sec:objects}

\subsection{Tracks and primary vertex}

Tracks are reconstructed in the ATLAS ID~\cite{PERF-2015-08}
and are used to reconstruct primary vertex candidates based on a pattern recognition algorithm and an adaptive vertex fitter with annealing~\cite{ATL-PHYS-PUB-2015-026}. Only events with at least one vertex are used in this search. If more than one vertex is reconstructed, the one with the highest sum of squared transverse momenta of associated tracks is chosen as the primary vertex (PV). Tracks are used to reconstruct leptons and jets. In addition, track-jets, track-subjets, and isolated secondary vertices use tracks that satisfy a \textit{loose} selection criteria: transverse momentum $\pt > 0.5\,\GeV$, longitudinal impact parameter relative to the PV multiplied by the sine of the polar angle $|z_0\sin{\theta}| < 5\,\text{mm}$, and transverse impact parameter relative to the beam line $|d_0|<3.5\,\text{mm}$~\cite{ATL-PHYS-PUB-2022-042}.

\subsection{Leptons}

Electron candidates are reconstructed from clusters of energy deposits in the electromagnetic calorimeter associated with reconstructed tracks in the ID~\cite{EGAM-2018-01,EGAM-2021-01,EGAM-2021-02}. Candidates are selected with $\pt > 10\, \GeV$ and $|\eta|<2.47$, excluding the calorimeter transition region $1.37<|\eta|<1.52$. Electrons must satisfy the \textit{tight} likelihood-based identification criterion defined in \Refn{\cite{EGAM-2021-01}} that is based on shower shape, track-quality, and track-to-cluster matching observables. Electrons are also required to satisfy a multivariate isolation criteria based on the transverse momentum of calorimeter topological clusters and tracks around the electron, and the properties of highly displaced tracks in the same region to distinguish real prompt electrons from electron candidates from hadronic jets, photon conversions, and heavy-flavor hadron decays (non-prompt electrons)~\cite{HIGG-2017-02}. Finally, the electron track is required to originate from the primary vertex by requiring $|z_0\sin{\theta}| < 0.5\,\text{mm}$ and $|d_0/\sigma(d_0)|<5$, where $\sigma(d_0)$ is the transverse impact parameter uncertainty computed relative to the PV.

Muon candidates are reconstructed from track segments in the various layers of the muon spectrometer and matched with tracks from the ID~\cite{MUON-2018-03}. The final muon candidates are refitted using the complete track information from both detector systems, and are required to satisfy $\pt > 10~\GeV$ and $|\eta|<2.5$. Muons are required to satisfy the \textit{medium} identification requirements defined in \Refn{\cite{MUON-2018-03}}, an analogous multivariate isolation criteria to that used for electrons, and $|z_0\sin{\theta}| < 0.5\,\text{mm}$ and $|d_0/\sigma(d_0)|<3$. In the $0\ell$ channel, the transverse momentum requirement for electrons and muons is lowered to $\pt > 7\,\GeV$ to remove backgrounds containing leptons. Muon candidates satisfying $\pt > 4\,\GeV$, $|\eta|<2.5$, and the \textit{medium} identification requirements, are used to identify semileptonic $b$-hadron decays and apply a \textit{muon-in-jet} correction~\cite{HIGG-2016-29}.

\subsection{Jets}

Jets are reconstructed using the \antikt algorithm~\cite{Cacciari:2008gp} implemented in the \Fastjet package~\cite{Cacciari:2005hq,Fastjet} with a radius parameter $R=0.4$. Jets are required to satisfy $\pt > 15\,\GeV$ and $|\eta| < 2.5$. The inputs to the clustering algorithm are determined with a particle flow algorithm that matches tracks to calorimeter topological clusters~\cite{PERF-2015-09}. A multivariate jet vertex tagger (JVT) identifies jets with $\pt<60~\GeV$ and $|\eta|<2.4$ as originating from the PV, suppressing jets from \pileup interactions~\cite{PERF-2014-03}. A dedicated measurement of the JVT efficiency is carried out using $Z\to \mu\mu$ events, which is used to improve the simulation and study the performance down to $\pT>15~\GeV$.

The uncalibrated jet four-momentum is estimated as the sum of the four-momenta of the constituent particle-flow inputs. The energy is corrected for \pileup effects, and simulations are used to calibrate the average jet energy response to the particle-jet level. A dedicated calibration is used for jets with $\pT>15~\GeV$ based on the global neural network calibration method to improve the energy resolution by correcting the jet energy scale as a function of the tracks and topological cluster shapes without changing the average scale~\cite{JETM-2022-01}. These sets of corrections account for variations coming from the flavor of the jet and the specific hadron composition arising from jet fragmentation. A residual \textit{in situ} calibration is applied to data to correct the residual difference between data and MC simulations~\cite{JETM-2018-05}.

An isolation criteria for jets is defined by reclustering calibrated jets~\cite{Nachman:2014kla} with an \antikt algorithm and radius parameter $R=0.8$. Jets that are the sole constituent of a reclustered jet are defined as \textit{isolated}. The jet is not considered isolated if electrons or muons are found in the ring region $0.4 < \Delta R < 0.8$ around a jet, and if such a lepton has a relative isolation, calculated based on the sum of energies of the calorimeter topological clusters within a radius of $\Delta R < 0.3$ of the lepton, less than 0.1.

\subsection{Large-$R$ track-jet and subjet}

The decay $a\to \bbbar$ with high $\pt^a/m_a$ can yield a single jet originating from the hadronization of the two $b$-quarks. In this merged regime, the standard jet reconstruction~\cite{PERF-2015-09} does not capture all the tracks from the hadronization of the two $b$-quarks, preventing efficient flavor tagging with the standard track collection associated with a jet. To capture a more representative collection of particles from a merged $a\to \bbbar$ decay, a track-jet with an extended track collection is associated with each isolated jet with $\pt > 20\,\GeV$ and $|\eta|<2.0$. The jet is required to be isolated so that there is no ambiguity on the origin of the additional tracks. Track-jets are reconstructed by reclustering jets and tracks using an \antikt algorithm with radius parameter $R=0.8$. The tracks are ghost-associated with the jets during the jet reconstruction~\cite{Cacciari:2007fd}. This process creates large-$R$ track-jets around each isolated jet. The same radius parameter was successfully used in a previous version of this search~\cite{HDBS-2018-47}. The four-momentum of the track-jet is estimated as the sum of the four-momenta of the constituent tracks.

An exclusive-$k_t$ (\exkt) clustering~\cite{Ellis:1993tq} of the track-jet tracks is used to reconstruct the flight axes of the two $b$-quark decays within a large-$R$ track-jet. The exclusive-$k_t$ algorithm is a sequential clustering algorithm that compares the relative $k_t$-distance $\min(p_{\text{T},i},p_{\text{T},j})\times \Delta R_{ij}$ between pairs of components $(i,j)$ and the beam distance $p_{\text{T},i}$. If the smallest value in the set is the beam distance, the component is removed. On the other hand, if the smallest value is the $k_t$-distance, then components $i$ and $j$ are clustered together into a pseudo-jet. The algorithm then iterates over the merged pseudo-jets and stops when two pseudo-jets remain, which are the track-subjets used to estimate the flight direction of the two particle-jets merged together. Studies based on boosted Higgs boson decays into a pair of $b$-quarks show that the exclusive-$k_t$ clustering identifies two subjets with the correct generator-level flavor in approximately 90\% of the jets~\cite{ATL-PHYS-PUB-2017-010}.

\subsection{Heavy-flavor tagging}
\label{sec:heavy_flavor_tagging}
The signal is characterized by the presence of multiple $b$-quarks in the final state, $4b$ or $6b$, which often can be challenging to identify using traditional methods since the decay products can be soft or overlapping in the detector. Heavy-flavor identification is extremely important in this search and several algorithms are used to identify $b$-hadron decays.

Two algorithms are used to classify the flavor of jets. \DEXTER is an algorithm used to identify jets coming from the overlapping hadronization of two $b$-quarks~\cite{ATL-PHYS-PUB-2022-042}. The \DEXTER algorithm uses the tracks from the associated large-$R$ track-jet to reconstruct secondary vertices, which provide a distinctive signature of merged $b$-decays when more than one is reconstructed inside the same jet or when they merge in a single secondary vertex with very high mass. Displaced tracks, secondary vertices, and the properties of the two \exkt jets are used in a deep-set neural network (NN) to classify the flavor of the jet. The NN exploits the presence of highly displaced tracks along the two flight axes and the reconstruction of multiple secondary vertices with sufficiently large mass to identify the merged hadronization of two $b$-quarks. Jets that satisfy the \DEXTER 60\% efficiency working point (WP) are classified as $B$-jets. The classification is further refined by defining a tighter 40\% WP\@. When the distinction between the two WPs is needed, jets satisfying the looser 60\% WP but not the 40\% WP are referred to as $B_l$, while jets satisfying the tighter 40\% WP are referred to as $B_t$. The mis-identification efficiency for jets with a single $b$-hadron is up to 2.5\% and 7\%, for the 40\% and 60\% working points, respectively, as measured using \ttbar events~\cite{ATL-PHYS-PUB-2022-042}. Contributions from other jet flavors are found to be negligible in this search. In the $2\ell$ channel, since the signal contains only low \pt\ jets, the \DEXTER\ algorithm is not used for jets with $\pt>200\,\GeV$.

The three-momentum of a $B$-jet is defined by summing the three-momentum of the isolated jet with up to one muon candidate per \exkt-associated track-subjet if $\Delta R(\mu,\text{track-subjets})<0.3$. The main goal of this \textit{muon-in-jet} correction is to include muons arising from semileptonic $b$-decays, which are expected to be close to the flight axes of the two $b$-decays represented by the \exkt axes. If multiple muon candidates satisfy this $\Delta R$ requirement, the highest \pt one is used. The mass of a $B$-jet is derived using a NN-based regression with a similar structure to the one used for the \DEXTER algorithm, but with an additional deep-set encoder for muons inside the track-jet. The NN regression is trained using a Huber loss~\cite{HuberLoss} with parameter $\delta=1$ using signal MC samples that contain merged $a\to \bbbar$ decays. The target of the regression is defined as the ratio of the generator-level mass, defined by clustering all stable generator-level particles with an \antikt algorithm with radius parameter $R=0.8$, and the track-jet mass.

Jets that are not classified as $B$-jets, either because they fail to satisfy the isolation criteria or fail to meet the 60\% \DEXTER WP, are classified as either $b$-jets or light-flavored jets (called light jets), using the \dlr algorithm~\cite{FTAG-2019-07}. The \dlr algorithm uses a deep NN combining information from several specialized low-level taggers using information from the interaction point, the impact parameter of tracks, and secondary vertices to determine the probability for a jet to be $b$-, $c$-, or light-flavored. The \dlr discriminant combines the probabilities to define WPs based on the efficiency to identify $b$-jets. Jets satisfying the 85\% WP are classified as $b$-jets, and otherwise as light jets. Four WPs are defined, corresponding to efficiencies of 85\%, 77\%, 70\%, and 60\%. The five intervals define a \textit{pseudo-continuous} (PC) $b$-tagging score for $b$-jets. The four-momentum of a $b$-jet is defined by summing the four-momentum of the jet with that of up to one muon candidate if $\Delta R(\mu,\text{jet})<0.3$.

Finally, soft secondary vertices (soft-v, or simply v) are used to identify heavy-flavor hadron production that is not energetic enough to form a jet. Soft vertices are reconstructed by clustering tracks using the \textsc{TC-LVT} algorithm~\cite{FTAG-2023-02}. The clustering is based on the $\Delta R$ between the tracks and the closest high-$\pt$, displaced seed tracks, defined as $\pt > 1.5\,\GeV$ and $|d_0/\sigma(d_0)|>0.5$. A secondary-vertex finding algorithm is run on all tracks inside a cluster, which rejects tracks until a single vertex can be fitted~\cite{ATL-PHYS-PUB-2017-011}. The four-momentum of the soft-v is estimated as the vector sum of the tracks that fit the single vertex. Soft vertices are required to satisfy a mass requirement $0.6 < m < 6\,\GeV$ and have $\pt > 3\,\GeV$.

The generator-level flavor of non-isolated reconstructed jets and of soft vertices is defined by $\Delta R$-matching of weakly-decaying $b$- and $c$-hadrons~\cite{FTAG-2019-07}. The generator-level flavor of isolated reconstructed jets is defined by ghost-association of $b$- and $c$-hadrons using an \antikt clustering algorithm with radius parameter $R=0.8$. Only $b$- and $c$-hadrons within $\Delta R<0.3$ of one of the two \exkt axes are counted.

The $B$-jet identification efficiency of the \DEXTER\ algorithm is measured in collider data by using \ttbar\ and $Z$+$g(\to \bbbar)$ events~\cite{ATL-PHYS-PUB-2022-042}. This measurement is used to correct the identification efficiency of heavy-flavor jets in simulation as a function of the jet $\pt$ and $\eta$ to match the one observed in data. The efficiency scale factors are in the range of $1.0-1.9$ with uncertainties in the range of $0.1-0.5$~\cite{ATL-PHYS-PUB-2022-042}. The \bjet identification efficiency of the \dlr\ algorithm is also measured using $\ttbar$, \Zjets, $W$+jets, and multijet events~\cite{FTAG-2018-01, FTAG-2020-08, FTAG-2019-02}. In addition, $\ttbar$ and \Zjets events are used to measure the identification efficiency for the \dlr algorithm for jets with $15<\pt<20~\GeV$~\cite{HDBS-2021-07} and the soft-v identification efficiency of the \textsc{TC-LVT} algorithm~\cite{FTAG-2023-02}. Discrepancies between the identification efficiency in simulation and in data stem from both mismodeling of the detector response and of the underlying physics processes.

\subsection{Overlap removal}

An overlap removal algorithm is applied to prevent double-counting of objects. Electrons are removed if they share their track with a muon. If an electron is within $\Delta R<0.2$ of a jet, the jet is removed. However, if the closest jet surviving is in the range of $0.2< \Delta R <0.4$ of a selected electron, then the electron is removed. Muons are usually rejected if they are separated from the nearest jet by $\Delta R < 0.4$, though the muon is still used in the definition of the four-momentum of $b$-jets (see \Sect{\ref{sec:heavy_flavor_tagging}}). However, if this jet has fewer than three associated tracks, the jet is removed instead to avoid an inefficiency for high-energy muons undergoing significant energy loss in the calorimeter. Soft vertices within $\Delta R<0.4$ of a jet or within $\Delta R < 0.8$ of a $B$-jet are removed.

\subsection{Construction of $b$-objects}
\label{sec:bojects}

\Fig{\ref{fig:jet_flow_diagram}} shows a summary of the different heavy-flavor algorithms used. The concept of a $b$-object is introduced to provide a unified approach to heavy-flavor identification. In the case of a $b$-jet or a soft-v, the four momentum of the $b$-object is defined as the four-momentum of the $b$-jet or of the soft-v, respectively. Two $b$-objects are defined per $B$-jet by projecting the four-momentum of the $B$-jet onto the four-momenta of the two \exkt jets.

\begin{figure}[tb]
\centering
\includegraphics[width=0.65\linewidth]{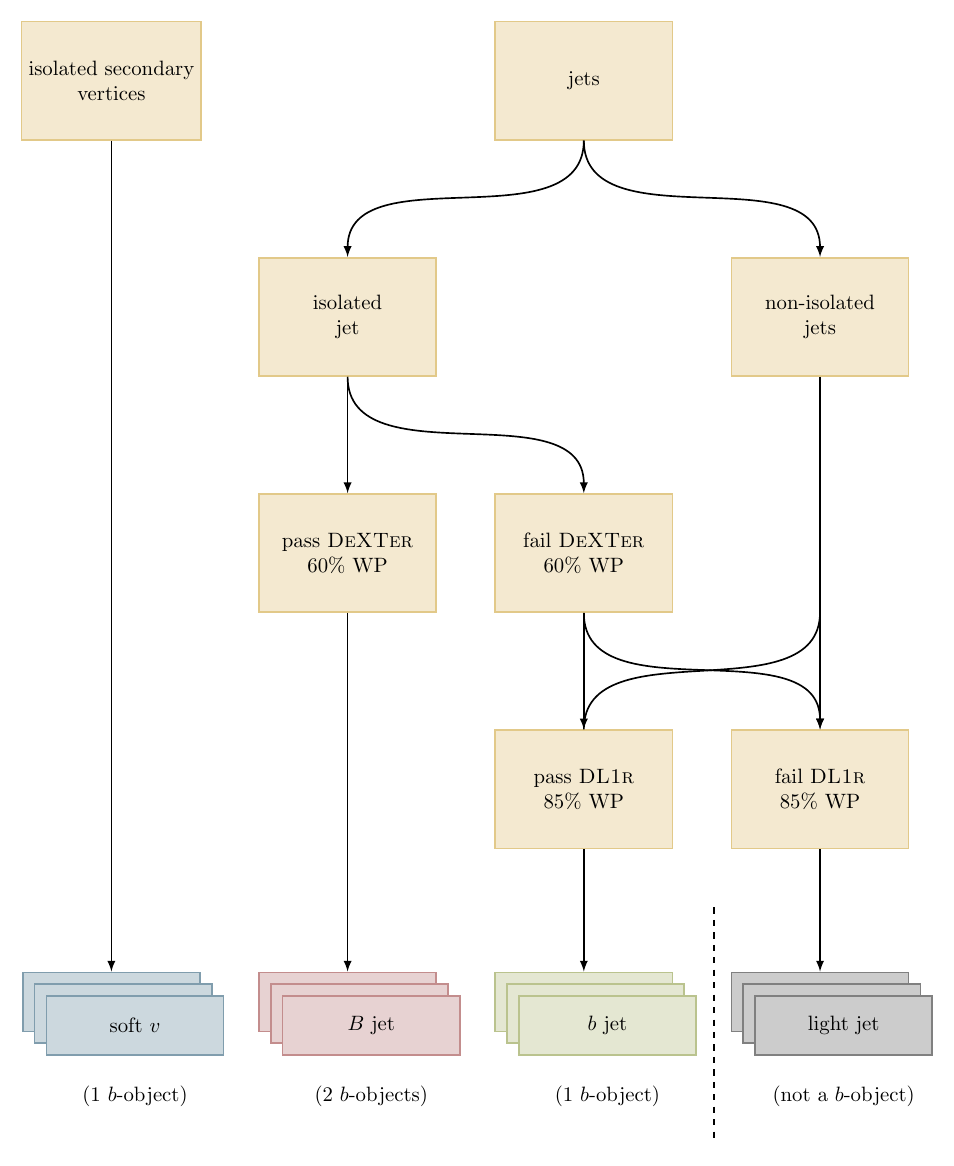}
\caption{Schematic overview summarizing the heavy-flavor algorithms used and the criteria used to define the $b$-objects.}
\label{fig:jet_flow_diagram}
\end{figure}

\subsection{Missing transverse momentum}

The missing transverse momentum \vecmet, with magnitude \met, is calculated as the negative vector sum of the $\vec{\pt}$ of all reconstructed objects and any additional tracks not associated with any reconstructed objects in the event, called the soft term~\cite{JETM-2020-03}. It measures the imbalance of the transverse momentum in the detector, for example from $Z\rightarrow \nu\nu$.

The \textit{effective missing transverse momentum}, \meteff, is defined as the magnitude of the vector:
\begin{equation}\nonumber
\meteffvec = \vecmet +\sum_{\text{leptons}}\vec{p_T}({\ell})
\end{equation}
The \meteff variable considers the leptons as part of the \MET, facilitating comparisons of the \MET variables in samples with different lepton multiplicities.

\section{Event selection}
\label{sec:events}

The analysis requires at least three $b$-objects, $n_b + 2n_B + n_v \geq 3$, as defined in \Sect{\ref{sec:bojects}}. Events with exactly three $b$-objects are used to derive corrections to the background modeling in simulation, as described in \Sect{\ref{sec:background}}. Events with at least four $b$-objects are used to define the signal regions (SR) and control regions (CR).

The analysis uses events with zero, one, or two charged leptons. The highest-$\pt$ lepton in events with at least one lepton is required to have $\pt>27~\GeV$ and to be matched to an object reconstructed at trigger level with $\Delta R<0.15$ to ensure good modeling of the trigger in simulation. Events with two same-flavor (SF) opposite-charge leptons are required to satisfy a requirement on the invariant mass of the two leptons $\mll > 50\,\GeV$. Events without charged leptons are required to satisfy the $\MET$ trigger requirement and have $\MET > 150~\GeV$ to ensure trigger efficiency in the plateau, where the trigger is fully efficient~\cite{TRIG-2019-01}. After applying these requirements, the event sample is predominantly comprised of \Zjets and $\ttbar$ events.

For the 2-lepton channel, regions are defined depending on the flavor of the leptons.  SRs are required to have two SF opposite-charge leptons satisfying $|m_{\ell\ell}-m_Z|<20\,\GeV$. The same condition is applied to CRs enriched in the \Zjets events. CRs enhanced in $\ttbar$ use events with different-flavor (DF) leptons. In the $H\to 2a/a_1a_2\to (\bbbar)(\bbbar)$ analyses, events are placed in SRs and CRs depending on how compatible a quadruplet of $b$-objects in the event is with the signal Higgs boson decay. The compatibility of each quadruplet in the event is assessed by a NN, called the \textit{quadruplet selection NN}. This NN is invariant under permutations of $b$-objects and permutations of $a$-bosons, and is parameterized as a function of the two $a$-boson-mass hypotheses: $m_{a_1}$ and $m_{a_2}$. Events in which no quadruplets compatible with the $H\to 2a/a_1a_2 \to (\bbbar)(\bbbar)$ decay are found are placed in CRs that are categorized based on the number of $b$-jets, $B$-jets, and soft-v in the event. Events with at least one compatible quadruplet are placed in SRs categorized based on the number of $b$-jets, $B$-jets, and soft-v used to form the most compatible quadruplet in the event.
For the $H\to a_1a_2\to 3a_1\to 3(\bbbar)$ analysis, where full reconstruction of all the objects in the final state is very rare, events are categorized according to the number of reconstructed $b$-jets, $B$-jets, and soft-v.

For the 0-lepton channel, only events with exactly two $B$-jets or with exactly one $B$-jet and two $b$-jets are used. Events with exactly one lepton are used to define a CR enriched in \ttbar events. Events with zero leptons are used in SRs and CRs depending on how compatible the only quadruplet of $b$-objects in the event is with a signal Higgs boson decay, based on the invariant mass of the quadruplet as described in \Sect{\ref{sec:events_0l}}. Events are placed in SRs categorized based on the number of jets ($n_j$), $b$-jets, and $B$-jets. A summary of the event selection requirements, and the signal and control region definitions are shown in \Tab{\ref{tab:eventselection}}.

\begin{table}[!tbp]
\caption{Summary of the event selection, and signal and control region definitions, for the $0\ell$ and $2\ell$ channels. The different variables are defined in \Sects{\ref{sec:objects}}{\ref{sec:events}}}.
\label{tab:eventselection}
\centering
\resizebox{\textwidth}{!}{%
\begin{tabular}{l | c | c | c}
\hline\toprule
\multirow{2}{*}{\textbf{Analysis channels}} & \textbf{0$\ell$} & \multicolumn{2}{c}{\textbf{2$\ell$}} \\
& $H\rightarrow 2a\rightarrow 4b$ & $H\rightarrow 2a/a_1a_2\rightarrow 4b$ & $H\rightarrow a_1a_2\rightarrow 6b$\\
\midrule
\multicolumn{4}{c}{\textbf{Common selection}}\\
\midrule Triggers & \met  & \multicolumn{2}{c}{Single lepton}\\
Leptons & $\pt > 7\,\GeV$ & \multicolumn{2}{c}{lead $\pt > 27\,\GeV$, sublead $\pt > 10\,\GeV$} \\
$b$-object multiplicity & $n_b = 2,\, n_B=1$ or $n_B = 2$ & \multicolumn{2}{c}{$n_b + 2 n_B + n_\text{v} \geq 4$} \\\midrule
\multicolumn{4}{c}{\textbf{Signal regions}}\\
\midrule
Leptons & $0$ leptons & 2 SFOS leptons & 2 SFOS leptons \\
\met & $\met > 150\,\GeV$ & -- & -- \\
$Z$ boson selection & -- & $|m_{\ell\ell}-m_{\Zboson}|<20~\GeV$ & $|m_{\ell\ell}-m_{\Zboson}|<20~\GeV$ \\
Higgs boson selection & $|m_{aa}-m_H| < 50~\GeV$ & High quadruplet NN score & -- \\
Multijet rejection & $\min \Delta\phi(\met, a\text{ cand.}) > 60^{\circ}$ & -- & -- \\
Categorization & $n_j, n_b, n_{B_t}, n_{B_l}$ & Quadruplet NN prediction & $n_{\text{v}}, n_b, n_{B_t}, n_{B_l}$\\
Binning & -- & Loose, Medium, Tight BDT bins 	&     Medium, Tight BDT bins \\
\midrule \multicolumn{4}{c}{\textbf{Control regions}}\\
\midrule
\textbf{Multijet} & \textit{ABCD method} & -- & -- \\
& $150 < m_{aa}-m_H < 250~\GeV$ & & \\
& $\min \Delta\phi(\met, a\text{ cand.}) < 30^{\circ}$ & & \\
\midrule
\textbf{Lost lepton} &  & -- & -- \\
\hspace{3mm} Signal veto & $|m_{aa}-m_H| > 50~\GeV$ & & \\
\hspace{3mm} Leptons & 1 lepton, $\pt>27\,\GeV$, $\Delta R(\ell,B)>0.8$ & & \\
\hspace{3mm} \met & $\meteff > 150~\GeV$ & & \\
\hspace{3mm} Categorization & $n_b, n_{B_t}, n_{B_l}$  & & \\
\midrule
\textbf{\Zjets enriched} & & & \\
\hspace{3mm} Signal veto & $|m_{aa}-m_H| > 50~\GeV$ & Low quadruplet NN score & Loose BDT bins \\
\hspace{3mm} Leptons & $2$ SFOS leptons, lead $\pt > 27~\GeV$ & 2 SFOS leptons & 2 SFOS leptons \\
\hspace{3mm} \met & $\met < 50~\GeV, \meteff > 30~\GeV$ & $\met < 60~\GeV$ & $\met < 60~\GeV$\\
\hspace{3mm} $Z$ boson selection & $|m_{\ell\ell}-m_{\Zboson}|<10\,\GeV$ & $|m_{\ell\ell}-m_{\Zboson}|<10\,\GeV$ & $|m_{\ell\ell}-m_{\Zboson}|<10\,\GeV$\\
\hspace{3mm} Categorization & $n_b, n_{B_t}, n_{B_l}$ & $n_\text{v}, n_b, n_{B_t}, n_{B_l}$ & $n_\text{v}, n_b, n_{B_t}, n_{B_l}$\\
\midrule
\textbf{\ttbar enriched} & -- & \multicolumn{2}{c}{}\\
\hspace{3mm} Leptons & & \multicolumn{2}{c}{2 DFOS leptons}\\
\hspace{3mm} Categorization &  & \multicolumn{2}{c}{$n_\text{v}, n_b, n_{B_t}, n_{B_l}$}\\
\bottomrule\hline
\end{tabular}
}
\end{table}

\subsection{$2\ell$ channel: $H\rightarrow 2a\rightarrow 4b$ and $H\rightarrow a_1a_2\rightarrow 4b$}
\label{sec:events_2l}

In the $4b$ case, a quadruplet selection NN is used to identify the $b$-objects in the final state. The signal $H\rightarrow 2a/a_1a_2\rightarrow 4b$ can result in different combinations of reconstructed $b$-objects: $b$-jets, $B$-jets, or soft vertices. The NN is trained to identify the best combination of objects under the signal hypothesis of a Higgs boson decay through two intermediate $a$-bosons, according to the NN structure shown in \Fig{\ref{fig:htnn_diagram}}. The diagram corresponds to an example input combination that highlights that different $b$-object inputs ($B$-jet, $b$-jet, and soft-v) have different encoder NNs in the leftmost column. The second column shows different encoder NNs for the two different $a$-bosons, such that each possible quadruplet and each possible pairing of a quadruplet goes through different NNs in different orders, yielding different compatibility scores that are used to select the best pairing.

\begin{figure}[!tbp]
\centering
\includegraphics[width=0.8\linewidth]{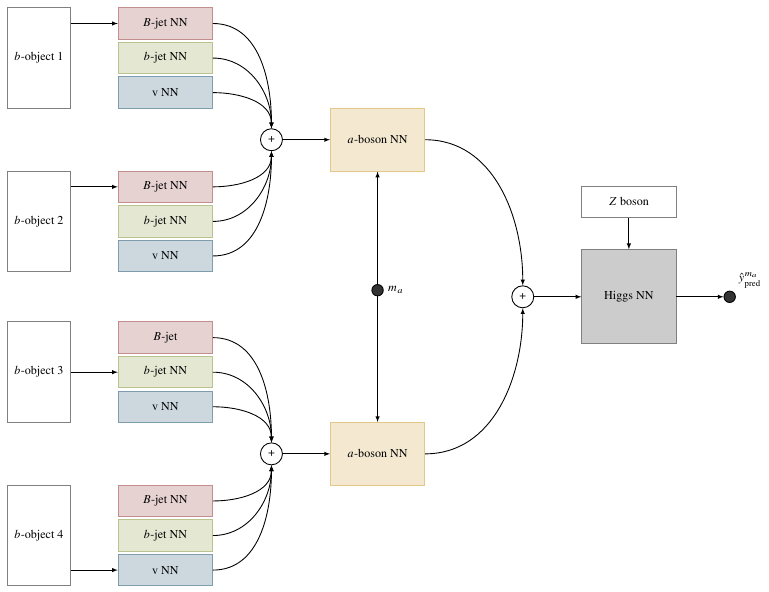}
\caption{Diagram of the NN used for an example of a quadruplet selection showing the case of a hypothesis with $B$-jets, $b$-jets, and soft-v. All filled blocks are dense multi-layer perceptrons (MLP) that share common weights between MLPs of the same color. The $a$-boson MLPs are parameterized as a function of the mass hypothesis $m_a$. Unfilled rectangles are inputs from the four $b$-objects and the \Zboson\ boson. The value of the quadruplet selection score, $\NNscore$, is used to classify the $b$-object quadruplet. In events with more than 4 $b$-objects, different quadruplets are tested to identify the one with highest likelihood to come from a Higgs boson decay.}
\label{fig:htnn_diagram}
\end{figure}

The NN contains fully connected encoders for $b$-jets, $B$-jets, and soft vertices. The encoded information is then used in a pair of deep-set NNs, which ensure that the overall NN score is invariant under permutations of $(\bbbar)\leftrightarrow (\bbbar)$ (Higgs boson deep sets) and $b\leftrightarrow \bar{b}$ ($a$-boson deep sets). The $a$-boson deep-sets training depends on the true $m_a$ value, since the reconstruction modes depend on the boost of the $a$-bosons. When considering models with additional $a$-bosons, $H\to a_1a_2$ permutation invariance is broken by assigning the different true mass values to each of the two $a$-boson encoders.

The NN is built from up to five $b$-objects, as shown in \Fig{\ref{fig:hypo_example}} for an example combination of objects. In cases where more than five objects are reconstructed, the $b$-objects with highest $\pt$ are used. Input combinations with more than one soft-v are not considered, since they do not contribute significantly to the acceptance. Kinematic variables characterizing the $\Zboson$ boson are also used in the NN\@. \Tab{\ref{tab:hypothesis_feature}} summarizes the input features used for the quadruplet selection NN\@. The input features include the four-momenta of each input object and the likelihood for their flavor to be properly reconstructed.

\begin{table}[!tbp]
\caption{List of features used as input for the quadruplet selection NN\@. The input objects are described in \Sect{\ref{sec:objects}}. The $Z$ boson candidate is reconstructed as the sum of the four-momenta of the two leptons.
}
\label{tab:hypothesis_feature}
\centering
\begin{tabular}{ll}
\toprule
\multicolumn{1}{c}{Feature}      & \multicolumn{1}{c}{Description}   \\
\midrule
\multicolumn{2}{c}{\DEXTER-tagged $B$-jet}  \\
\midrule
$\pt(B)$         &   Jet transverse momentum \\
$m_B$         &   Track jet mass                           \\
$\eta(B)$          &   Jet pseudorapidity                                \\
$\phi(B)$          &   Jet azimuthal angle                                \\
$B_l$ or $B_t$& Satisfies loose or tight \DEXTER requirement \\
\midrule
\multicolumn{2}{c}{\dlr-tagged $b$-jet}                     \\     \midrule
$\pt(b)$         &   Jet transverse momentum                           \\
$m_b$         &   Jet mass                           \\
$\eta(b)$          &   Jet pseudorapidity                                \\
$\phi(b)$          &   Jet azimuthal angle                                \\
PC score & The pseudo-continuous \dlr score \\
\midrule
\multicolumn{2}{c}{Soft secondary vertex v} \\     \midrule
$m_\text{v}$ & Track mass of the secondary vertex \\
$\pt(\text{v})$         &   Secondary vertex transverse momentum                           \\
$\eta(\text{v})$      &     Secondary vertex pseudorapidity          \\
$ \phi(\text{v})$   &     Secondary vertex azimuthal angle          \\
$L_{3D}$ &  Decay length relative to the PV\\
$S_{L_{3D}}$ &  Decay length significance\\
\midrule
\multicolumn{2}{c}{$Z$ boson candidate}                     \\  \midrule
$\pt(Z)$         &   $Z$ boson candidate transverse momentum                           \\
$\eta(Z)$          &   $Z$ boson candidate pseudorapidity                                \\
$\phi(Z)$          &   $Z$ boson candidate azimuthal angle                                \\
$m_{\ell\ell}$          &   $Z$ boson candidate mass                                \\
\bottomrule
\end{tabular}
\end{table}

\begin{figure}[!tpb]
\centering
\includegraphics[width=\linewidth]{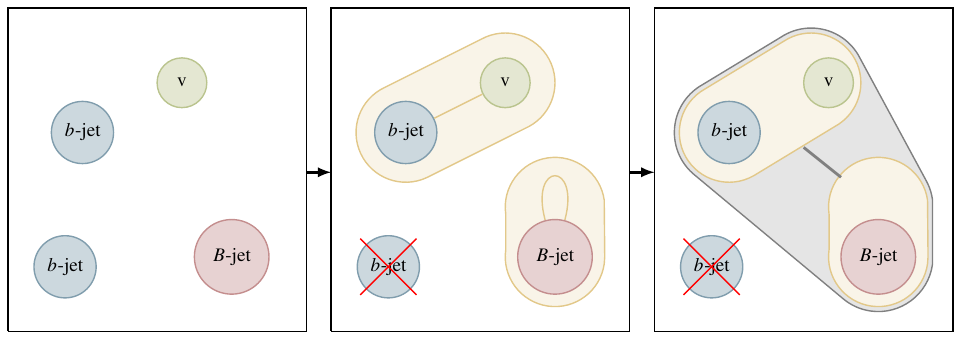}
\caption{Diagram depicting an example of how the NN selects a quadruplet. Each $b$-object is represented as an MLP ($b$-object MLP). The $a$-boson is represented as a pairing between $b$-objects, which also corresponds to an MLP ($a$-boson MLP). All physically motivated pairings are considered, including loop pairings representing merged reconstructed objects. The $a$-bosons themselves form new pairings representing the fully reconstructed Higgs boson, which also corresponds to an MLP (Higgs MLP). The value of this last pairing is used to select the quadruplet.
The middle diagram shows the two reconstructed $a$-boson candidates and the rightmost diagram shows the final Higgs boson candidate.}
\label{fig:hypo_example}
\end{figure}

The NN is trained with all $H \rightarrow 2a \rightarrow 4b$ signal simulation samples. In addition, it also uses background events, \Zjets\ and \ttbar\, to provide training samples with broader quadruplet $\pt$ and $m$ distributions. The samples are split into training and validation samples, and the validation sample is used to measure the performance and check for overtraining. All permutations satisfying the selection criteria described above are included in the training and the correct permutation is identified based on the generator-level event history information. The NN is trained by minimizing a binary cross-entropy loss in which the correct permutation has label one and all wrong permutations have label zero. In the case of background events, all permutations have label zero.

The reconstructed event hypothesis with the largest score and the value of the NN score, $\NNscore$, are used for event classification.
The accuracy of associating $b$-partons to $b$-objects ranges from 60\% to 98\% depending on the type and properties of the reconstructed object and the signal.

\paragraph{Signal regions:} Five SRs are defined based on the objects of the selected quadruplet and the value of the score $\NNscore$. For events with quadruplets formed with two $B$-jets ($2B$), the score is required to be larger than $\NNscore> 0.5$. For the cases where the quadruplet is formed with one $B$-jet and two $b$-jets ($1B2b$); with one $B$-jet, one $b$-jet and one v ($1B1b1$v); with three $b$-jets and one v ($3b1$v); and with four $b$-jets ($4b$); the score is required to be larger than $\NNscore>0.05$. The choice of this \NNscore requirement is intended to maintain high acceptance while ensuring accurate reconstruction of the Higgs boson decay for signal samples across the mass range probed. The \NNscore\ requirements are not optimized to reject background since a dedicated discriminator is used based on the properties of the selected quadruplet. \Fig{\ref{fig:NNscore}} shows example distributions of the $\NNscore$ observable for the $2B$ and $1B2b$ cases.

\begin{figure}[!tbp]
\centering
\subfloat[]{\label{fig:2B_NN}
\includegraphics[width=0.4\linewidth]{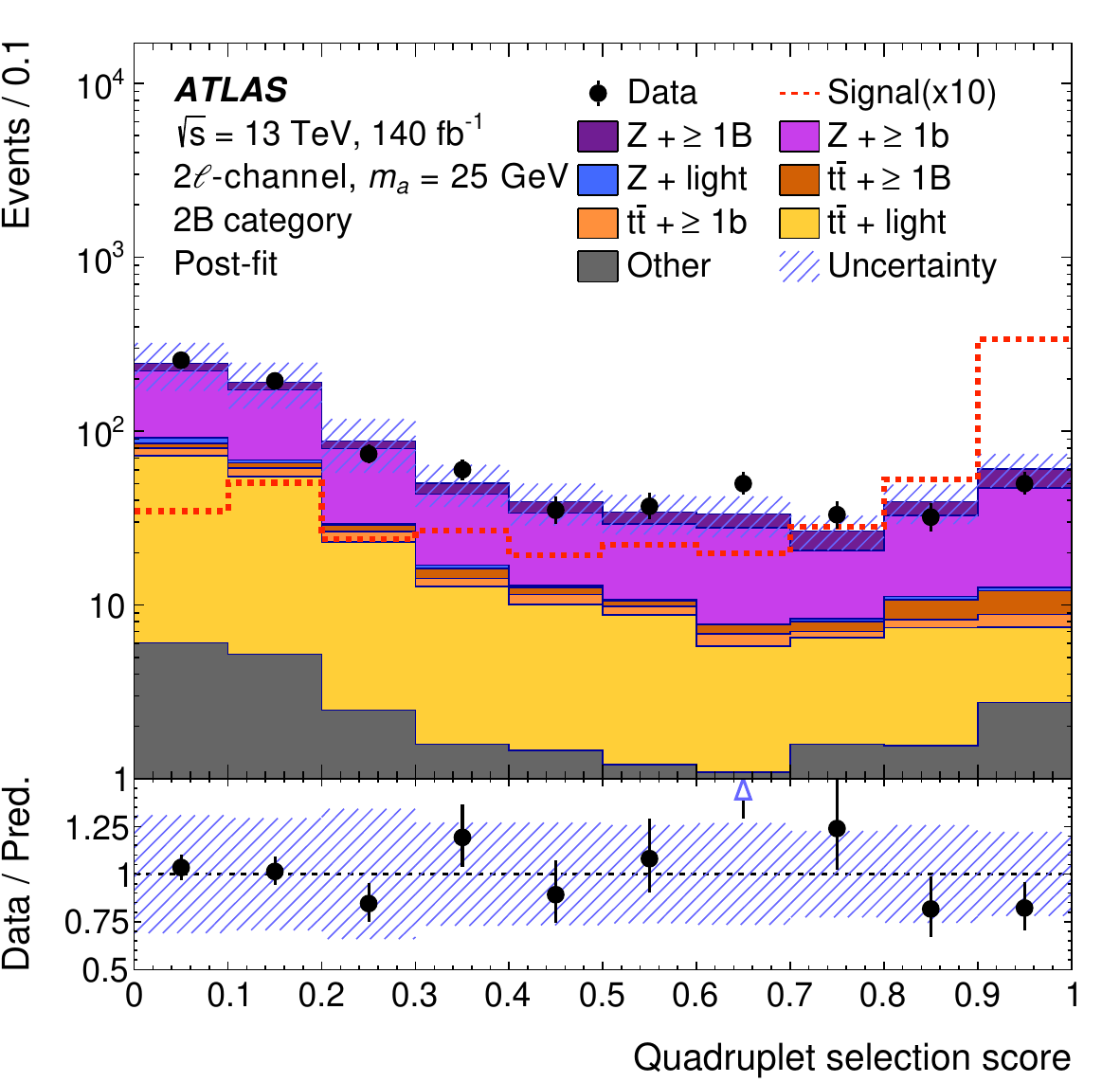}}%
\subfloat[]{\label{fig:1B2b_NN}
\includegraphics[width=0.4\linewidth]{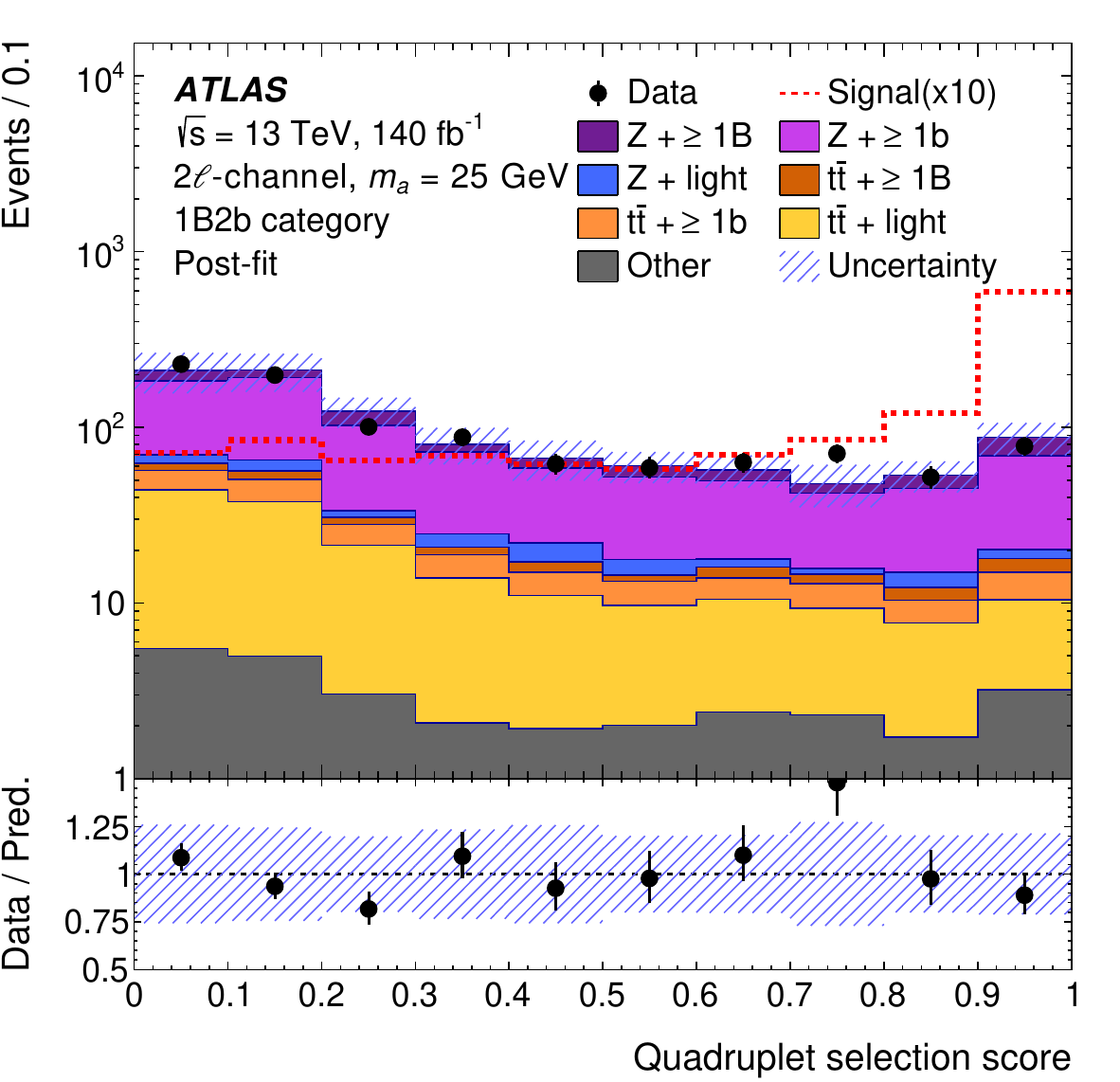}}%
\caption{The quadruplet selection score \NNscore~distribution assuming the signal hypothesis $m_{a} = 25~\GeV$ for the $b$-object categories \protect\subref{fig:2B_NN} $2B$ and \protect\subref{fig:1B2b_NN} $1B2b$. Signal events for the $m_{a} = 25~\GeV$ mass hypothesis are also shown, assuming $\mathcal{B}=1$ and scaling this signal by a factor of ten. The category labeled ``Other'' includes processes with small contributions to the total yield (diboson, $\ttZ$, and $\ttW$). The lower panels show the ratio of the observed to the estimated SM background. The hashed area represents the total uncertainty in the background.}
\label{fig:NNscore}
\end{figure}

\paragraph{Control regions:} Six CRs for each of the main backgrounds \Zjets and \ttbar are used based on the $b$-object multiplicity, as summarized in \Tab{\ref{tab:4b:cr_summary}}. The regions distinguish between events containing jets satisfying the loose and tight \DEXTER tagging requirements to enhance the sensitivity to different heavy-flavor background components. The control regions including the combination of $b$-objects corresponding to $1B2b$ and $1B1b1$v are combined due to the low expected number of events. The \Zjets control regions are defined by inverting the requirements on the quadruplet selection NN score $\NNscore$. Additional requirements are applied to the \Zjets control region to increase the purity: $|m_{\ell\ell}-m_Z|<10~\GeV$ and $\MET < 60\,\GeV$. The same categorization is used to build the control regions enriched in $\ttbar$ events using events with two charged leptons of DF\@. In this case, no quadruplet selection is applied, since the signal has SF leptons, and no additional mass cut is imposed, since the region is already very pure in $\ttbar$ events.

\begin{table}[!tbp]
\caption{Overview of the $b$-object multiplicity regions used for the control regions in the search targeting the $4b$ final state. The subscripts in $B_l$ and $B_t$ refer to the \DEXTER $B$-jet loose and tight working points, respectively.}
\label{tab:4b:cr_summary}
\centering
\begin{tabular}{lccc}
\toprule
$2\ell$ $4b$ control regions & $B$-jets & $b$-jets & Soft v\\
\midrule
$B_tB_l$    & $\geq 2 B_t$ or $1 B_t + 1 B_l$ & -- & -- \\
$2B_l$      & $\geq 2 B_l$    & -- & -- \\
$1B_t2b$ (includes $1B_t1b1$v) & $1 B_t$ & $\geq 2$ (or $\geq 1$ &  and $\geq$ 1) \\
$1B_l2b$ (includes $1B_l1b1$v) & $1 B_l$ & $\geq 2$ (or $\geq 1$ &  and $\geq$ 1) \\
$4b$        & $0$   & $\geq 4$  & -- \\
$3b1$v      & $0$   & $\geq 3$  & $\geq$ 1 \\
\bottomrule
\end{tabular}
\end{table}

\subsection{$2\ell$ channel: $H\rightarrow a_1a_2 \rightarrow 3a_1 \rightarrow 6b$}
\label{sec:events_2l_a1a2}

The $6b$ search targets signals where the Higgs boson decays into two $a$-bosons with different mass $H \rightarrow a_1 a_2$ and the heavier one can then decay into the lighter one, $a_2\rightarrow 2a_1$, resulting in six $b$-quarks in the final state  $H \rightarrow a_1 a_2 \rightarrow 3a_1 \rightarrow 6b$. In only approximately 1\% of the signal decays, the six $b$-objects would be reconstructed and if the full reconstruction of the Higgs boson cascade decay were required, the signal efficiency would be low. Instead, the analysis is performed with partially reconstructed events that are classified solely based on the number and type of reconstructed $b$-objects, and no attempt to identify all the $b$-objects in the event is made. Moreover, the regions are not defined exclusively to account for the rare cases where more than four $b$-objects are reconstructed. \Tab{\ref{tab:6b-bdt:regions summary}} summarizes the $b$-object multiplicity used to form the event regions when searching for signals with six $b$-quarks in the final state.

\begin{table}[!tbp]
\caption{Overview of the multiplicity regions designed to target the $6b$ final state for both signal and control regions. No requirement on the number of soft vertices is applied for the $2B0b$, $2B1b$, and $1B2b$ regions.}
\label{tab:6b-bdt:regions summary}
\centering
\begin{tabular}{lccc}
\toprule
$2\ell$ $6b$ regions & $B$-jets & $b$-jets & Soft vertices \\
\midrule
2B0b  & $\geq 2$ & 0 & -- \\
2B1b & $\geq 2$ & $\geq 1$ & --  \\
1B2b  & $1$ & $\geq 2$ & --  \\
4b0v     & $0$ & $\geq 4$ & $0$  \\
4b1v & $0$ & $\geq 4$ & $\geq 1$  \\
3b1v  & $0$ & $3$ & $\geq 1$  \\
\bottomrule
\end{tabular}
\end{table}

The SRs and the \Zjets CRs are defined by the boosted decision tree (BDT) discriminators described in \Sect{\ref{sec:discrim_2l_a1a2}}. To increase the purity of \Zjets events and suppress \ttbar contributions in the CRs, stricter requirements are applied on the dilepton invariant mass $|m_{\ell\ell}-m_Z|<10~\GeV$ and the missing transverse momentum $\met < 60~\GeV$. The \ttbar control regions are defined using the same $b$-object multiplicity categories shown in \Tab{\ref{tab:6b-bdt:regions summary}}. As in the case of $4b$, events with different-flavor leptons are selected, and no additional selection is applied on the invariant mass of the leptons $m_{\ell\ell}$.

\subsection{$0\ell$ channel: $H\rightarrow 2a\rightarrow 4b$}
\label{sec:events_0l}

Since it uses a relatively simpler signal region selection and background discrimination strategy, the $0\ell$ channel is used only to search for $H\rightarrow 2a\rightarrow 4b$ decays. Events in the $0\ell$ channel are sorted into eight regions depending on the total number of jets ($j$), the $B$-jet multiplicity and whether they are tight or loose ($B_t$ or $B_l$), and the $b$-jet multiplicity, as shown in \Tab{\ref{tab:0l4b_regions_summary}}. The regions with two $b$-jets are also required to have $\Delta\phi(b_1,b_2)<100^{\circ}$ to suppress contributions from QCD multijet production.

\begin{table}[!tbp]
\caption{Overview of the signal regions for the $0\ell$ 4b final state. The requirements on the number of jets are exclusive. %
}
\label{tab:0l4b_regions_summary}
\centering
\begin{tabular}{lcccc}
\toprule
$0\ell$ $4b$ Regions & $B$-jets & $b$-jets & light-jets & Additional selections\\
\midrule
$2B_t (2j)$     & $2 B_t$           & $0$ & $0$ & --  \\
$B_tB_l (2j)$   & $1 B_t + 1B_l$    & $0$ & $0$ & --  \\
$2B_l (2j)$     & $2 B_l$           & $0$ & $0$ & --  \\
$2B_t (3j)$     & $2 B_t$           & $0$ & $1$ & --  \\
$B_tB_l (3j)$   & $1 B_t + 1B_l$    & $0$ & $1$ & --  \\
$2B_l (3j)$     & $2 B_l$           & $0$ & $1$ & --  \\
$1B_t2b$        & $1 B_t$           & $2$ & $0$ & $\Delta\phi(b_1,b_2)<100^\circ$  \\
$1B_l2b$        & $1 B_l$ & $2$ & $0$ & $\Delta\phi(b_1,b_2)<100^\circ$  \\
\bottomrule
\end{tabular}
\end{table}

In each category, the heavy-flavor jets are used to reconstruct the Higgs boson. This parton-jet matching is nearly fully accurate due to the large Higgs boson boost imposed by the \MET trigger requirement. The $2B$ channels target $a$-boson masses in the range of $m_a\simeq 10-20~\GeV$, while the $1B2b$ channel is more sensitive to $m_a\simeq 20-30~\GeV$. The Higgs boson mass $m_H \approx m_{aa}$ is defined as the invariant mass of the $a$-candidate jets, depending on the final state. For the final state channel with two $B$-jets, each $B$-jet is an $a$-candidate, and thus $m_{aa}$ is defined as $m_{2B}$. For the final state channels with one $B$-jet and two $b$-jets, the $B$-jet is an $a$-candidate and the two $b$-jets combined are an $a$-candidate, and thus $m_{aa}$ is defined as $m_{1B2b}$. Events are categorized into regions based on the number of leptons and \meteff.

\paragraph{Signal regions:} Events are required to be selected with the \MET trigger and satisfy $\met > 150~\GeV$ to ensure high trigger efficiency. Events are also required to have no charged leptons with $\pt > 7~\GeV$ and satisfy $|m_{aa}-125|<50\,\GeV$. An additional selection requirement on the minimum azimuthal distance between the \met\ vector and the two reconstructed $a$-bosons, $\min \Delta\phi(\met, a\text{ cand.}) > 60^{\circ}$, is imposed to suppress multijet backgrounds.

\paragraph{Control regions:} Events in the $1\ell$ CR are selected with single-lepton triggers and requiring exactly one charged lepton with $\pt > 27~\GeV$ matched to the object reconstructed at trigger level. The lepton is also required to have $\Delta R(\ell,B)>0.8$, since most sources of backgrounds do not have leptons produced close to the jets. In order for the CR to be as similar to the SR as possible, events are required to satisfy $\meteff > 150\,\GeV$. Finally, events must have $|m_{aa}-125|>50\,\GeV$ to remove contributions from the signal $W(\rightarrow\ell\nu)H$ production mode. The processes in the $1\ell$ CR are also background processes in the SR when the lepton is outside of acceptance region or when it is not successfully reconstructed and identified (lost-lepton background).

A CR with $2\ell$ is used for results using the $0\ell$ channel only, mainly to study the sensitivity of this channel alone. This region contains events selected with the single-lepton trigger with exactly two charged leptons with $\pt > 15~\GeV$. The leading lepton is additionally required to satisfy $\pt > 27\,\GeV$ and to match to an object reconstructed at trigger level. Leptons are required to have $\Delta R(\ell,B)>0.8$ and events are required to satisfy $\meteff > 30\,\GeV$. Events are also required to have $|m_{aa}-125|>50\,\GeV$ to remove contributions from the $Z(\rightarrow\ell\ell)H$ production mode and are required to have $|m_{\ell\ell}-m_Z| < 10\,\GeV$ and $\met< 50\,\GeV$ to improve the purity of $Z\to \ell\ell$ events.

\section{Signal and background discrimination}

\subsection{$2\ell$ channel: $H\rightarrow 2a\rightarrow 4b$ and $H\rightarrow a_1a_2\rightarrow 4b$}
\label{sec:discrim_2l}

As described in \Sect{\ref{sec:events_2l}}, the $2\ell$ quadruplet selection NN selects the best combination of four $b$-objects to reconstruct the decay chain $H\to 2a/a_1a_2\to 4b$. The score of the best hypothesis provides strong discrimination between signal and background since the main sources of background do not have a similar cascade decay. The Higgs boson is constructed from the two $a$-bosons and the $Z$ boson is constructed from the two leptons. Additional kinematic properties further discriminate signal from background, and these variables are used as inputs to a BDT\@.

The BDT is trained using the input features summarized in \Tab{\ref{tab:bdt:input_vars}} separately for each mass hypothesis and for each region. Signal events are characterized by the presence of a resonance resulting from the Higgs boson decay $H \rightarrow 2a/a_1a_2 \rightarrow 4b$, where these four $b$-objects are determined by the quadruplet selection NN\@. Several variables are used to reconstruct the particles from the signal decay chain and probe their kinematic properties. These include the kinematics of the $a$-bosons, particularly the transverse momentum $\pt(a)$ and the reduced mass, defined as:
\begin{equation}\nonumber
m_{a}^{\text{red}} = m_a - m_a^{\text{gen}}
\end{equation}
where $m_a^{\text{gen}}$ corresponds to the mass hypothesis under consideration. Other variables include the transverse momentum of the Higgs boson $\pt(H)$ and the reduced Higgs boson mass $m_{H}^{\text{red}}$ reconstructed from the two $a$-bosons, defined as:
\begin{equation}\nonumber
m_{H}^{\text{red}} = (m_{H}- m_{H}^{\text{gen}}) - (m_{a_1} - m_{a_1}^{\text{gen}}) - (m_{a_2} - m_{a_2}^{\text{gen}}),
\end{equation}
where $m_H^{\text{gen}}=125~\GeV$ corresponds to the Higgs boson mass, and $m_{a_1}^{\text{gen}}$ and $m_{a_2}^{\text{gen}}$ to the $a$-boson mass hypotheses being considered, where these can be equal to each other or correspond to two different $a$-bosons. The ordering $a_{1,2}$ of the two reconstructed $a$-bosons is determined by the quadruplet selection NN, which ensures consistent ordering in the case of models with different $a$-bosons, and consistent ordering based on reconstruction channel in models with a single $a$-boson. The reduced mass $m_{H}^{\text{red}}$ has better resolution than $m_{H}$ due to correlations in the energy response of the jets.

\FloatBarrier

\begin{table}[!tbp]
\caption{List of input variables for the multivariate BDT discriminant used to distinguish signal and background in the $2\ell$ channel.}
\label{tab:bdt:input_vars}
\centering
\begin{tabular}{p{0.2\linewidth}p{0.7\linewidth}}
\toprule
\multicolumn{1}{c}{Feature}      & \multicolumn{1}{c}{Description}   \\
\midrule
\multicolumn{2}{c}{$2\ell$ $4b$ Variables}  \\
\midrule
$\pt(a_1)$, $\pt(a_2)$ & $a$-boson transverse momenta \\
$m_{a_1}^{\text{red}}$, $m_{a_2}^{\text{red}}$ & $a$-boson reduced masses \\
$m_{\text{H}}^{\text{red}}$ & Higgs boson reduced mass \\
$p_{\text{T}}(H)$ & Higgs boson transverse momentum \\
$\cos\theta^*$ & Cosine of the $a$-boson polar angle in the Higgs boson rest frame \\
$\cos\theta_{\text{CS}}$ & Cosine of the angle between leptons in the Collins-Soper frame~\cite{Collins:1977iv} \\
$\pt(Z)$ & $Z$ boson transverse momentum  \\
$m_{\ell\ell}$ & $Z$ boson invariant mass \\
$|\Delta \phi (Z, H)|$ & Azimuthal angle between the Higgs boson and the $Z$ boson \\
$|\Delta \eta (Z, H)|$ & Pseudorapidity separation between the Higgs boson and the $Z$ boson \\
\NNscore & Quadruplet selection NN score \\
\met    & Missing transverse energy\\
\midrule
\multicolumn{2}{c}{$2\ell$ $6b$ Variables}  \\
\midrule
$m_{\text{incl.}}$ & Invariant mass of all $b$-objects  \\
$\Delta R (B, B)$ & $\Delta R$ between the two leading $B$-jets \\
$\Delta R (B, bb)$ & $\Delta R$ between the leading $B$-jet and the 4-momentum sum of the two leading $b$-jets \\
$\Delta R^{\text{average}} (b,b)$ & $\Delta R$ between two $b$-jets, averaged over all possible pairs of $b$-jets \\
$\Delta R (\exkt,\exkt)$ & $\Delta R$ between the two \exkt associated track-subjets for the leading $B$-jet \\
$\cos\theta^* \ (2B)$ & Cosine of the sum of the two leading $B$-jets in the $ZH$ rest frame \\
$\cos\theta^* \ (1B2b)$ & Cosine of the sum of the leading $B$-jet and the two leading $b$-jets in the $ZH$ rest frame\\
$\cos(\theta^*) \ (\text{incl.})$ & Cosine of the sum of all $b$-objects ($B$-jets, $b$-jets and soft vertices) in the $ZH$ rest frame \\
\met    & Missing transverse energy\\
$H_\text{T}^{\text{had}}$   & Scalar sum of the $\pt$ of all jets\\
$\Delta R (\ell, \ell)$ & $\Delta R$ between the two leptons \\
\bottomrule
\end{tabular}
\end{table}

Additional kinematic variables show differences between signal and background and are used as input variables for the BDT\@. The cosine of the polar angle of the $a$-boson in the rest frame of the Higgs boson, $\cos\theta^*$, is sensitive to the spin-$0$ nature of the Higgs boson. The spin-$1$ nature of the $Z$ boson is probed by the cosine of the angle between the two leptons ($\ell_1$ and $\ell_2$) in the Collins-Soper (SC) frame~\cite{Collins:1977iv} $\cos\theta_{\text{CS}}$, defined as
\begin{equation}\nonumber
\cos\theta_{\text{CS}} = 2\frac{\ell_1^+\ell_2^--\ell_1^-\ell_{2}^{+}}{\mll\sqrt{\mll^2+\pt^2(Z)}},
\end{equation}
where $\ell^{\pm}_i = (E_i\pm p_{z,i})/\sqrt{2}$ with $E_i$ and $p_{z,i}$ corresponding to the energy and momentum along the $z$-direction of the lepton $i=1,2$, and where $\pt(Z)$ is the $Z$ boson transverse momentum. The variables capturing the kinematics of the $Z$ boson are $\pt(Z)$ and the $Z$ boson mass $\mll$. In addition, the kinematics of the $ZH$ events, where the $Z$ boson is recoiling against the Higgs boson in the case of the signal, is probed by the separation in $\phi$ and $\eta$ between the $Z$ boson and the Higgs boson, $\Delta\phi(H,Z)$ and $\Delta\eta(H,Z)$, respectively. Finally, the $\met$ variable is used to distinguish the \ttbar background, which contains neutrinos in the final state, from the signal.

\Fig{\ref{fig:kine_mass_SR_1B2b_25}} shows several key kinematic distributions used as inputs for the BDT\@. The BDT training is performed using the \textsc{xgboost} framework~\cite{XGboost}. The output score is used to define three bins (loose, medium, and tight), which are optimized for each signal mass hypothesis and each region based on the expected signal significance. The loose and medium regions are enriched in backgrounds and are used to validate the agreement between the data and the background predictions for the BDT input variables in several regions that probe different background composition. Good agreement between data and the background prediction is observed.

\begin{figure}[!tbp]
\centering
\subfloat[]{\label{fig:1B2b_higgs_m}
\includegraphics[width=0.33\linewidth]{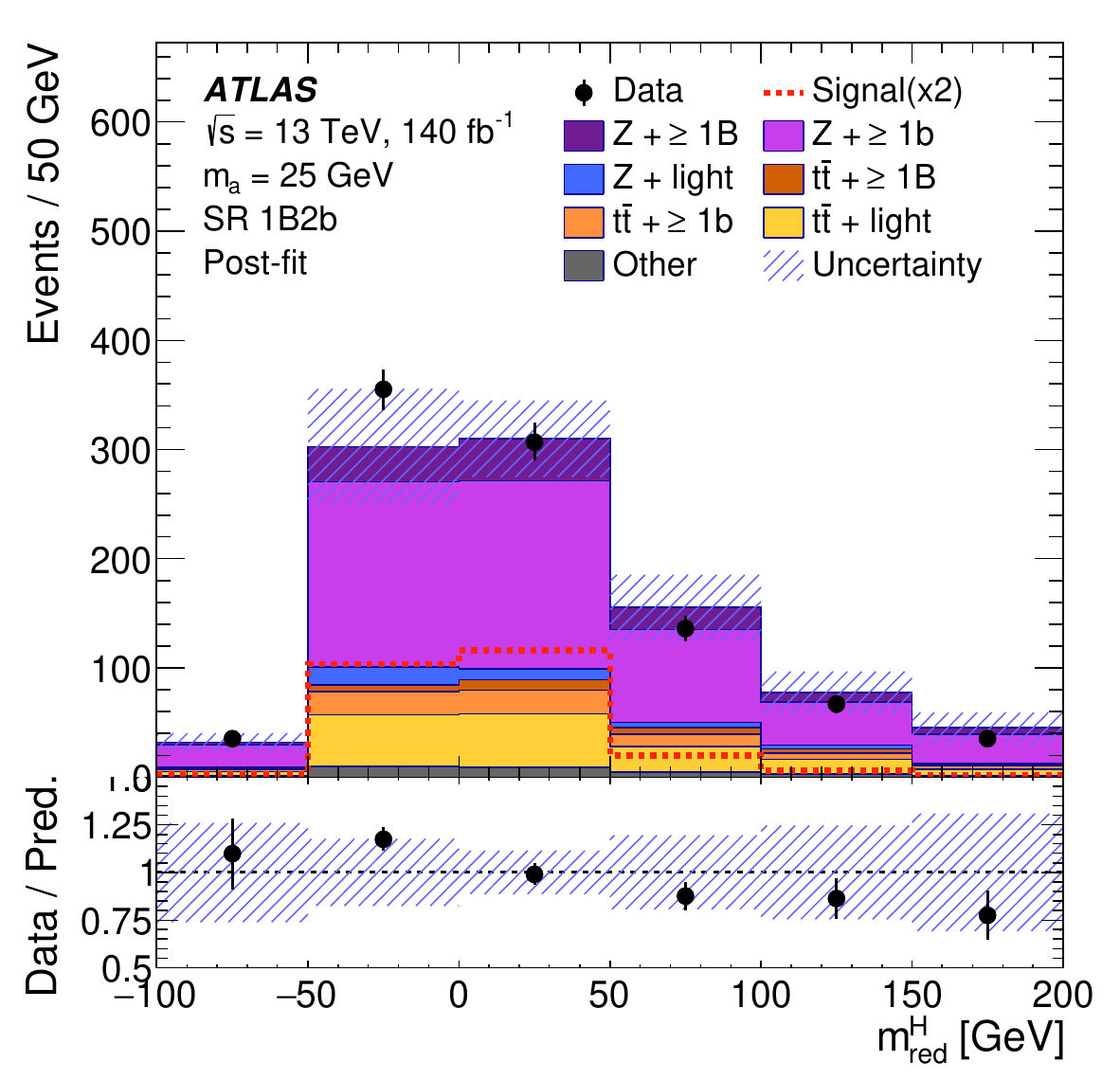}}
\subfloat[]{\label{fig:1B2b_a1_m}
\includegraphics[width=0.33\linewidth]{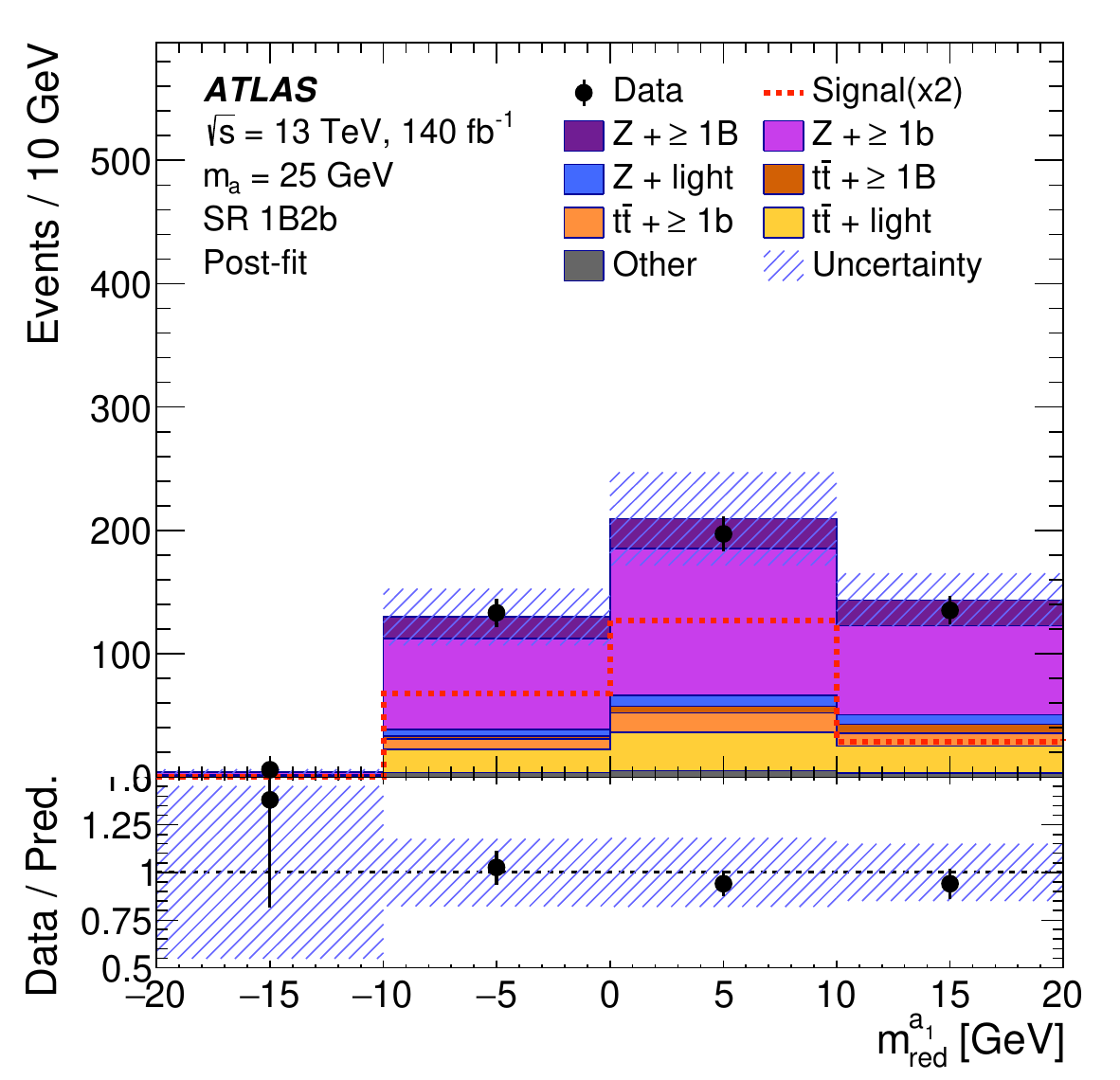}}
\subfloat[]{\label{fig:1B2b_a2_m}
\includegraphics[width=0.33\linewidth]{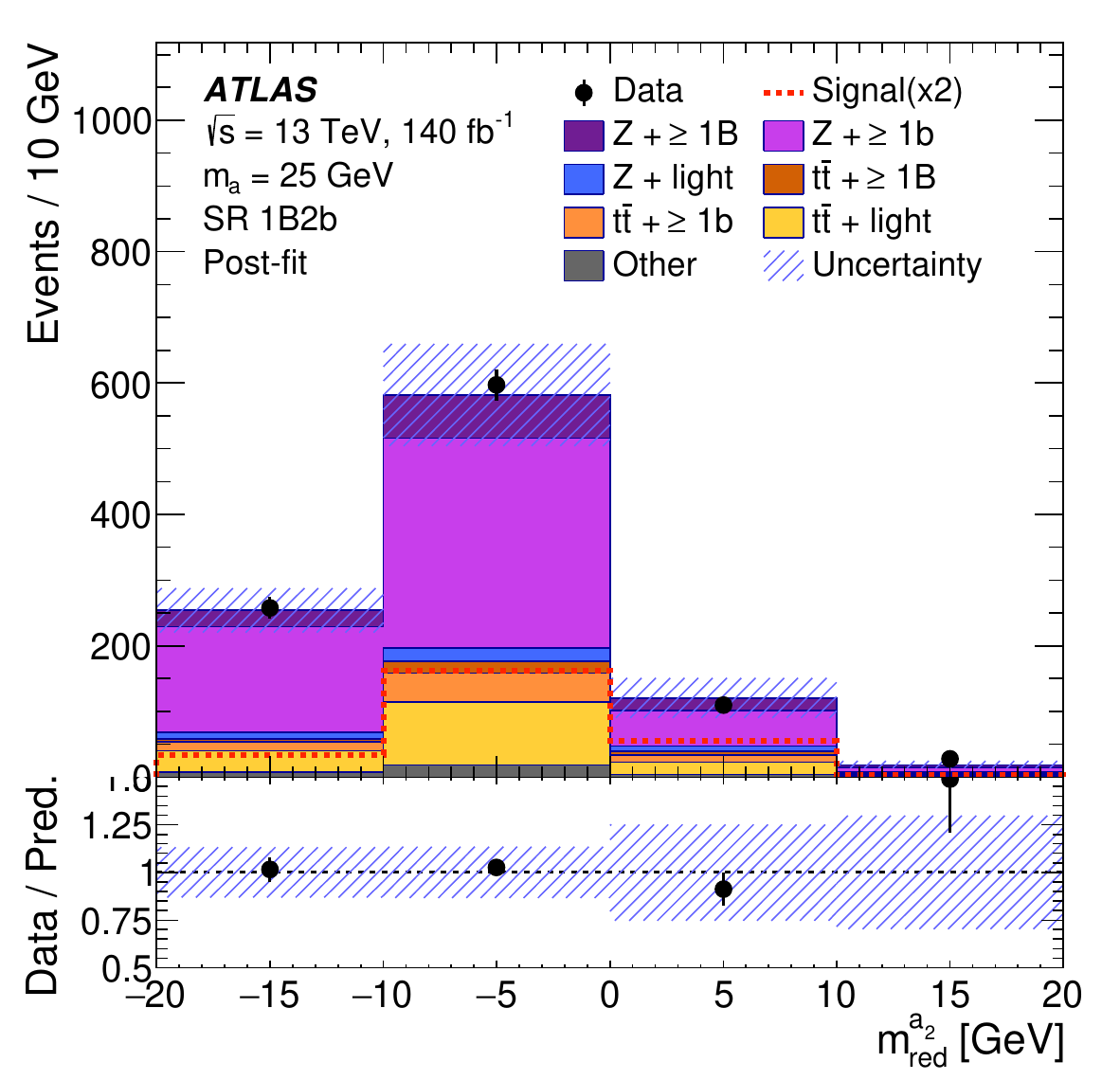}}\\
\caption{The kinematic distributions for events in the $1B2b$ signal region assuming the $a$-boson signal mass hypothesis $m_{a} = 25~\GeV$: \protect\subref{fig:1B2b_higgs_m} the reduced mass of the reconstructed Higgs boson, \protect\subref{fig:1B2b_a1_m} the reduced mass of the reconstructed $a_1$-boson, and \protect\subref{fig:1B2b_a2_m} the reduced mass of the reconstructed $a_2$-boson. The SM background is compared with the expected signal events assuming $\mathcal{B}(H \rightarrow 2a \rightarrow 4b) = 1$ and scaling such an expected signal by a factor of two. These are the highest-ranked variables in the BDT used to discriminate between signal and background. The category labeled ``Other'' includes processes with small contributions to the total yield (diboson, $\ttZ$, and $\ttW$). The lower panels show the ratio of the observed to the estimated SM background. The hashed area represents the total uncertainty in the background.}
\label{fig:kine_mass_SR_1B2b_25}
\end{figure}

\FloatBarrier

\subsection{$2\ell$ channel: $H\rightarrow a_1a_2 \rightarrow 3a_1 \rightarrow 6b$}
\label{sec:discrim_2l_a1a2}

As described in \Sect{\ref{sec:events_2l_a1a2}}, six regions are used to search for $H\rightarrow a_1a_2 \rightarrow 3a_1 \rightarrow 6b$ decays based on the number of $b$-objects in the final state. Even though there is no attempt to identify all the objects in the final state, a similar strategy is employed to discriminate signal and background by combining kinematic variables into a multivariate BDT discriminant in each region. The Higgs boson is constructed from the sum of the $b$-objects, while the $Z$ boson is determined from the two leptons. The input variables are summarized in \Tab{\ref{tab:bdt:input_vars}} and \Tab{\ref{tab:6b-bdt:variables summary}} shows the combination of variables used in each region.

\begin{table}[!tbp]
\caption{List of variables used for the multivariate BDT discriminant in the $6b$ regions.}
\label{tab:6b-bdt:variables summary}
\centering
\begin{tabular}{lcccccc}
\toprule
Variable &  2B0b &  2B1b & 1B2b & 4b & 4b1v & 3b1v \\
\midrule
$m_{\text{incl.}}$ & \checkmark & \checkmark & \checkmark & \checkmark & \checkmark & \checkmark \\
$\Delta R (B, B)$ & \checkmark & \checkmark & -- & -- & -- & -- \\
$\Delta R (B, bb)$ & -- & -- & \checkmark & -- & -- & -- \\
$\Delta R^{\text{average}} (b, b)$ & -- & -- & -- & \checkmark & \checkmark & \checkmark \\
$\Delta R (\exkt,\exkt)$  & \checkmark & -- & \checkmark & -- & -- & -- \\
$\cos(\theta^*) \ (2B)$ & \checkmark & -- & -- & -- & -- & -- \\
$\cos(\theta^*) \ (1B2b)$ & -- & -- & \checkmark & -- & -- & -- \\
$\cos(\theta^*) \ (\text{incl.})$ & -- & -- & -- & \checkmark & -- & \checkmark \\
$\met$  & \checkmark & \checkmark & \checkmark & \checkmark & \checkmark & \checkmark \\
$H_\text{T}^{\text{had}}$ & \checkmark & -- & \checkmark & \checkmark & -- & \checkmark \\
$\Delta R (\ell, \ell)$ & \checkmark & \checkmark & \checkmark & \checkmark & \checkmark & \checkmark \\
\bottomrule
\multicolumn{7}{c}{} \\
\end{tabular}
\end{table}

\FloatBarrier

The kinematic variables include the invariant mass of all reconstructed $b$-objects, $m_{\text{incl.}}$, which peaks around the Higgs boson mass for the signal. Additional variables probe differences between the angular separation between $b$-objects in signal and background, which are typically more separated for backgrounds such as \ttbar since the $b$-objects originate from the decays of different particles. The separation variables depend on the $b$-objects available in the final state and if multiple $b$-objects are available, the ones with the highest transverse momentum are used. If $B$-jets are available, the angular separation between two $B$-jets $\Delta R (B, B)$, between a $B$-jet and the sum of two $b$-jets $\Delta R (B, bb)$, or between the two \exkt track-subjects in the $B$-jet $\Delta R (\exkt,\exkt)$ are used. In the case of multiple $b$-jets, the angular separation between the $b$-jets, averaged over all combinations of $b$-jets, $\Delta R^{\text{average}} (b,b)$, is also used as an input variable.

In addition, the variable $\cos\theta^*$ can also be used to distinguish signal and backgrounds since it probes the spin-$0$ nature of the Higgs boson. In this case, the polar angle of the Higgs boson is used in the rest frame of the parent $ZH$ system. Three variables are used to discriminate signal from background based on the properties of the backgrounds. These include the $\met$ variable, which is expected to be larger in the case of the \ttbar background, due to the presence of neutrinos, and the $H_\text{T}^{\text{had}}$ variable, defined as the scalar sum of the \pt\ of all jets, which is also expected to be larger in \ttbar events due to the larger mass and hadronic activity. Finally, the separation between the leptons $\Delta R (\ell, \ell)$ tends to be smaller for signal than for the \Zjets background since the $Z$ boson tends to have lower transverse momentum in the latter case. The BDT is trained similarly to the $4b$ regions (see \Sect{\ref{sec:discrim_2l}}).

\section{Background estimate}
\label{sec:background}

\subsection{Background estimate for the $2\ell$ channel}
\label{sec:background_2l}

The backgrounds in the $2\ell$ channel are described using MC simulations that are corrected to data for the two main sources of background events: \Zjets and $\ttbar$. The correction factors are derived using data samples enhanced in these two types of background processes following a similar procedure to several other ATLAS analyses~\cite{TOPQ-2020-10,HDBS-2018-51}.

Corrections to $\ttbar$ are derived using events that have two DF opposite-charge leptons and at least two $b$-jets. The corrections are also applied to the single top $tW$ process, since it cannot be distinguished from $\ttbar$ when additional jets are produced. For \Zjets, corrections are derived using events with two leptons with the same flavor and opposite-sign charge (SFOS) and at least two jets. Events are also required to satisfy $n_b+2n_B+n_v=3$, $|m_{\ell\ell}-m_Z| < 20\,\GeV$, and $\met< 60\,\GeV$ to reduce $\ttbar$ contributions.

A first step, only applied to the $\ttbar$ and $tW$ processes, corrects the rate of production of additional heavy flavor (HF) jets, which is known to be mismodeled at high multiplicities of heavy-flavor jets~\cite{TOPQ-2017-12}. This correction is derived from the distribution of the sum of the PC $b$-tagging scores for all jets in the event. The ratio of the observed data and predictions are used to reweight the $\ttbar$ and $tW$ events as a function of this distribution, and \Fig{\ref{fig:sumPCBTag_comparison}} shows the variable after the correction is applied.

\begin{figure}[!tbp]
\centering
\includegraphics[width=0.5\linewidth]{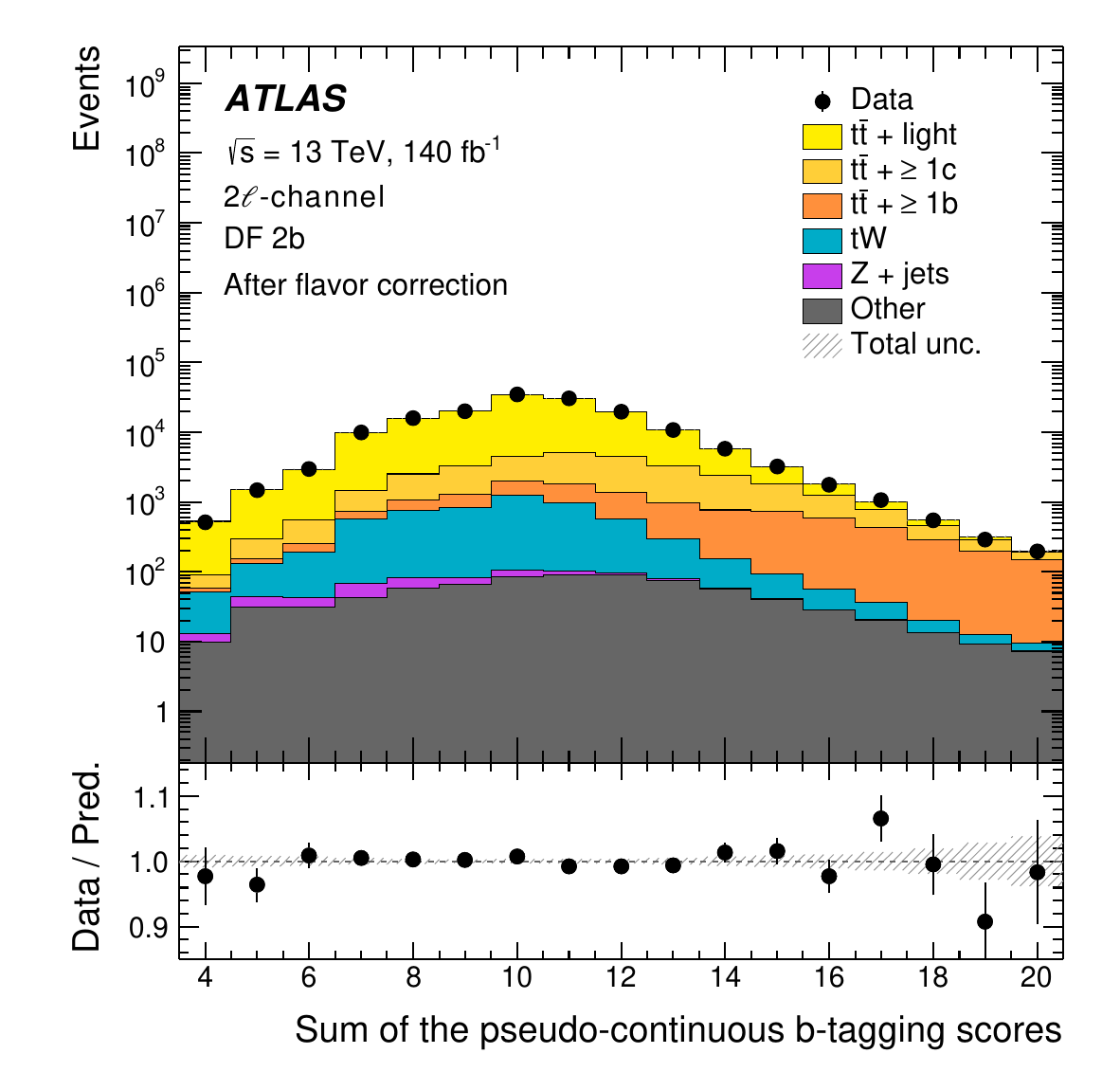}%
\caption{Comparison of data and simulation for the sum of the PC $b$-tagging score of all jets per event, after applying a correction to account for the different fractions of \ttbar components with additional heavy-flavor jets. The lower panel shows the ratio of the observed to the estimated SM background. The shaded area represents the combination of statistical and flavor-tagging uncertainties. The category labeled ``Other'' includes processes with small contributions to the total yield (diboson, $\ttZ$, and $\ttW$). The correction is less than 5\% in the bulk of the distribution, increasing to 20--30\% in the tail.}
\label{fig:sumPCBTag_comparison}
\end{figure}

The second step improves the modeling of kinematic properties, particularly at high jet multiplicities and high jet $\pt$, to match observed distributions in data. For both $\ttbar$ and \Zjets events, the jet multiplicity distribution is corrected in simulation to match the data. An additional correction to improve the modeling of the $\pt$ of objects in the event is applied using different kinematic variables for $\ttbar$ and \Zjets events. For the $\ttbar$ background, the variable used for the correction is a modified version of \HT, the scalar sum of the $\pt$ of all jets and leptons. The reduced \HT or \HTred variable is then defined as the \HT corrected by the average $\pt$ contributed by each additional jet, where the correction is applied to reduce the dependence on the jet multiplicity. The variable \HTred is defined as:
\begin{equation}\nonumber
\HTred = H_\text{T} - (n_\text{j}-2)\Delta \HT,
\end{equation}

where $n_\text{j}$ is the number of jets in the event, and $\Delta \HT$ is the approximate contribution of each additional jet to the \HT, defined as:
\begin{equation}\nonumber
\Delta \HT = \left\langle\frac{\langle \HT^{n_\text{j}}\rangle-\langle \HT^{n_\text{j}=2}\rangle}{n_\text{j}-2}\right\rangle_{n_\text{j}=3,4,...,\geq8},
\end{equation}
where $\HT^{n_\text{j}}$ is the \HT value for an event with jet multiplicity $n_\text{j}$. For the case of \Zjets, the variable used for the correction is the transverse momentum of the $Z$ boson $\pt^{Z}$. The resulting distributions comparing data and simulation after applying the kinematic corrections on the jet multiplicity and on \HTred or $\pt^{Z}$ are displayed in Figures~\ref{fig:modelling_kinematics}~\subref{fig:kin_ttbar} and~\ref{fig:modelling_kinematics}~\subref{fig:kin_zjets} for $\ttbar$ and \Zjets, respectively.

\begin{figure}[!tbp]
\centering
\subfloat[]{\label{fig:kin_ttbar}
\includegraphics[width=0.5\linewidth]{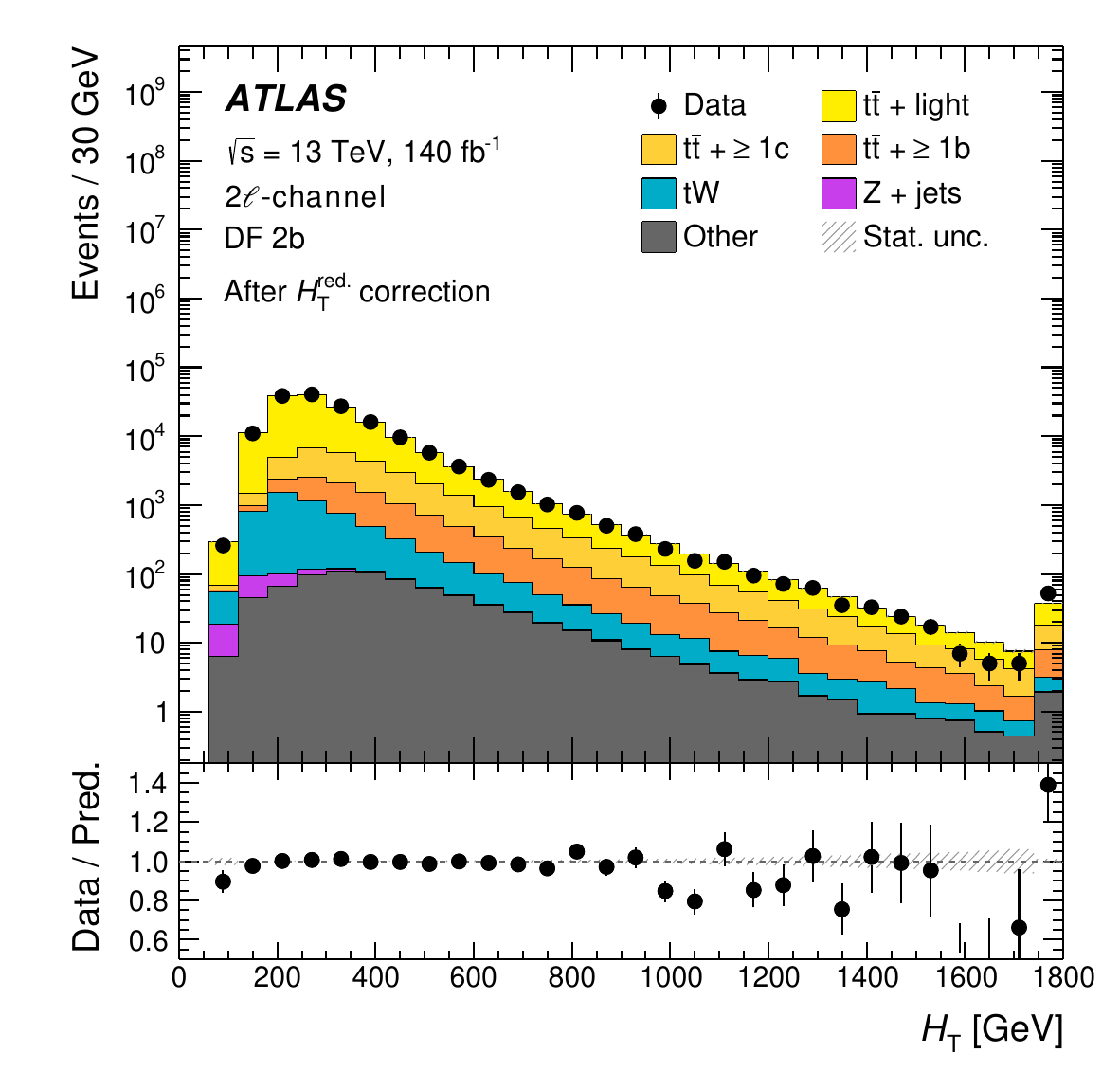}}%
\subfloat[]{\label{fig:kin_zjets}
\includegraphics[width=0.5\linewidth]{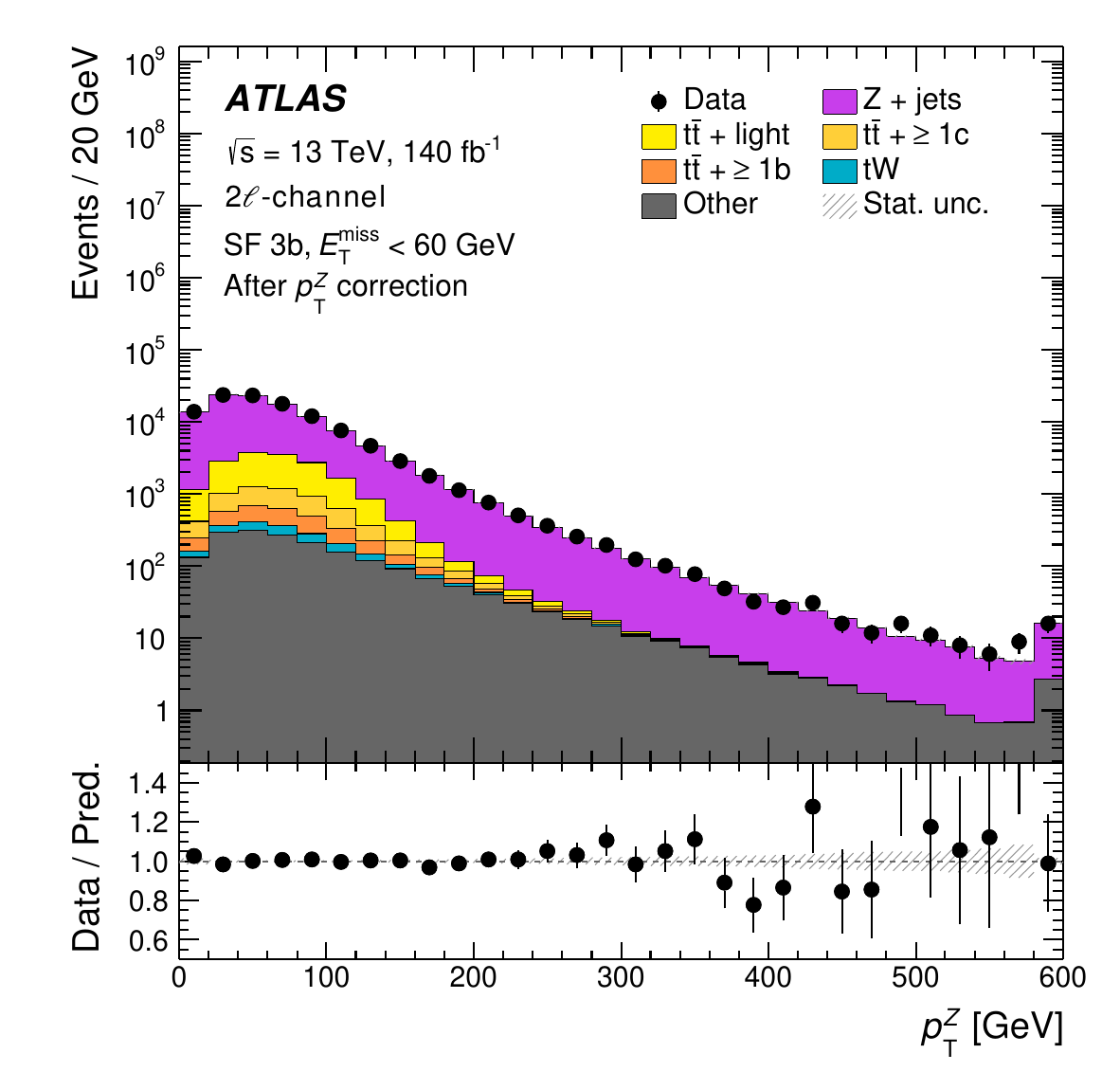}}\\
\caption{Comparison of data and simulation for \protect\subref{fig:kin_ttbar} $H_\text{T}$ distribution used for the modeling of the \ttbar background and \protect\subref{fig:kin_zjets} $p_\text{T}^Z$ distribution used for the modeling of the \Zjets background, after applying corrections. The lower panels show the ratio of the observed to the estimated SM background. The shaded area represents the combination of statistical and flavor-tagging uncertainties. Overflow events are included in the last bin. The category labeled ``Other'' includes processes with small contributions to the total yield (diboson, $\ttZ$, and $\ttW$).}
\label{fig:modelling_kinematics}
\end{figure}

\FloatBarrier

\subsection{Background estimate for the $0\ell$ channel}
\label{sec:background_0l}

For the $0\ell$ channel, the backgrounds can be classified according to the source of \met. The first source is \met from neutrinos, which do not interact with the detector, mainly from $Z$ + jets with $Z \rightarrow \nu\nu$. Another source of \met arises from leptons failing reconstruction, identification, or isolation requirements, referred to as ``lost-leptons,'' primarily from $t\bar{t}$ + jets, $W$ + jets, and single-top events. Finally, \met also arises from the mis-measurement of the jet energy, which can occasionally cause high \met that is typically aligned with the jet. This corresponds to the QCD multijet background.

Several methods are used to estimate the background contributions to the SRs, depending on the source of \met. As described in \Sect{\ref{sec:events_0l}}, CRs enhanced in background processes are used to normalize the dominant background predictions to data. The $2\ell$ CR is used to normalize the $Z+\mathrm{HF}$ background that has \met from neutrinos, while the $1\ell$ CR is used to normalize the backgrounds with lost-leptons.

The QCD multijet background is estimated from data using an ``ABCD'' method, where requirements are applied on two variables separating signal and background defining four regions. The background yield in the SR is estimated by using the other three CRs. The two variables used are the minimum difference between the azimuthal angle of the \met vector and the two $a$-candidates, $\min \Delta\phi(\met, a)$, and the invariant mass of the two $a$-candidates that reconstruct the Higgs boson, $m_{aa}$, where the $a$-candidates are formed from a $B$-jet or two $b$-jets, depending on the final state. The signal tends to have higher $\min \Delta\phi(\met, a)$ compared with the background, and a requirement of $\min \Delta\phi(\met, a) > 60^{\circ}$ is used to define the SR, while events with $\min \Delta\phi(\met, a) < 30^{\circ}$ are used in the CRs. Events in the SR are additionally required to satisfy $75~\GeV < m_{aa} < 175~\GeV$, while events with $275~\GeV < m_{aa} < 375~\GeV$ are used in the CRs. The mass range for the SR is chosen to have high signal efficiency and the range for the CRs are chosen to be of the same size as the SR\@. Validation regions next to the SRs are used to validate the background estimate. A schematic of the regions is shown in \Fig{\ref{fig:ABCD}}.

\begin{figure}[!tbp]
\centering
\includegraphics[width=0.5\linewidth]{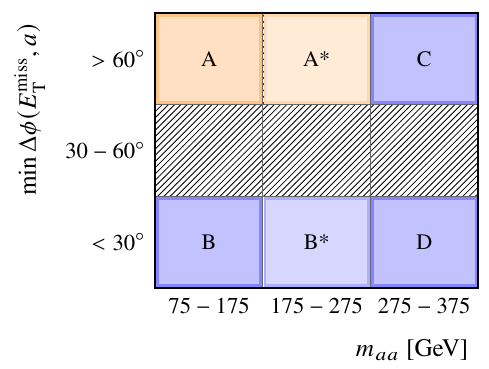}
\caption{Diagram of the four regions used in the ``ABCD'' method to estimate the QCD multijet background for the $0\ell$ channel. The regions A* and B* are used to validate the estimate and derive uncertainties.}
\label{fig:ABCD}
\end{figure}

The QCD multijet background in each of the CRs: B, C, and D is estimated as the difference between event yields in data and the background prediction for all other processes. The QCD multijet background estimate in the SR A is determined as A $=$ B $\times$ (C/D). The QCD multijet background estimate is validated in a sample that is similar to the SR but enhanced in background events (validation region A*, using control regions B*, C, and D). The validation region shows good agreement between data and the background prediction, indicating that the degree of correlation between the two variables used in the ABCD estimate does not have a significant effect on the result. The estimate obtained with the validation region is used to derive an uncertainty in this background contribution, as described in \Sect{\ref{sec:systematics_modeling}}.

\section{Systematic uncertainties}
\label{sec:systematics}

\subsection{Experimental uncertainties}

The uncertainty in the integrated luminosity for the full \RunTwo data is 0.83\%~\cite{DAPR-2021-01}. An uncertainty associated with the modeling of the \pileup in simulation covers the difference between the predicted and measured fiducial inelastic cross-section~\cite{STDM-2015-05}.

Uncertainties associated with charged leptons arise from the trigger, reconstruction, identification, and isolation, and the momentum scale and resolution. Efficiencies, and scale and resolution calibrations are measured using $Z\to \ell^+\ell^-$ and $J/\psi \to \ell^+\ell^-$ events~\cite{EGAM-2018-01,MUON-2018-03}. Systematic uncertainties in these measurements are applied to electrons and muons, accounting for 22 independent variations.

Uncertainties associated with the \met trigger are propagated from a dedicated measurement of data-to-simulation scale factors~\cite{TRIG-2019-01}. The trigger efficiency as a function of the offline \met for an inclusive sample is compared with events with at least four jets. The differences between the efficiency in these two samples are incorporated as an uncertainty as a function of offline \met and these differences are below 2\%.

The jet energy scale uncertainty is derived by combining information from test-beam data, LHC collision data and simulation~\cite{JETM-2018-05}. The jet energy resolution uncertainty is derived from simulation and dijet $\pt$-balance measurements as a function of jet $\pt$ and rapidity~\cite{JETM-2018-05}. The uncertainty in the JVT efficiency to identify and remove jets from \pileup is measured in $Z\to \mu^+\mu^-$ events in data~\cite{PERF-2014-03}. This measurement is extended to cover low-$\pt$ jets with $15~\GeV < \pt < 20~\GeV$, with uncertainties in the range of $1-3\%$. Additional uncertainties are considered related to the jet flavor, \pileup corrections, and high-$\pt$ jets. Systematic uncertainties in these measurements account for 32 independent variations.

The uncertainties in the efficiency to tag jets using \dlr are derived using measurements in data as a function of $\pt$ for $b$-jets, $c$-jets, and light-jets separately. The efficiency to correctly tag $b$-jets is measured using dileptonic \ttbar events~\cite{FTAG-2018-01}. The mis-tag rate for $c$-jets is measured using hadronically decaying \Wboson\ bosons from single-lepton \ttbar events~\cite{FTAG-2020-08}. The mis-tag rate from light jets is measured using a negative-tag method in \Zjets events~\cite{FTAG-2019-02}. Several uncertainty components are considered for the calibration of the four working points used, resulting in 45, 20, and 20 uncorrelated variations for $b$-, $c$-, and light jets, respectively.

Uncertainties in the efficiency to tag jets using \DEXTER are derived using calibration measurements in data as a function of the $\pt$ of the jet for $B$-jets and $b$-jets~\cite{ATL-PHYS-PUB-2022-042}. The other flavors have negligible contributions to the analysis. Efficiencies to correctly tag $B$-jets and mis-tag $b$-jets are derived using \ttbar and $Z$+$g(\to \bbbar)$ events. The efficiency measurement uses gluon splitting $g\to \bbbar$, since it is the main SM source of $b$-quark pairs at low $\pt$. This measurement is extended to cover high-$\pt$ jets with $200~\GeV < \pt < 500~\GeV$ using the same calibration methods~\cite{ATL-PHYS-PUB-2022-042}, which is particularly important for the $0\ell$ channel. The \DEXTER efficiencies are measured in several bins of \pt and generator-level flavor. All sources of systematic uncertainty are described in Ref.~\cite{ATL-PHYS-PUB-2022-042}. The covariance matrix from a simultaneous fit to all bins in jet $\pt$, working point range, and generator-level flavor of the jets are diagonalized into 12 uncorrelated eigenvectors that are used as uncertainty variations.

Uncertainties in the reconstruction efficiency for $b$-decays and the mis-tag rate for the soft-v identified using the TC-LVT tagger are derived using \ttbar events, as described in Ref.~\cite{FTAG-2023-02}. The uncertainties are derived as a function of the actual \pileup and are accounted for as three independent sources. Since the soft-v are reconstructed from tracks, uncertainties in the selection efficiency and the mis-reconstruction of tracks are propagated to the soft-v four momentum. The tracking uncertainties are derived from studies of the material in the ID and the physics models used in the \GEANT detector simulation~\cite{ATL-PHYS-PUB-2015-051,PERF-2015-08} and are included as 12 independent variations.

A large-$R$ track-jet mass scale correction is determined using $Z+\text{jets}$ events with one $b$-jet and one $B$-jet. These events have large HF jet multiplicity but are not part of any signal or control region. The data-to-simulation mass scale correction is derived from a fit to the observed large-$R$ track-jet mass distribution. The correction required is below 5\% and the uncertainty in the fit is propagated to the analysis.

All uncertainties related to the energy scales or resolution of the reconstructed objects are propagated to the \met calculation. Additional uncertainties are derived using measurements in $Z\to\ell\ell$ events in data~\cite{JETM-2020-03}. Three additional independent uncertainties associated with the scale and resolution of the soft term are also included.

\subsection{Modeling uncertainties}
\label{sec:systematics_modeling}

The systematic uncertainties affecting the modeling of the signal and the main backgrounds, \ttbar and \Zjets, are due to effects such as missing higher-order terms in the perturbative QCD calculations, and the choice of PDF set, PS and hadronization models. Several uncertainties are derived using alternative simulated MC samples, which are listed in \Tab{\ref{tab:modeling_samples}}.

For the \ttbar background, an uncertainty related to the amount of initial-state radiation (ISR) is estimated by simultaneously varying the renormalization and factorization scales in the ME and in the PS~\cite{ATL-PHYS-PUB-2020-023}, and an uncertainty related to the final-state radiation (FSR) is estimated by varying scales in the PS\@. For the \Zjets background, an uncertainty is derived from the envelope of seven scale variations of the renormalization $\mu_{\mathrm{R}}$ and factorization $\mu_{\mathrm{F}}$ scales covering the pairwise combinations of increasing and decreasing the scales by a factor of two. The effect of PDF variations~\cite{Butterworth:2015oua} on the shape of distributions and the uncertainty in the value of $\alpha_s$ are also considered.

For the \ttbar background, an uncertainty associated with the choice of PS and hadronization model is assessed by comparing the nominal sample generated with {\POWHEGBOX[v2]}+\PYTHIA[8] to a sample generated with {\POWHEGBOX[v2]}+\HERWIG[7]. A comparison with {\MGNLO}+\PYTHIA[8] is also used to assess the uncertainty arising from both the ME corrections and the choice of procedure matching ME and PS at NLO.

For the \Zjets and \Wjets backgrounds, the nominal sample generated with \SHERPA is compared with an alternative sample with {\MGNLO}+\PYTHIA[8], which is produced with up to three additional partons at NLO accuracy~\cite{PMGR-2021-01}. The uncertainty in the heavy flavor components is derived by comparing the nominal \Zjets sample with alternatives with varied settings for the overlap between matrix element and parton shower emissions and for the resummation scale~\cite{Cascioli:2013gfa}. An uncertainty in the ME matching (CKKW) is estimated by varying the scale used to distinguish between jets from the ME and the PS~\cite{Hoeche:2009rj,Catani:2001cc}. An uncertainty in the resummation scale (QSF) is estimated by varying the scale used for the resummation of soft gluon emission by a factor of two~\cite{Bothmann:2016nao}.

Uncertainties are derived for the modeling corrections applied to the \Zjets and \ttbar backgrounds, described in \Sect{\ref{sec:background_2l}}. The uncertainties in the parameters of the functions used to improve the modeling in simulation are propagated to the final result of the analysis.

In the case of the QCD multijet background, described in \Sect{\ref{sec:background_0l}}, several sources of uncertainties are considered, including the full statistical and systematic uncertainties associated with the contributions from other background sources and the differences between the estimate obtained with the validation region compared with the SR\@. Even though the uncertainty in this background component is around 100\%, this background is not a significant contributor in the SRs and has a negligible impact on the result.

\section{Statistical treatment}
\label{sec:statistics}

Event yields in each of the control and signal regions are combined to test for the presence of a signal. The statistical analysis is based on a binned likelihood function $\mathcal{L}(\mu, \theta)$ constructed as a product of Poisson probability terms over all BDT bins in the signal region and every control region considered. This function depends on the signal-strength parameter $\mu$, a multiplicative factor to the theoretical signal production cross-section for $ZH$ production times the branching fraction of the Higgs boson to the final state considered in each model $\mathcal{B}(H\to 2a\to 4b)$, $\mathcal{B}(H\to a_1a_2\to 4b)$, or $\mathcal{B}(H\to a_1a_2\to 6b)$ and $\theta$, a set of nuisance parameters (NP) that encode the effect of systematic uncertainties in the signal and background expectations and that are implemented in the likelihood function as Gaussian distribution constraint terms. Therefore, the total expected number of events in a given bin depends on $\mu$ and $\theta$. The NPs $\theta$ allow variations of the expectations for signal and background according to the corresponding systematic uncertainties. Their fitted values correspond to the deviations from the nominal expectations that globally provide the best fit to the data. This procedure reduces the impact of systematic uncertainties in the search sensitivity by taking advantage of the highly populated background-dominated channels included in the likelihood fit. It requires a good understanding of the systematic effects affecting the shapes of the discriminant distributions, for which detailed validation studies of the fitting procedure are performed using the simulation. Statistical uncertainties in each bin of the discriminant distributions are also taken into account by dedicated parameters in the fit.

Systematic uncertainties are pruned away and smoothed to reduce the time for fitting, make the minimization more stable, and deal with unphysical effects due to statistical fluctuations in the uncertainties. The shape and the normalization of systematic uncertainties are pruned separately per region and sample. The pruning algorithm removes systematic uncertainties that have an impact on the normalization below one percent and that cause a negligibly small impact on the shape of the final discriminant (all bins are within one percent of the nominal shape). Smoothing is applied to the systematic variations to mitigate the effects of statistical fluctuations.

The test statistic $\tilde{q}_{\mu}$ is defined as the profile likelihood ratio, $\tilde{q}_{\mu} = -2 \text{ln}(\mathcal{L}(\mu, \hat{\hat{\theta}}_\mu)/\mathcal{L}(\hat{\mu}, \hat{\theta}))$, where $\hat{\mu}$ and $\hat{\theta}$ are the values of the parameters that maximize the likelihood function (with the constraint $0\leq\hat{\mu}\leq\mu$), and $\hat{\hat{\theta}}_\mu$ are the values of the NP that conditionally maximize the likelihood function for a given value of $\mu$. The test statistic $\tilde{q}_{\mu}$ is implemented in the \textsc{RooStats} package~\cite{Moneta:2010pm} and is used to measure the compatibility of the observed data with the signal plus background hypothesis (i.e., the exclusion test). The primary statistical analysis package used is \textsc{TRExFitter}~\cite{TRexFitter}, which makes use of the \textsc{RooFit}~\cite{RooFit} and \textsc{RooStats}~\cite{Moneta:2010pm} packages. The $p$-value representing the compatibility of the data with the background-only hypothesis is estimated by integrating the distribution of $q_0$ from signal plus background pseudo-experiments approximated using the definition and asymptotic formulae given in Ref.~\cite{Cowan:2010js}, above the observed value. The observed $p$-value is checked for each explored signal scenario. In the absence of any significant excess above the background expectation, upper limits on the signal production cross-section for each signal scenario are derived using $\tilde{q}_{\mu}$ in the CLs method~\cite{Cowan:2010js}. For a given signal hypothesis, values of the branching fractions of exotic Higgs boson decays times the ratio of the production cross-section over the SM prediction (parameterized by $\mu$) yielding CLs$<$0.05, where CLs is computed using the asymptotic approximation, are excluded at 95\% CL\@. The limits computed with the asymptotic approximation are compared with calculations based on pseudo-experiments for several representative cases and found to have relative differences below 10\%.

Statistical uncertainties in each region are taken into account by dedicated NPs with Poisson constraint terms.
For the \Zjets background,  the $Z+{\geq}1b$ and ${\geq}1B$ components are normalized by one common unconstrained NP that is instead mainly determined using the dedicated \Zjets CRs. Extensive studies on the possibility of using separate parameters for additional $Z+\mathrm{HF}$ categories, e.g., separate parameters for $Z+{\geq}1b$ and $Z+{\geq}1B$, were carried out. However, due to the lack of CRs with enough purity in additional $Z+\mathrm{HF}$ categories, the fit does not have the ability to constrain them separately. The uncertainty in the relative fraction of each $Z+\mathrm{HF}$  background component  is accounted for by modeling uncertainties instead. The common \Zjets background normalization parameter is fitted in the range of 1.04--1.18 with a typical uncertainty of about 25\%.
For the $\ttbar$+jets background, the overall $\ttbar$ normalization, and the fractions of the $\ttbar+{\geq}1b$ and $\ttbar+{\geq}1B$ components are normalized by three distinct unconstrained NPs. They are mainly determined in the dedicated \ttbar CRs that are split by the reconstructed $b$-object multiplicity. The overall $\ttbar$ normalization parameter is in the range of 0.96--1.03 depending on the mass hypothesis with a typical uncertainty of 20\%, while the $\ttbar+{\geq}1b$ and $\ttbar+{\geq}1B$ normalization parameters are in the range of 0.87--1.17 and 1.08--1.39, with typical uncertainties of approximately 65\% and 80\%, respectively. In the $0\ell$ channel, since the description of backgrounds coming from lost-leptons is modeled with different methods (see \Sect{\ref{sec:background_0l}}), an independent normalization parameter is introduced with a fitted value in the range of 1.25--1.28 and a typical uncertainty of 16\%.

\subsection{Combination of $0\ell$ and $2\ell$ channels}

The analyses targeting the $0\ell$ and $2\ell$ $H\rightarrow 2a\rightarrow 4b$ channels are combined in the final result. The combination of the two channels is made easier by the fact that both analyses use the same modeling of background samples and the same systematic uncertainty model. The SRs of the $0\ell$ and $2\ell$ channels are orthogonal thanks to the lepton vetoes of these regions and the strictly looser lepton definition ($\pt > 7~\GeV$ compared with $\pt > 10~\GeV$) of the $0\ell$ channel compared with the $2\ell$ channel. The $2\ell$ CRs of the $0\ell$ and $2\ell$ channels overlap, so when the combination is performed the \Zjets normalization parameter in the $0\ell$ channel is constrained by the CRs of the $2\ell$ channel. Most common systematic uncertainties are correlated between the two channels. \Tab{\ref{tab:uncorrelated_sys}} shows the few de-correlated systematic uncertainties along with the rationale for their treatment.

\begin{table}[ht]
\caption{Summary of de-correlated systematic uncertainties used to combine the $0\ell$ and $2\ell$ channels.}
\label{tab:uncorrelated_sys}
\centering
\begin{tabular}{llcc}
\toprule
\multicolumn{2}{l}{Systematic uncertainties} & Channels affected & Motivation for de-correlation \\
\midrule
\multicolumn{2}{l}{QCD multijet uncertainty} & \multirow{5}{*}{$0\ell$} & \multirow{5}{*}{Only impacts $0\ell$ channel} \\
\multicolumn{2}{l}{$\met$ trigger scale factor} & & \\
\multicolumn{2}{l}{High-$\pt$ \DEXTER flavor tagging} & & \\
\multicolumn{2}{l}{$W$ + jets modeling} & & \\
\multicolumn{2}{l}{Single top modeling} & & \\
\midrule
\multicolumn{2}{l}{Low-$\pt$ \dlr flavor tagging} & \multirow{3}{*}{$2\ell$} & \multirow{3}{*}{Only impacts $2\ell$ channel} \\
\multicolumn{2}{l}{Low-$\pt$ JVT} & & \\
\multicolumn{2}{l}{TC-LVT flavor tagging} & & \\
\midrule
\multicolumn{2}{l}{$W/Z$+HF modeling } & \multirow{2}{*}{Both} & \multirow{2}{*}{Different kinematic phase-space} \\
\multicolumn{2}{l}{$t\bar{t}$ background modeling } & & \\
\bottomrule
\end{tabular}
\end{table}

\section{Results}

The validation of the background modeling is performed in several different signal-depleted regions. In the $2\ell$ channel, the validation is performed after the quadruplet-selection NN using signal-depleted regions corresponding to low BDT discriminator score values. Good agreement is observed as function of several key kinematic variables such as the reconstructed $a$-boson and Higgs boson masses, as shown in \Fig{\ref{fig:kine_mass_SR_1B2b_25}}. Further validation is performed using events with different-flavor lepton pairs or same-flavor lepton pairs away from the $Z$ boson mass, which are dominated by processes including top quarks. As described in \Sect{\ref{sec:events}} (Tables~\ref{tab:eventselection},~\ref{tab:4b:cr_summary},~\ref{tab:6b-bdt:regions summary}, and~\ref{tab:0l4b_regions_summary}), these regions are split by the number of $B$-jets and $b$-jets to validate the production of heavy-flavor in background processes. Good agreement is observed in all regions, as shown in Figures~\ref{fig:results_0l_yields} and~ \ref{fig:results_2l_yields}.

\Fig{\ref{fig:sr_unblinded_ZHaa_ma25}} shows the distributions of the multivariate discriminant for the different object categories for the $2\ell$ channel. The signal mass hypothesis $m_a = 25~\GeV$ has the largest difference between the data and the background prediction in the tightest BDT bins among all masses tested in the $H \rightarrow 2a \rightarrow 4b$ model. The background-rich bins are consistent between the data and the background prediction. \Fig{\ref{fig:results_a1a2_yields}} shows the yields for the signal and control regions for the searches for the $H \rightarrow a_1a_2 \rightarrow 4b$ and $H \rightarrow a_1a_2 \rightarrow 3 a_1 \rightarrow 6b$ processes. Upper limits on the branching fraction $ \mathcal{B}(H \rightarrow 2a/a_1a_2 \rightarrow 4b/6b)$ times the ratio of the $ZH$ production cross-section to the SM prediction \sigratio are determined at 95\% CL using the CLs technique~\cite{Cowan:2010js,Read:2002hq,Junk:1999kv} and the $\tilde{q}_{\mu}$ test statistic described in Section~\ref{sec:statistics}. The SM Higgs boson production cross-section used is $\sigma_{\text{SM}}(ZH) = 0.88\,\text{pb}$~\cite{LHCHXSWG3}.

\begin{figure}[!tbp]
\centering
\includegraphics[width=0.6\textwidth]{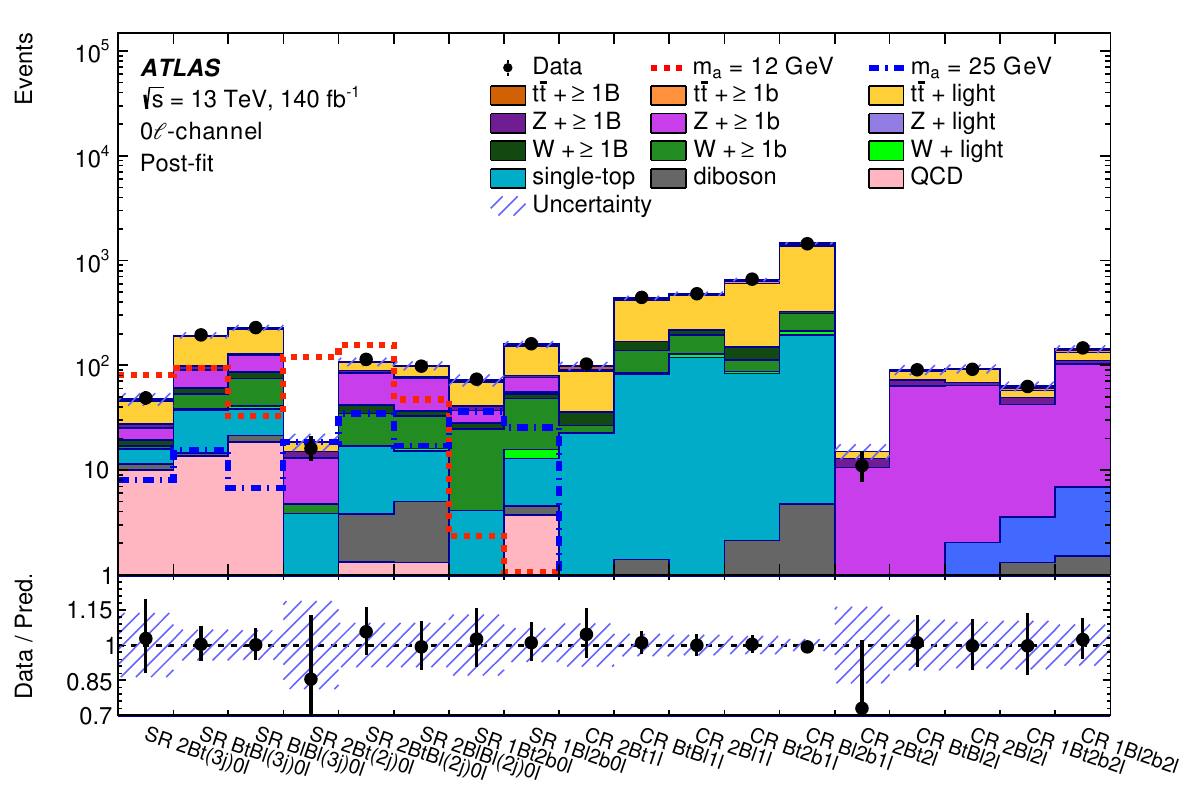}
\caption{Observed yields for the signal and control regions in the $0\ell$ channel after the background-only fit to data. The signal for the $m_a$ = 12~\GeV and 25~\GeV mass hypotheses are overlaid (dashed line) assuming a branching fraction $\mathcal{B}(H \rightarrow 2a \rightarrow 4b) = 1$ and the SM rate of production of Higgs bosons. The lower panels show the ratio of the observed to the estimated SM background. The hashed area represents the total uncertainty of the background. The category labeled ``Other'' includes processes with small contributions to the total yield (diboson, $\ttZ$, and $\ttW$).}
\label{fig:results_0l_yields}
\end{figure}

\begin{figure}[!tbp]
\centering
\subfloat[][]{\includegraphics[width=0.5\textwidth]{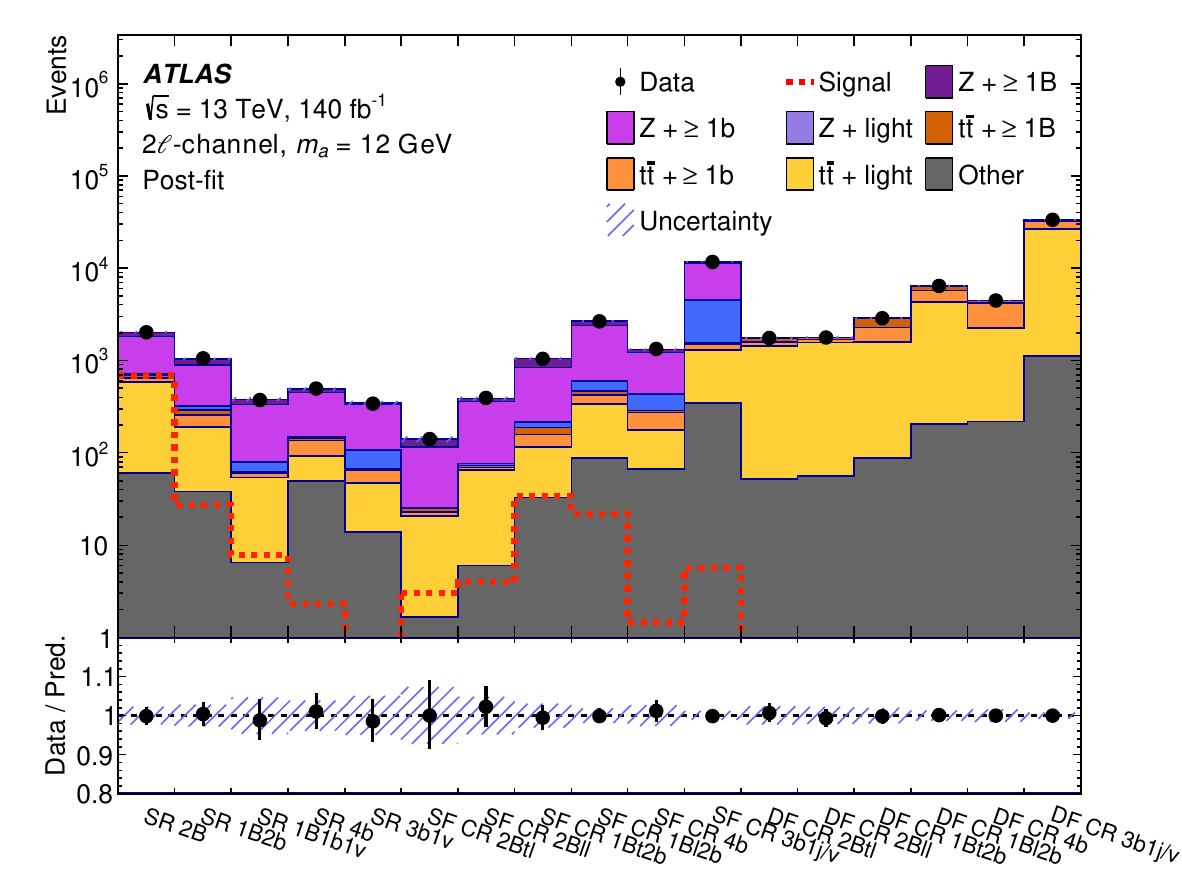}\label{fig:results_2l_yields_12}}
\subfloat[][]{\includegraphics[width=0.5\textwidth]{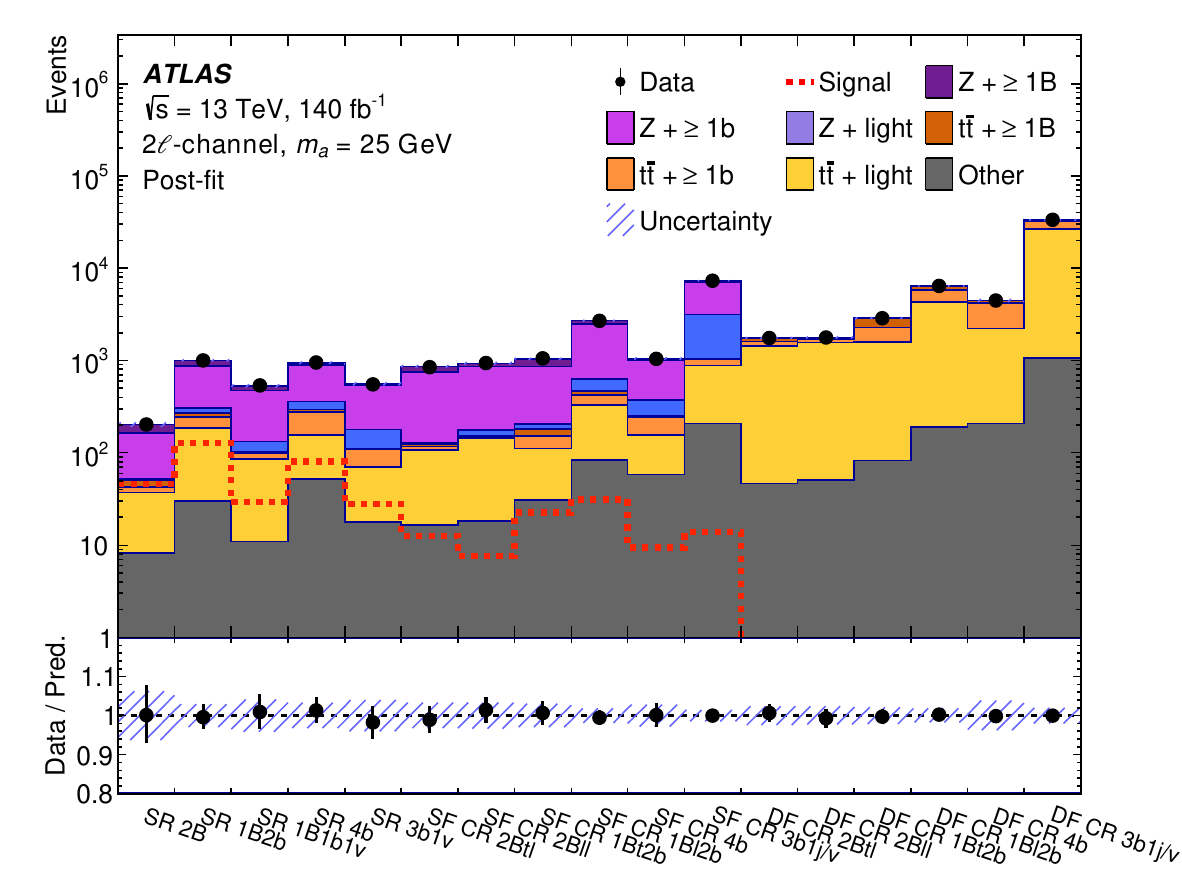}\label{fig:results_2l_yields_25}}
\caption{Observed yields for the signal and control regions in the $2\ell$ channel for the \protect\subref{fig:results_2l_yields_12} $m_a$ = 12~\GeV mass hypothesis and the \protect\subref{fig:results_2l_yields_25} $m_a$ = 25~\GeV mass hypothesis after the background-only fit to data. The signal is overlaid (dashed line) assuming a branching fraction $\mathcal{B}(H \rightarrow 2a \rightarrow 4b) = 1$ and the SM rate of production of Higgs bosons. The lower panels show the ratio of the observed to the estimated SM background. The hashed area represents the total uncertainty of the background. The category labeled ``Other'' includes processes with small contributions to the total yield (diboson, $\ttZ$, and $\ttW$).}
\label{fig:results_2l_yields}
\end{figure}

\begin{figure}[!tbp]
\centering
\subfloat[][]{\label{SR_2B}
\includegraphics[width=0.3\textwidth]{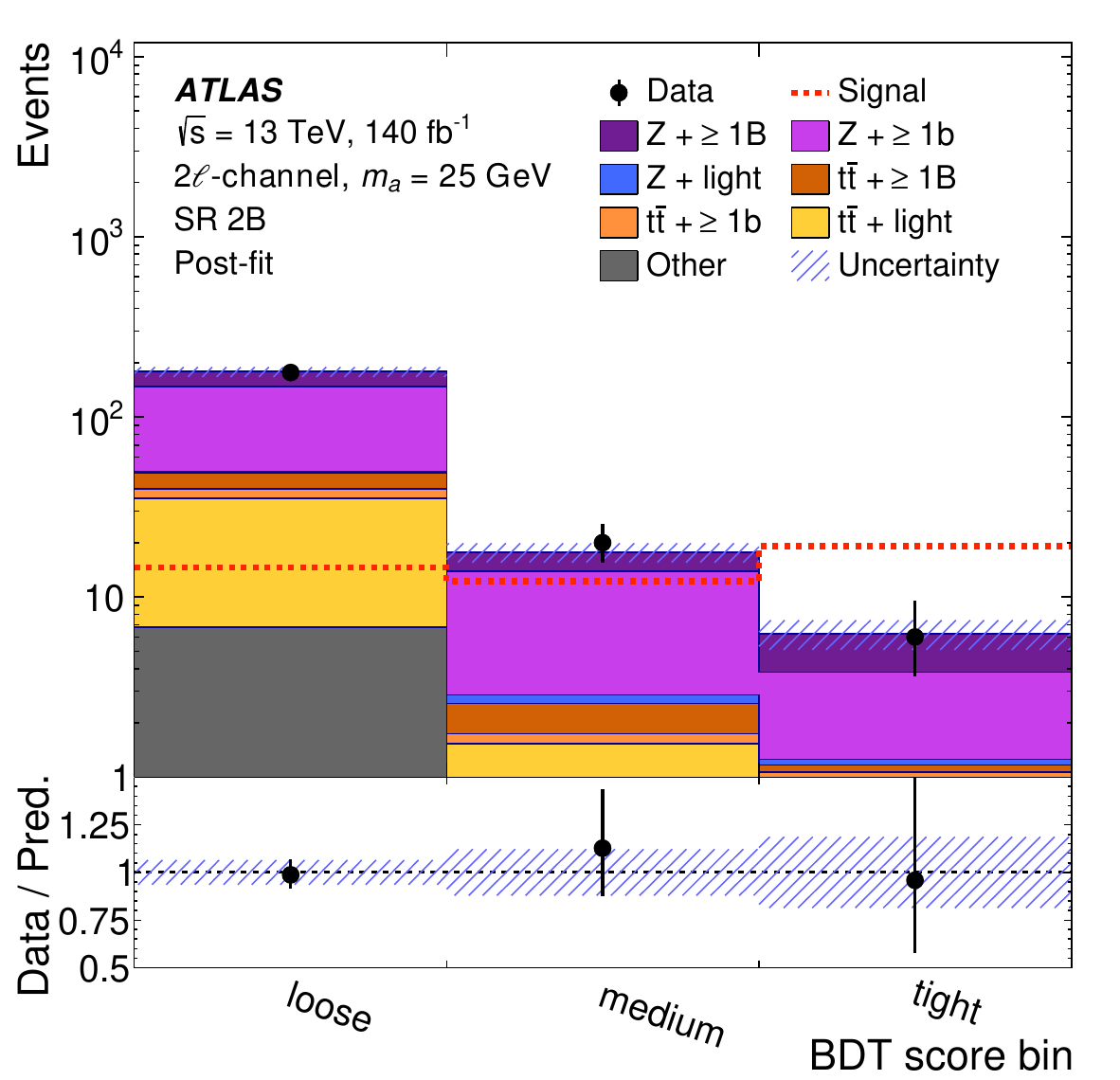}}
\subfloat[][]{\label{SR_1B2b}
\includegraphics[width=0.3\textwidth]{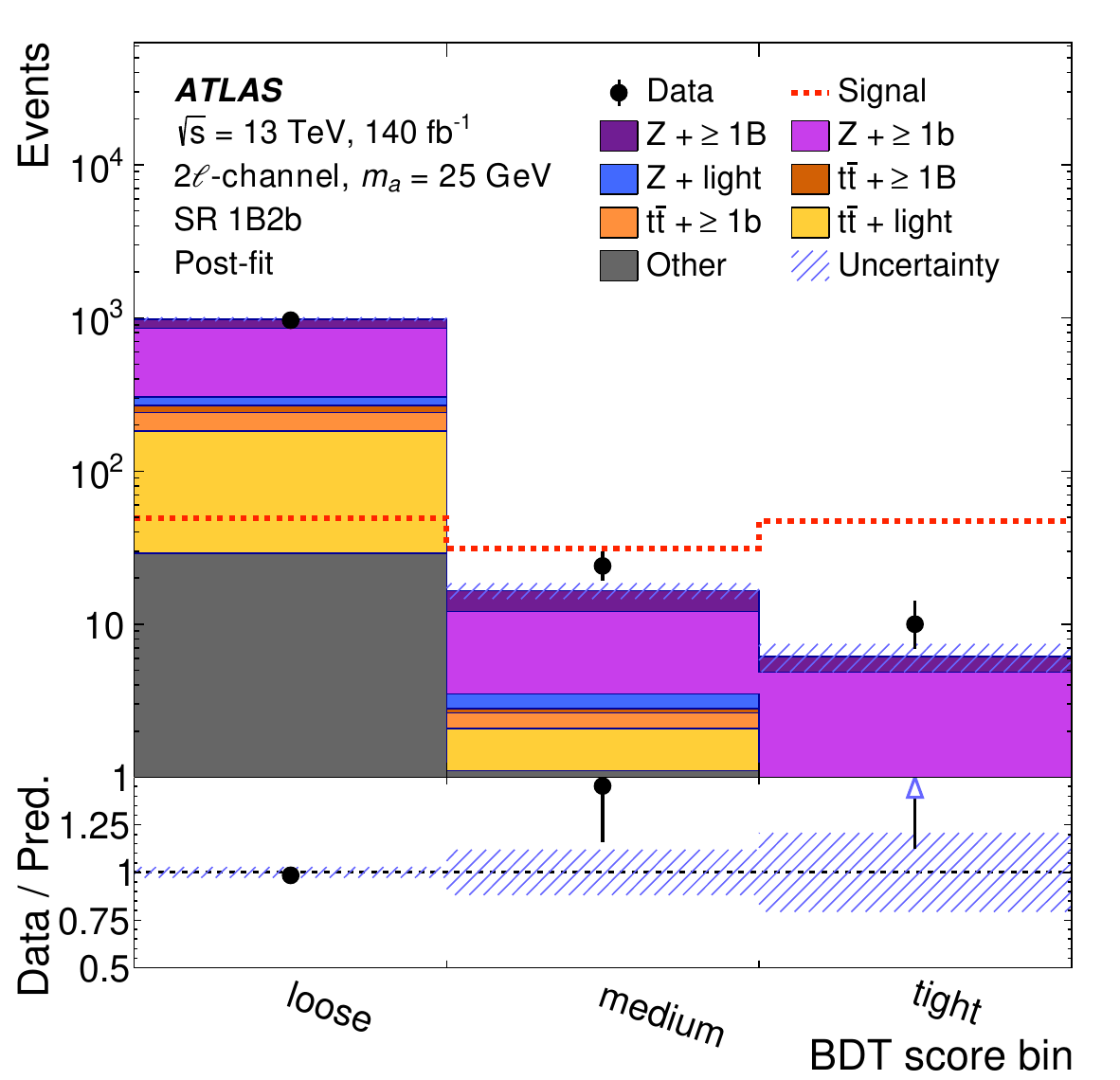}}
\subfloat[][]{\label{SR_1B1b1v}
\includegraphics[width=0.3\textwidth]{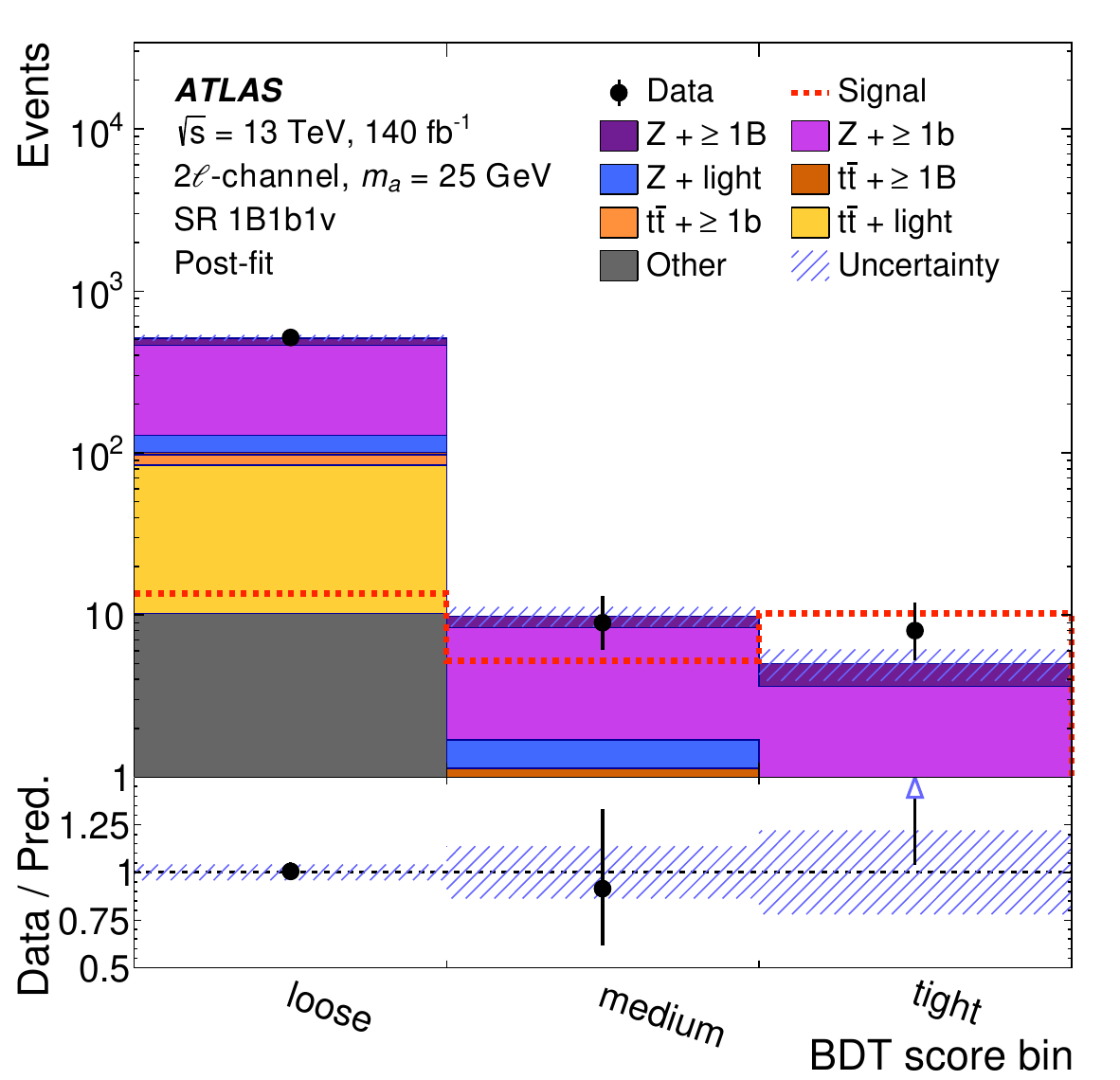}}\\
\subfloat[][]{\label{SR_4b}
\includegraphics[width=0.3\textwidth]{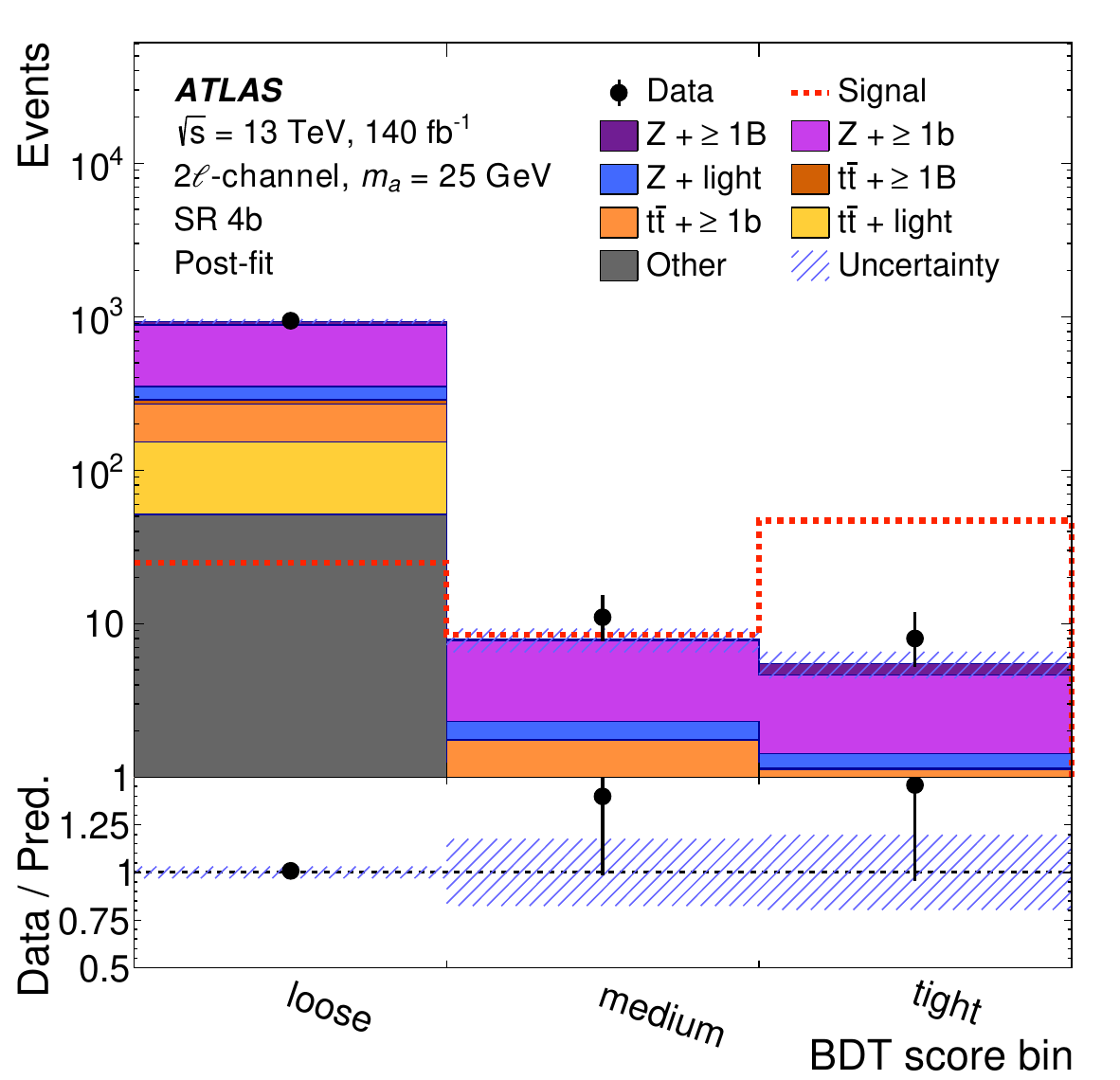}}
\subfloat[][]{\label{SR_3b1v}
\includegraphics[width=0.3\textwidth]{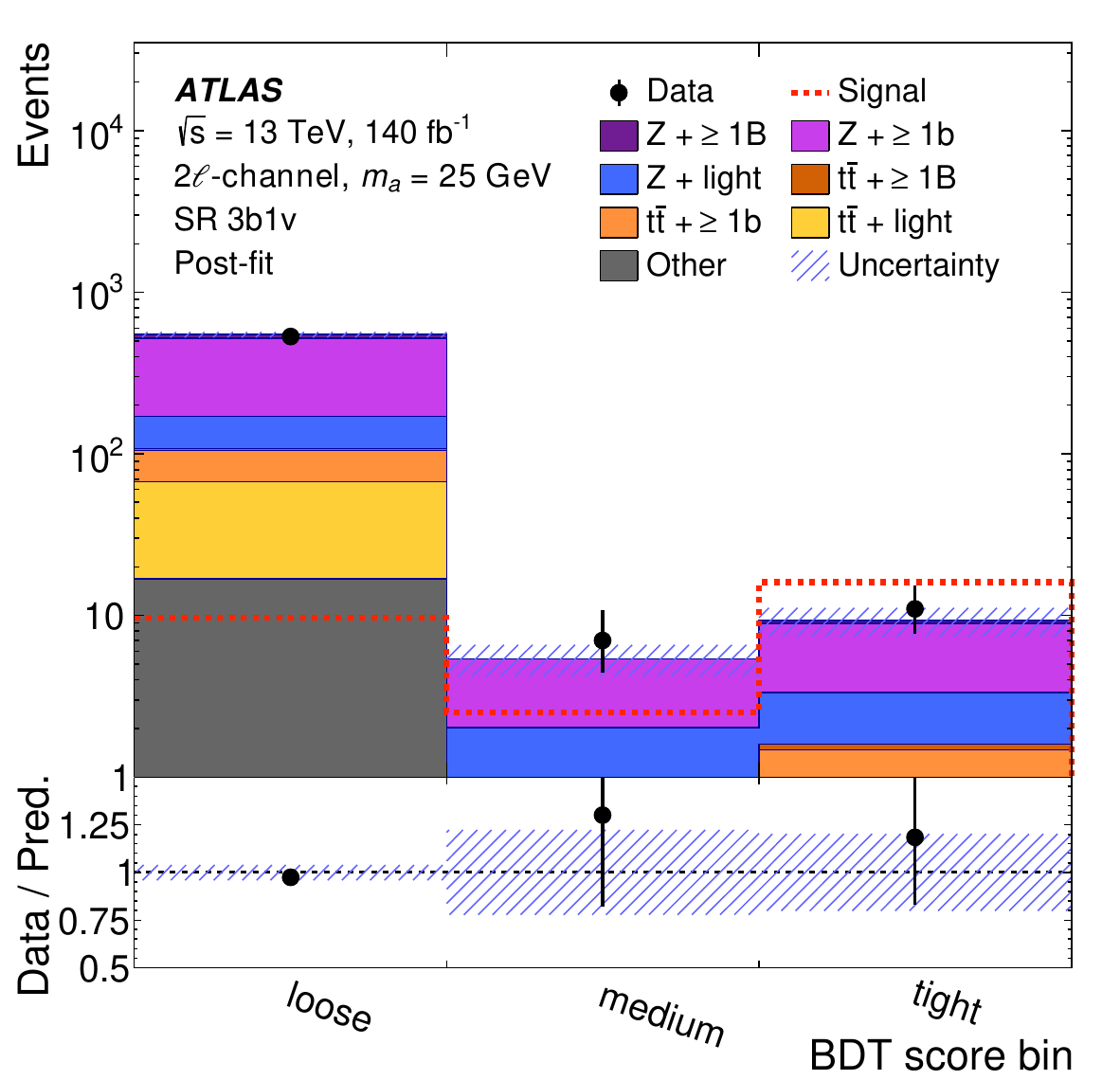}}
\caption[.]{Observed yields for the signal regions in the $2\ell$ channel for the $m_a$ = 25~\GeV mass hypothesis after the background-only fit to data in the \protect\subref{SR_2B} $2B$, \protect\subref{SR_1B2b} $1B2b$, \protect\subref{SR_1B1b1v} $1B1b1$v, \protect\subref{SR_4b} $4b$, and \protect\subref{SR_3b1v} $3b1$v regions. The signal is overlaid (dashed line) assuming a branching fraction $\mathcal{B}(H \rightarrow 2a \rightarrow 4b) = 1$ and the SM rate of production of Higgs bosons. The lower panels show the ratio of the observed to the estimated SM background. The hashed area represents the total uncertainty of the background. The category labeled ``Other'' includes processes with small contributions to the total yield (diboson, $\ttZ$, and $\ttW$).}
\label{fig:sr_unblinded_ZHaa_ma25}

\end{figure}

\begin{figure}[!tbp]
\centering
\subfloat[][]{\label{ma1ma2_2030}
\includegraphics[width=0.5\textwidth]{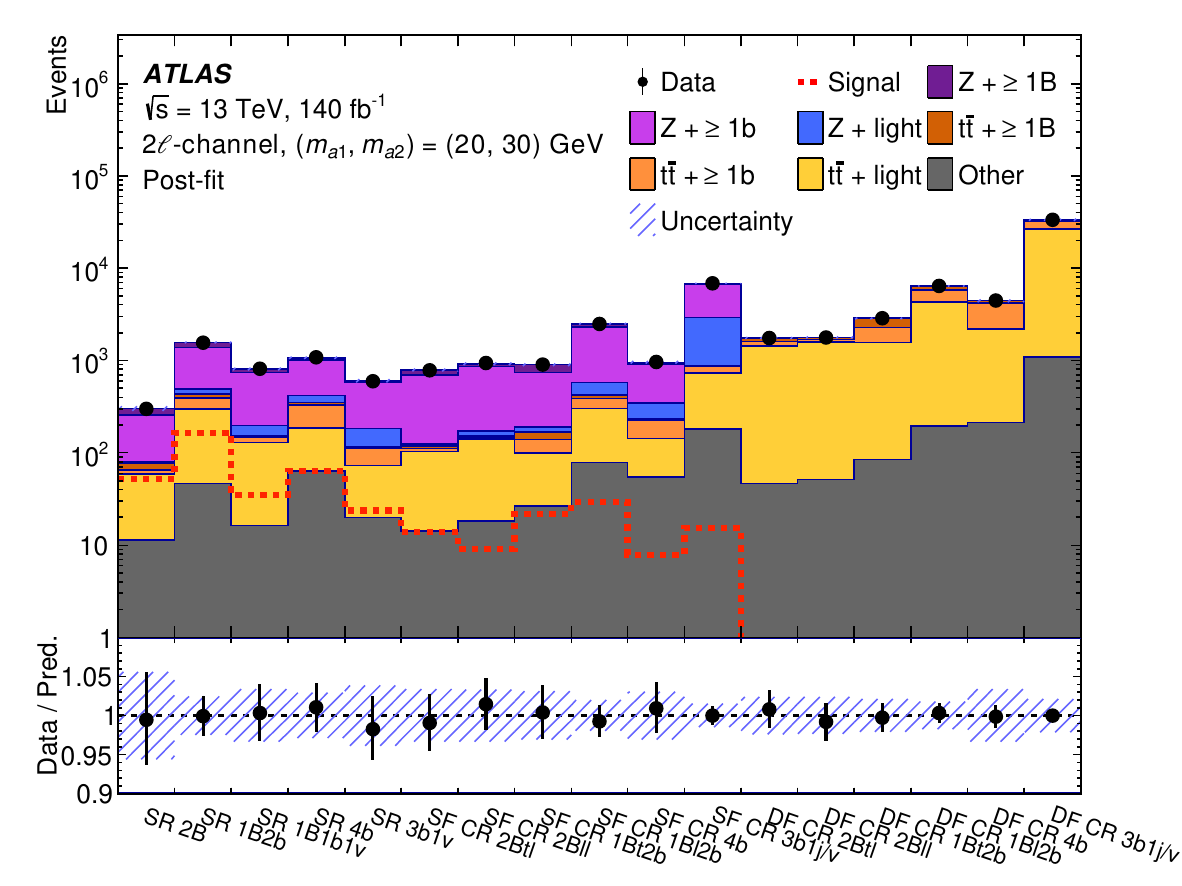}}
\subfloat[][]{\label{a1a2_3080}
\includegraphics[width=0.5\textwidth]{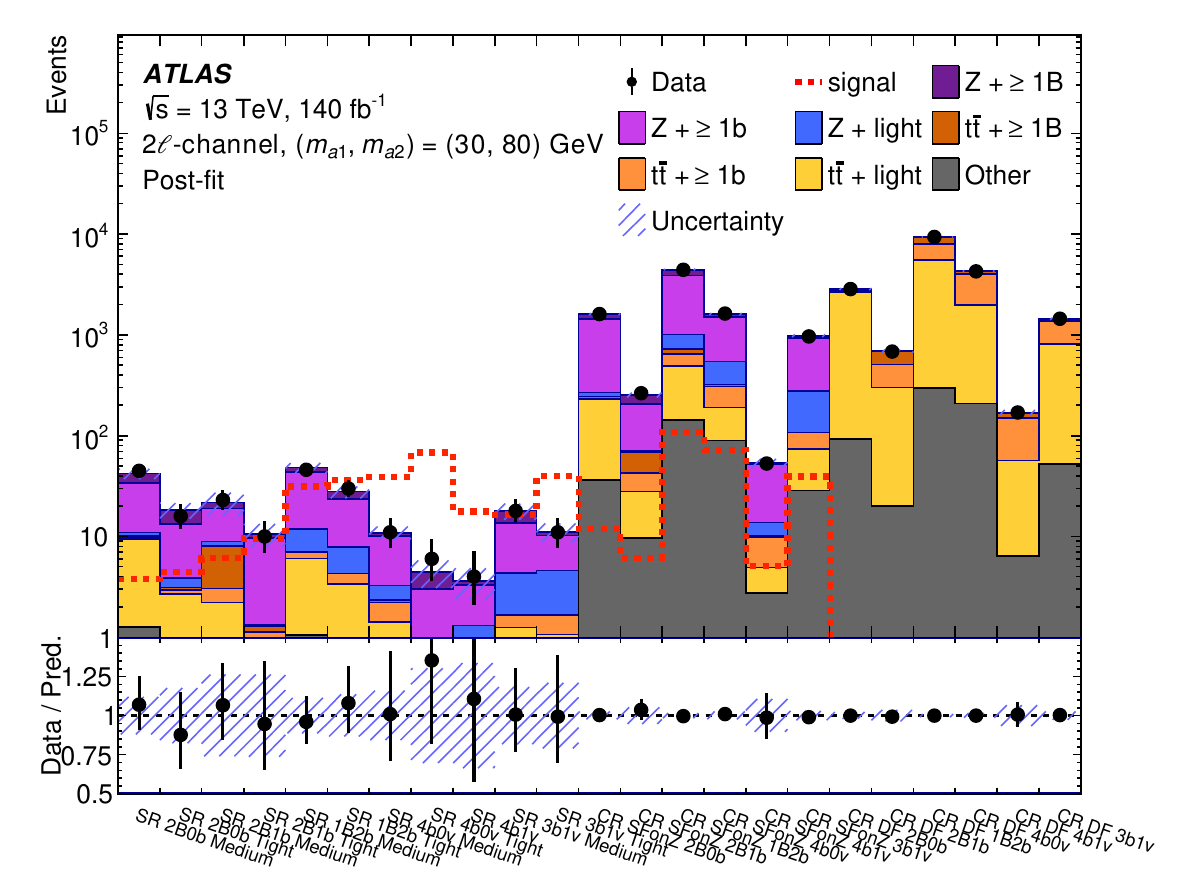}}
\caption{Observed yields for the signal and control regions in the $2\ell$ channel after the background-only fit to data for signals with two new particles for the case \protect\subref{ma1ma2_2030} $(m_{a_1}, m_{a_2})=(20,30)~\GeV$, which gives rise to a signal with $4b$ in the final state, and \protect\subref{a1a2_3080} $(m_{a_1}, m_{a_2})=(30,80)~\GeV$, which gives rise to a signal with $6b$ in the final state. The signal is overlaid (dashed line) assuming a branching fraction $\mathcal{B}(H \rightarrow a_1a_2 \rightarrow 4b/6b) = 1$ and the SM rate of production of Higgs bosons. The lower panels show the ratio of the observed to the estimated SM background. The hashed area represents the total uncertainty of the background. The category labeled ``Other'' includes processes with small contributions to the total yield (diboson, $\ttZ$, and $\ttW$).}
\label{fig:results_a1a2_yields}
\end{figure}

\Tab{\ref{tab:systematics_impacts}} shows the impact of dominant sources of systematic uncertainties in the observed upper limit for different signal mass hypotheses. The impact is estimated by the difference between the expected upper limit when only a group of uncertainties is considered and the case where no systematic uncertainty is considered. The $B$-jet identification efficiency uncertainty is an important source of systematic uncertainty at low value of masses due to the difficulty in measuring the identification efficiency of merged $B$-jets with low mass and low $\pt$. Modeling uncertainties are also important due to the challenges of modeling associated heavy flavor (defined as $b$-quarks, unless explicitly mentioned) production of \Zjets and \ttbar. \Fig{\ref{fig:comb_result}} compares the observed upper limits with the expected limits assuming the median value of $\tilde{q}_{\mu}$ under the background-only hypothesis for the $H \rightarrow 2a \rightarrow 4b$ process. The $0\ell$ and $2\ell$ channels are combined in the mass range of $12~\GeV < m_a<30~\GeV$. This is the first time the $0\ell$ channel is explored and provides an approximately 15\% improvement on the limit at the $12$ and $16~\GeV$ mass points, which is the region targeted by the channel. \Fig{\ref{fig:limits_a1a2_6b_xsec_unblinded}} shows the corresponding upper limits for the $H\rightarrow a_{1}a_{2} \rightarrow 4b$ and $6b$ cases. The values of $m_a$ used for signal simulation and for hypothesis testing were chosen so that the difference between neighboring points is much smaller than the mass resolution of the reconstructed $a$-boson and Higgs boson, which varies between about 10~\GeV and 50~\GeV depending on the value of $m_a$ and the type of $b$-objects used in the selected quadruplet. Interpolation tests are performed by calculating expected limits using Asimov samples~\cite{Cowan:2010js} with different values of generated $m_a$ than the one being tested. Due to the limited mass resolution, only three points are used when testing models with two different $a$-bosons for the $H\rightarrow a_{1}a_{2} \rightarrow 4b$ case. The phase-space region satisfying $m_{a_1} + m_{a_2} < m_H$, $m_{a_1} > m_{a_2}$, and $m_{a_1}, m_{a_2} \geq 12~\GeV$ is a narrow wedge that is fully probed by the three mass points used. Four points are similarly chosen to test models with two different $a$-bosons for the $H\rightarrow a_{1}a_{2} \rightarrow 6b$ case.

The limits in the low-mass regime ($m_a < 20\,\GeV$), where the $b\bar{b}$ system is collimated into a single object, are mostly driven by the 2B category, while limits in the high-mass regime ($m_a > 30\,\GeV$) are mostly driven by the 4b category. The largest deviations correspond to the $H \rightarrow 2a/a_{1}a_{2} \rightarrow 4b$ case in the $2\ell$ channel, with a local (global) significance of 2.83 (2.04) for $m_a = 25~\GeV$ and 3.28 (2.57) for $(m_{a_1},m_{a_2}) = (50,70)~\GeV$. The global $p$-value is computed using the formula from Ref.~\cite{Gross:2010qma}, where a reference $p$-value of 0.5 is used to count the crossings across all tested values of the mass of the new scalar $m_a$. The value is then extrapolated to the observed local $p$-value. For models with different $m_{a_1}$ and $m_{a_2}$, the method is extended using the Euler characteristic instead of up crossings, as described in Ref~\cite{Vitells:2011da}.

This search increases the sensitivity of previous ATLAS searches for $H \rightarrow 2a \rightarrow 4b$~\cite{HIGG-2016-01,HIGG-2017-05,HDBS-2018-47} by a factor of about 5--15 depending on the signal mass hypothesis by including a larger data sample, and by using novel identification algorithms for low-mass merged $B$-jets and including soft secondary vertices reconstructed outside of jets. These techniques enhance the sensitivity at low $m_a$ and to high multiplicity final states. The analysis considers the production mode $pp\rightarrow ZH (Z\rightarrow \nu\nu)$ for the first time in a search for exotic decays of the Higgs boson. In addition, this is also the first search for exotic decays of the Higgs boson in the final state $H \rightarrow a_1a_2 \rightarrow 4b/6b$ over the mass range $12 < m_{a} < 60~\GeV$.

\begin{table}[!tbp]
\caption{The upper limit on $\sigratio \times \mathcal{B}(H \rightarrow 2a \rightarrow 4b)$ is determined for the cases without systematic uncertainties (stat-only), when all uncertainties are applied (observed) and when only a group of systematic uncertainties is applied. The larger the difference between the limit obtained for a group of uncertainties and the stat-only limit, the more important the uncertainty group is for the final result. The entries with ``-'' change by less than 1\% relative to the stat-only limit. Impacts are shown for the $m_a = 12~\GeV$ and $25~\GeV$ mass points for the $0\ell$ channel, the $2\ell$ channel, and the combination of the $0\ell$ and $2\ell$ channels. The grouping ``Tracks'' includes large-$R$ track-jets. The trigger uncertainties are included in the groupings of corresponding objects.}
\centering
\begin{tabular}{p{0.3\linewidth}|p{0.08\linewidth}p{0.08\linewidth}p{0.08\linewidth}|p{0.08\linewidth}p{0.08\linewidth}p{0.08\linewidth}}
\toprule
\multirow{3}{*}{Uncertainty source} & \multicolumn{6}{c}{Upper limit on $\sigratio \times \mathcal{B}(H \rightarrow 2a \rightarrow 4b)$} \\ \cline{2-7}
& \multicolumn{3}{c|}{$m_a = 12~\GeV$} & \multicolumn{3}{c}{$m_a = 25~\GeV$} \\ \cline{2-7}
& \textbf{0$\ell$} & \textbf{2$\ell$} & \textbf{0$\ell$ \& 2$\ell$} & \textbf{0$\ell$} & \textbf{2$\ell$} & \textbf{0$\ell$ \& 2$\ell$} \\
\midrule
\textbf{Stat-only limit} & 0.060 & 0.029 & 0.025 & 0.35 & 0.19 & 0.19 \\
\midrule
\textbf{Observed limit} & 0.12 & 0.039 & 0.034 & 0.89 & 0.21 & 0.21 \\
\midrule
\hspace{3 mm}\textbf{MC statistics} & 0.065 & 0.030 & 0.026 & 0.37 & -- & 0.19 \\
\midrule
\hspace{3 mm}\textbf{Systematics} & 0.12 & 0.038 & 0.033 & 0.88 & 0.21 & 0.21 \\
\midrule
\hspace{6 mm}\textbf{Experimental} & 0.076 & 0.032 & 0.028 & 0.51 & 0.20 & 0.20 \\
\hspace{9 mm}Luminosity and \pileup & 0.062 & -- & -- & 0.36 & -- & -- \\
\hspace{9 mm}\dlr tagging & -- & -- & -- & 0.35 & 0.20 & 0.20 \\
\hspace{9 mm}\DEXTER tagging & 0.067 & 0.030 & 0.026 & 0.39 & -- & -- \\
\hspace{9 mm}Electrons & -- & -- & -- & -- & -- & -- \\
\hspace{9 mm}Muons & -- & -- & -- & -- & -- & -- \\
\hspace{9 mm}Jets & 0.063 & 0.029 & 0.025 & 0.39 & -- & -- \\
\hspace{9 mm}\met & 0.063 & -- & 0.025 & 0.37 & -- & -- \\
\hspace{9 mm}Tracks & 0.066 & 0.029 & 0.026 & 0.42 & -- & 0.20 \\
\hspace{9 mm}Soft-v &   & -- & -- &   & -- & -- \\
\midrule
\hspace{6 mm}\textbf{Modeling} & 0.088 & 0.034 & 0.030 & 0.60 & 0.20 & 0.20 \\
\hspace{9 mm}Top ($t\bar{t}$ and single top) & 0.065 & -- & 0.025 & 0.41 & -- & 0.20 \\
\hspace{9 mm}$W/Z$ + jets & 0.071 & 0.033 & 0.029 & 0.45 & 0.19 & 0.20 \\
\hspace{9 mm}Signal & 0.066 & 0.029 & 0.026 & 0.37 & 0.20 & 0.20 \\
\hspace{9 mm}QCD & 0.063 &   & -- & 0.35 &   & -- \\
\bottomrule
\end{tabular}
\label{tab:systematics_impacts}
\end{table}

\begin{figure}[!tbp]
\centering
\includegraphics[width=0.7\textwidth]{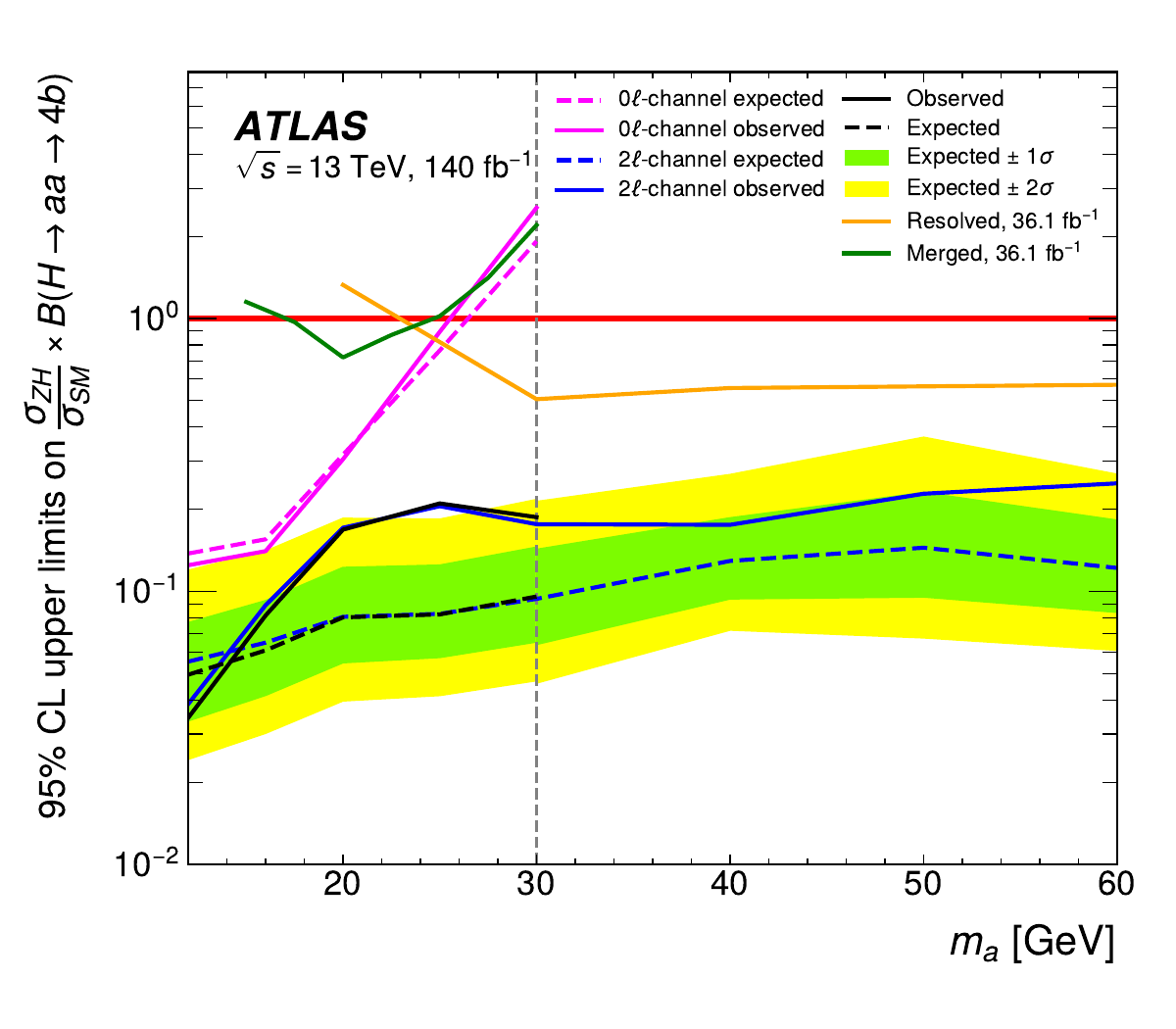}
\caption{The observed (solid) 95\% CL upper limits on $\sigratio \times \mathcal{B}(H \rightarrow 2a \rightarrow 4b)$ as a function of $m_a$ and the expected (dashed) limits under the background-only hypothesis showing the $0\ell$ and $2\ell$ separately and their combination. The vertical dashed line at 30~\GeV indicates the transition point and for higher masses only the $2\ell$ channel is used. For the combined result, the inner and outer shaded bands show the $\pm 1\sigma$ and $\pm 2\sigma$ uncertainties of the expected limits. The mass hypothesis $m_a$ is probed between 12 and 60~\GeV. The previous results from the ATLAS Collaboration using a partial \RunTwo data sample focusing on the resolved~\cite{HIGG-2017-05} and merged~\cite{HDBS-2018-47} regimes are also shown.}
\label{fig:comb_result}
\end{figure}

\begin{figure}[!tbp]
\centering
\subfloat[][]{\label{limit_a1a2_4b}
\includegraphics[width=0.49\textwidth]{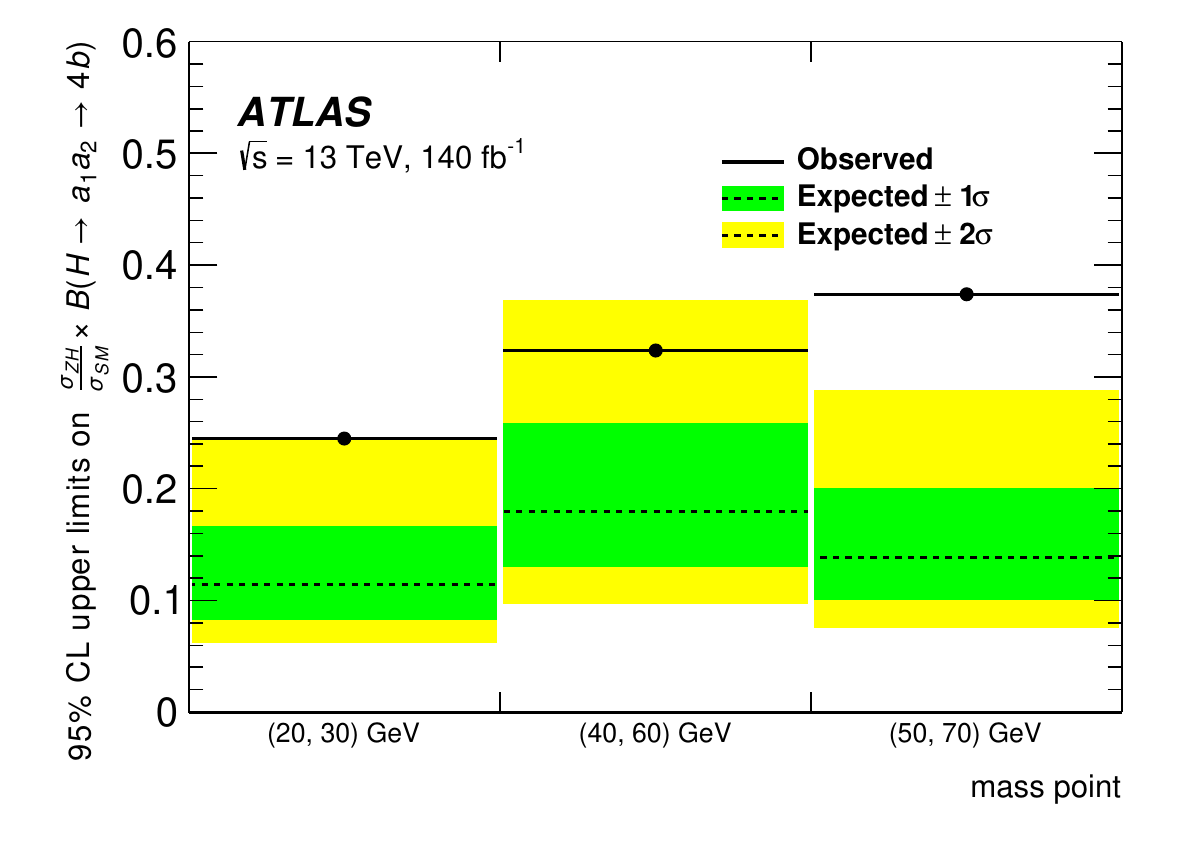}}
\subfloat[][]{\label{limit_a1a2_6b}
\includegraphics[width=0.49\textwidth]{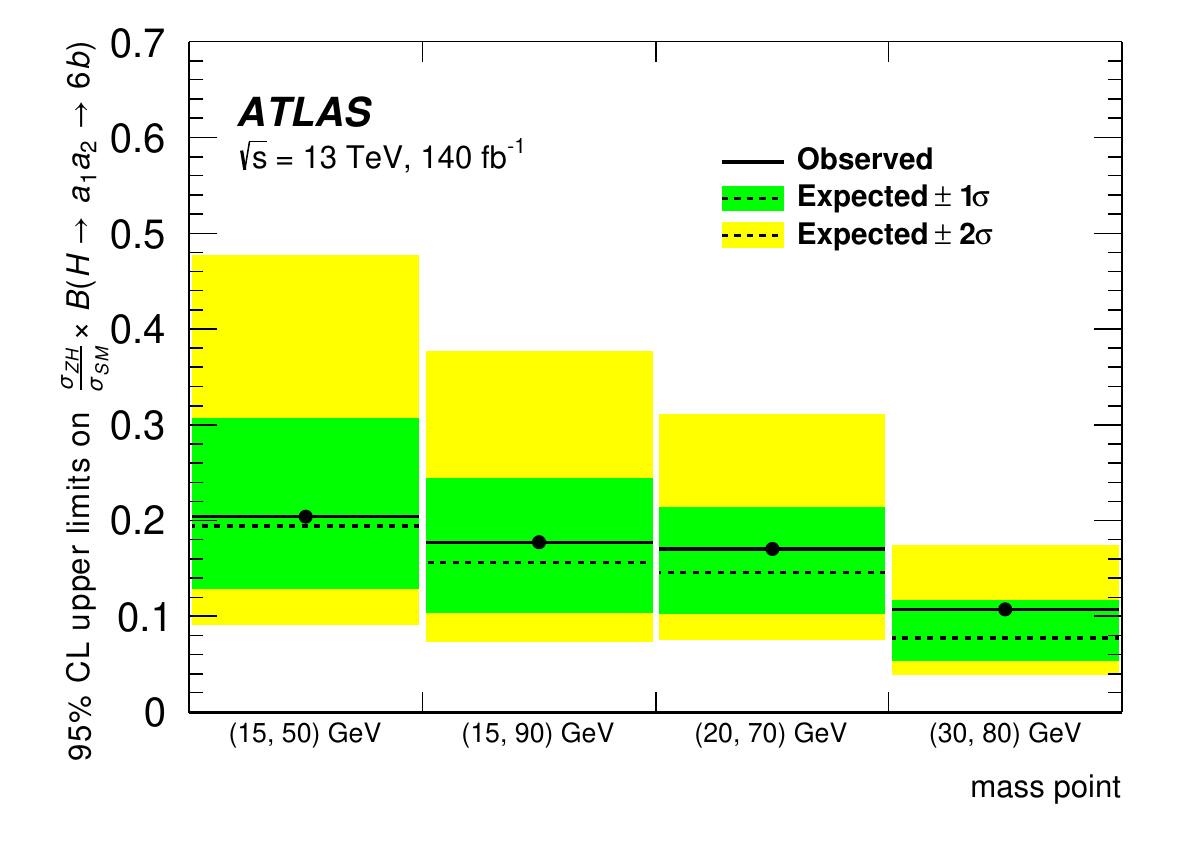}}
\caption{The observed (solid) 95\% CL upper limits on \protect\subref{limit_a1a2_4b} $\sigratio \times \mathcal{B}(H\rightarrow a_{1}a_{2} \rightarrow 4b)$ and \protect\subref{limit_a1a2_6b} $\sigratio \times \mathcal{B}(H\rightarrow a_{1}a_{2} \rightarrow 6b)$ for different signal mass hypotheses $(m_{a_{1}}, m_{a_{2}})$. The inner and outer shaded bands show the $\pm 1\sigma$ and $\pm 2\sigma$ uncertainties of the expected limits.}
\label{fig:limits_a1a2_6b_xsec_unblinded}
\end{figure}

\FloatBarrier

\section{Conclusion}

This paper presents a search for exotic decays of the Higgs boson into a pair of scalar particles, $H\rightarrow~2a$,
where each $a$-boson decays into two $b$-quarks. It also considers the case of two new $a$-bosons $H~\rightarrow~a_1a_2~\rightarrow~4b$ and $H~\rightarrow~a_1a_2~\rightarrow~3a_1~\rightarrow~6b$. The analysis targets the Higgs boson produced in association with a $Z$ boson ($ZH$) and is performed using the 140 \ifb\ of proton--proton collision data at a center-of-mass energy of $13$ \TeV recorded by the ATLAS detector at the LHC between 2015 and 2018.
The search makes use of heavy-flavor-tagging techniques to target collimated $a\rightarrow b\bar{b}$ decays and soft secondary vertices reconstructed outside of jets. The analysis categorizes events depending on the number of leptons and the strategy for identifying heavy-flavor jets: one or two $B$-jets, three or four $b$-jets, and one soft secondary vertex.

The largest deviations correspond to the $H \rightarrow 2a/a_{1}a_{2} \rightarrow 4b$ case in the $2\ell$ channel, with a local (global) significance of 2.83 (2.04) for $m_a = 25~\GeV$ and 3.28 (2.57) for $(m_{a_1},m_{a_2}) = (50,70)~\GeV$. The global significance is estimated independently for each model. As no significant excess above the SM background expectation is observed, upper limits at 95\% CL are set of $4\% - 25\%$ on $\sigratio \times \mathcal{B}(H\rightarrow 2a \rightarrow 4b)$ for scalar mass values in the range of $12~\GeV \leq m_{a} \leq 60~\GeV$, of $24\% - 38\%$ on $\sigratio \times \mathcal{B}(H \rightarrow a_1a_2 \rightarrow 4b)$, and of $10\% - 20\%$ on $\sigratio \times \mathcal{B}(H \rightarrow a_1a_2 \rightarrow 3a_1 \rightarrow 6b)$ for masses in the ranges of $15\leq m_{a_1}\leq 50~\GeV$ and $30\leq m_{a_2}\leq 80~\GeV$.  These results contribute to the broad program of searches for $H\rightarrow 2a/a_1a_2$ decays.

\section*{Acknowledgements}

%
%

%
%

We thank CERN for the very successful operation of the LHC and its injectors, as well as the support staff at
CERN and at our institutions worldwide without whom ATLAS could not be operated efficiently.

The crucial computing support from all WLCG partners is acknowledged gratefully, in particular from CERN, the ATLAS Tier-1 facilities at TRIUMF/SFU (Canada), NDGF (Denmark, Norway, Sweden), CC-IN2P3 (France), KIT/GridKA (Germany), INFN-CNAF (Italy), NL-T1 (Netherlands), PIC (Spain), RAL (UK) and BNL (USA), the Tier-2 facilities worldwide and large non-WLCG resource providers. Major contributors of computing resources are listed in Ref.~\cite{ATL-SOFT-PUB-2025-001}.

We gratefully acknowledge the support of ANPCyT, Argentina; YerPhI, Armenia; ARC, Australia; BMWFW and FWF, Austria; ANAS, Azerbaijan; CNPq and FAPESP, Brazil; NSERC, NRC and CFI, Canada; CERN; ANID, Chile; CAS, MOST and NSFC, China; Minciencias, Colombia; MEYS CR, Czech Republic; DNRF and DNSRC, Denmark; IN2P3-CNRS and CEA-DRF/IRFU, France; SRNSFG, Georgia; BMFTR, HGF and MPG, Germany; GSRI, Greece; RGC and Hong Kong SAR, China; ICHEP and Academy of Sciences and Humanities, Israel; INFN, Italy; MEXT and JSPS, Japan; CNRST, Morocco; NWO, Netherlands; RCN, Norway; MNiSW, Poland; FCT, Portugal; MNE/IFA, Romania; MSTDI, Serbia; MSSR, Slovakia; ARIS and MVZI, Slovenia; DSI/NRF, South Africa; MICIU/AEI, Spain; SRC and Wallenberg Foundation, Sweden; SERI, SNSF and Cantons of Bern and Geneva, Switzerland; NSTC, Taipei; TENMAK, T\"urkiye; STFC/UKRI, United Kingdom; DOE and NSF, United States of America.

Individual groups and members have received support from BCKDF, CANARIE, CRC and DRAC, Canada; CERN-CZ, FORTE and PRIMUS, Czech Republic; COST, ERC, ERDF, Horizon 2020, ICSC-NextGenerationEU and Marie Sk{\l}odowska-Curie Actions, European Union; Investissements d'Avenir Labex, Investissements d'Avenir Idex and ANR, France; DFG and AvH Foundation, Germany; Herakleitos, Thales and Aristeia programmes co-financed by EU-ESF and the Greek NSRF, Greece; BSF-NSF and MINERVA, Israel; NCN and NAWA, Poland; La Caixa Banking Foundation, CERCA Programme Generalitat de Catalunya and PROMETEO and GenT Programmes Generalitat Valenciana, Spain; G\"{o}ran Gustafssons Stiftelse, Sweden; The Royal Society and Leverhulme Trust, United Kingdom.

In addition, individual members wish to acknowledge support from Armenia: Yerevan Physics Institute (FAPERJ); CERN: European Organization for Nuclear Research (CERN DOCT); Chile: Agencia Nacional de Investigaci\'on y Desarrollo (FONDECYT 1230812, FONDECYT 1240864, Fondecyt 3240661); China: Chinese Ministry of Science and Technology (MOST-2023YFA1605700, MOST-2023YFA1609300), National Natural Science Foundation of China (NSFC - 12175119, NSFC 12275265); Czech Republic: Czech Science Foundation (GACR - 24-11373S), Ministry of Education Youth and Sports (ERC-CZ-LL2327, FORTE CZ.02.01.01/00/22\_008/0004632), PRIMUS Research Programme (PRIMUS/21/SCI/017); EU: H2020 European Research Council (ERC - 101002463); European Union: European Research Council (BARD No. 101116429, ERC - 948254, ERC 101089007), European Regional Development Fund (SMASH COFUND 101081355, SLO ERDF), Horizon 2020 Framework Programme (MUCCA - CHIST-ERA-19-XAI-00), European Union, Future Artificial Intelligence Research (FAIR-NextGenerationEU PE00000013), Horizon 2020 (EuroHPC - EHPC-DEV-2024D11-051), Italian Center for High Performance Computing, Big Data and Quantum Computing (ICSC, NextGenerationEU); France: Agence Nationale de la Recherche (ANR-21-CE31-0013, ANR-21-CE31-0022, ANR-22-EDIR-0002); Germany: Baden-Württemberg Stiftung (BW Stiftung-Postdoc Eliteprogramme), Deutsche Forschungsgemeinschaft (DFG - 469666862, DFG - CR 312/5-2); China: Research Grants Council (GRF); Italy: Istituto Nazionale di Fisica Nucleare (ICSC, NextGenerationEU), Ministero dell'Università e della Ricerca (NextGenEU 153D23001490006 M4C2.1.1, NextGenEU I53D23000820006 M4C2.1.1, NextGenEU I53D23001490006 M4C2.1.1); Japan: Japan Society for the Promotion of Science (JSPS KAKENHI JP22H01227, JSPS KAKENHI JP22H04944, JSPS KAKENHI JP22KK0227, JSPS KAKENHI JP23KK0245, JSPS KAKENHI JP24K23939); Norway: Research Council of Norway (RCN-314472); Poland: Ministry of Science and Higher Education (IDUB AGH, POB8, D4 no 9722), Polish National Science Centre (NCN 2021/42/E/ST2/00350, NCN OPUS 2023/51/B/ST2/02507, NCN OPUS nr 2022/47/B/ST2/03059, NCN UMO-2019/34/E/ST2/00393, UMO-2020/37/B/ST2/01043, UMO-2022/47/O/ST2/00148, UMO-2023/49/B/ST2/04085, UMO-2023/51/B/ST2/00920, UMO-2024/53/N/ST2/00869); Portugal: Foundation for Science and Technology (FCT); Spain: Generalitat Valenciana (Artemisa, FEDER, IDIFEDER/2018/048), Ministry of Science and Innovation (MCIN \& NextGenEU PCI2022-135018-2, MICIN \& FEDER PID2021-125273NB, RYC2019-028510-I, RYC2020-030254-I, RYC2021-031273-I, RYC2022-038164-I); Sweden: Carl Trygger Foundation (Carl Trygger Foundation CTS 22:2312), Swedish Research Council (Swedish Research Council 2023-04654, VR 2021-03651, VR 2022-03845, VR 2022-04683, VR 2023-03403, VR 2024-05451), Knut and Alice Wallenberg Foundation (KAW 2018.0458, KAW 2022.0358, KAW 2023.0366); Switzerland: Swiss National Science Foundation (SNSF - PCEFP2\_194658); United Kingdom: Leverhulme Trust (Leverhulme Trust RPG-2020-004), Royal Society (NIF-R1-231091); United States of America: U.S. Department of Energy (ECA DE-AC02-76SF00515), Neubauer Family Foundation.

%
%


%
\clearpage
\printbibliography
\clearpage
 
\begin{flushleft}
\hypersetup{urlcolor=black}
{\Large The ATLAS Collaboration}

\bigskip

\AtlasOrcid[0000-0002-6665-4934]{G.~Aad}$^\textrm{\scriptsize 104}$,
\AtlasOrcid[0000-0001-7616-1554]{E.~Aakvaag}$^\textrm{\scriptsize 17}$,
\AtlasOrcid[0000-0002-5888-2734]{B.~Abbott}$^\textrm{\scriptsize 123}$,
\AtlasOrcid[0000-0002-0287-5869]{S.~Abdelhameed}$^\textrm{\scriptsize 119a}$,
\AtlasOrcid[0000-0002-1002-1652]{K.~Abeling}$^\textrm{\scriptsize 55}$,
\AtlasOrcid[0000-0001-5763-2760]{N.J.~Abicht}$^\textrm{\scriptsize 49}$,
\AtlasOrcid[0000-0002-8496-9294]{S.H.~Abidi}$^\textrm{\scriptsize 30}$,
\AtlasOrcid[0009-0003-6578-220X]{M.~Aboelela}$^\textrm{\scriptsize 45}$,
\AtlasOrcid[0000-0002-9987-2292]{A.~Aboulhorma}$^\textrm{\scriptsize 36e}$,
\AtlasOrcid[0000-0001-5329-6640]{H.~Abramowicz}$^\textrm{\scriptsize 157}$
% ...
\bigskip
\\

% ...
$^{17}$Department for Physics and Technology, University of Bergen, Bergen; Norway.\\
% ...
$^{30}$Physics Department, Brookhaven National Laboratory, Upton NY; United States of America.\\
% ...
$^{36}$$^{(a)}$Facult\'e des Sciences Ain Chock, Universit\'e Hassan II de Casablanca;$^{(b)}$Facult\'{e} des Sciences, Universit\'{e} Ibn-Tofail, K\'{e}nitra;$^{(c)}$Facult\'e des Sciences Semlalia, Universit\'e Cadi Ayyad, LPHEA-Marrakech;$^{(d)}$LPMR, Facult\'e des Sciences, Universit\'e Mohamed Premier, Oujda;$^{(e)}$Facult\'e des sciences, Universit\'e Mohammed V, Rabat;$^{(f)}$Institute of Applied Physics, Mohammed VI Polytechnic University, Ben Guerir; Morocco.\\
% ...
$^{45}$Physics Department, Southern Methodist University, Dallas TX; United States of America.\\
% ...
$^{49}$Fakult\"{a}t Physik , Technische Universit{\"a}t Dortmund, Dortmund; Germany.\\
% ...
$^{55}$II. Physikalisches Institut, Georg-August-Universit\"{a}t G\"ottingen, G\"ottingen; Germany.\\
% ...
$^{104}$CPPM, Aix-Marseille Universit\'e, CNRS/IN2P3, Marseille; France.\\
% ...
$^{119}$$^{(a)}$New York University Abu Dhabi, Abu Dhabi;$^{(b)}$United Arab Emirates University, Al Ain; United Arab Emirates.\\
% ...
$^{123}$Homer L. Dodge Department of Physics and Astronomy, University of Oklahoma, Norman OK; United States of America.\\
% ...
$^{157}$Raymond and Beverly Sackler School of Physics and Astronomy, Tel Aviv University, Tel Aviv; Israel.\\
% ...
$^{a}$ Also at Affiliated with an institute formerly covered by a cooperation agreement with CERN.\\
$^{b}$ Also at An-Najah National University, Nablus; Palestine.\\
$^{c}$ Also at Borough of Manhattan Community College, City University of New York, New York NY; United States of America.\\
$^{d}$ Also at Center for Interdisciplinary Research and Innovation (CIRI-AUTH), Thessaloniki; Greece.\\
$^{e}$ Also at Centre of Physics of the Universities of Minho and Porto (CF-UM-UP); Portugal.\\
$^{f}$ Also at CERN, Geneva; Switzerland.\\
% ...

\end{flushleft}


%

%
%

%

%
%
%
%

\end{document}